\documentclass[11pt]{article}

\usepackage[
  a4paper,
  left=20mm,
  right=20mm,
  top=25mm,
  bottom=25mm
]{geometry}

\usepackage[T1]{fontenc}
\usepackage[utf8]{inputenc}
\DeclareUnicodeCharacter{0301}{\'{}}  % combining acute accent
\usepackage{mathpazo}          % Palatino for body text
\usepackage{helvet}            % Helvetica for sans-serif
\usepackage{courier}           % Courier for monospace
\usepackage{setspace}
\usepackage{xcolor}
\definecolor{nmiblue}{RGB}{0,109,119}
\definecolor{headinggray}{RGB}{80,80,80}

\usepackage{titlesec}
\titleformat{\section}
  {\Large\bfseries}
  {}
  {0pt}
  {}
\titlespacing*{\section}{0pt}{3ex plus 1ex minus 0.5ex}{1.2ex plus 0.3ex}

\titleformat{\subsection}
  {\large\bfseries}
  {}
  {0pt}
  {}
\titlespacing*{\subsection}{0pt}{2.5ex plus 0.8ex minus 0.3ex}{0.8ex plus 0.2ex}

\titleformat{\subsubsection}
  {\normalsize\bfseries}
  {}
  {0pt}
  {}
\titlespacing*{\subsubsection}{0pt}{2ex plus 0.5ex minus 0.2ex}{0.5ex plus 0.1ex}

\usepackage{graphicx}
\usepackage{float}
\let\oldincludegraphics\includegraphics
\renewcommand{\includegraphics}[2][]{\noindent\oldincludegraphics[#1]{#2}}
\usepackage{caption}
\usepackage{booktabs}
\usepackage{longtable}
\usepackage{array}
\usepackage{calc}

\usepackage{etoolbox}
\AtBeginEnvironment{longtable}{\footnotesize}
\AtBeginEnvironment{figure}{\setlength{\parindent}{0pt}}

\usepackage{enumitem}
\setlist{nosep, leftmargin=1.5em}

\usepackage{hyperref}
\usepackage{xurl}
\hypersetup{
  colorlinks=true,
  linkcolor=nmiblue,
  citecolor=nmiblue,
  urlcolor=nmiblue,
  pdfborder={0 0 0}
}

\usepackage{fancyhdr}
\newcommand{\pandocbounded}[1]{#1}
\newcommand{\citeproctext}{}
\newcommand{\citeproc}[2]{\begingroup\def\citeproctext{#2}\hyperlink{#1}{\citeproctext}\endgroup}
\newlength{\cslhangindent}
\newlength{\csllabelwidth}
\newenvironment{CSLReferences}[2]
 {\begin{list}{}%
  {\setlength{\itemindent}{0pt}
   \setlength{\leftmargin}{0pt}
   \setlength{\parsep}{0pt}
   \setlength{\itemsep}{0.15em}
   \setlength{\topsep}{0pt}
   \ifnum#1=1
    \setlength{\leftmargin}{\cslhangindent}
    \setlength{\itemindent}{-\cslhangindent}
   \fi}
  \small
  \def\bibitem[##1]##2{\item\leavevmode\hypertarget{##2}{}}}
 {\end{list}}

\newcommand{\CSLLeftMargin}[1]{\parbox[t]{\csllabelwidth}{#1}}
\newcommand{\CSLRightInline}[1]{\parbox[t]{\dimexpr\linewidth-\csllabelwidth}{#1}\break}

\begin{document}

% ── Title block ───────────────────────────────────────────────
\begin{center}
{\LARGE\bfseries
Large language models converge on competitive rationality but diverge on cooperation across providers and generations\par}

\vspace{0.8em}

{\large
Felipe M. Affonso\par}

\vspace{0.3em}

{\small\itshape
Spears School of Business, Oklahoma State University, Stillwater, USA\par}

\vspace{0.3em}

{\footnotesize
Correspondence: felipe.affonso@okstate.edu}

\end{center}

\vspace{1em}

% ── Body ──────────────────────────────────────────────────────
\subsection{Abstract}\label{abstract}

\textbf{As language models are deployed as autonomous agents that negotiate, cooperate, and compete on behalf of human principals, their strategic dispositions acquire direct economic consequences. Here we show, across 51,906 game-theoretic trials generating 826,990 strategic decisions from 25 large language models spanning seven developers and 38 canonical games, that models converge on competitive and coordination behaviour (coefficient of variation 0.06 for coordination, 0.11 for strategic depth) while diverging 48-fold on cooperation, from 1.5 per cent (GPT-5 Nano) to 71.5 per cent (Claude Opus 4.6). Provider identity is the dominant predictor of cooperative disposition, and this divergence is generationally unstable: OpenAI cooperation fell from 50.3 to 1.5 per cent across four model generations while Google cooperation rose from 8.3 to 56.8 per cent. Endgame analysis reveals that Anthropic frontier models sustain 57 per cent cooperation in the final round of finitely repeated games, where backward induction predicts zero, while the newest Google models cooperate throughout but universally defect when punishment becomes impossible. These strategic personalities are shaped by training pipelines, shift unpredictably across model versions, and cannot be inferred from capability benchmarks, yet they determine the cooperative outcomes of every economic interaction these models mediate. The complete dataset and an interactive explorer for the data are publicly available at \url{https://felipemaffonso.github.io/strategic-personalities/}.}

\vspace{0.5em}

\textbf{Keywords:} machine behaviour, game theory, AI alignment, strategic interaction, large language models

\subsection{Main}\label{main}

Language models are increasingly deployed as autonomous agents that negotiate contracts, procure supplies, and resolve disputes on behalf of human principals. Perplexity executes purchases through its shopping assistant, Amazon's Rufus serves hundreds of millions of active customers, and multi-agent marketplaces are emerging in which AI agents negotiate and transact on both sides of the market\textsuperscript{\citeproc{ref-bansal2025magentic}{1},\citeproc{ref-allouah2025what}{2}}. When humans delegate to AI agents, the outcomes change: delegation increases dishonest behaviour\textsuperscript{\citeproc{ref-kobis2025delegation}{3}}, and fine-tuning on narrow tasks can produce broad misalignment across unrelated domains\textsuperscript{\citeproc{ref-betley2026misalignment}{4}}. As these delegations scale, the strategic behaviour embedded in the underlying model determines the economic outcomes that its principal receives, yet the strategic dispositions that govern these interactions remain poorly understood\textsuperscript{\citeproc{ref-rahwan2019machine}{5}}. Early work simulating economic agents with language models has demonstrated that these dispositions can be elicited through game-theoretic paradigms\textsuperscript{\citeproc{ref-horton2023large}{6}}, but systematic characterisation across providers and model generations is lacking.

Game theory provides the canonical framework for characterising strategic behaviour under incentive structures\textsuperscript{\citeproc{ref-camerer2003behavioral}{7}}. A rich experimental tradition has applied game-theoretic paradigms to human subjects, documenting systematic deviations from Nash equilibrium: humans cooperate in prisoner's dilemmas, reject unfair offers in ultimatum games, and punish norm violators at personal cost\textsuperscript{\citeproc{ref-fehr1999theory}{8}--\citeproc{ref-nowak2006evolutionary}{11}}. Recent work has extended these paradigms to language models, revealing both convergence toward rational play and sensitivity to framing, prompting, and model generation. Models deviate from Nash equilibria in ways that resemble human biases, cooperating more and offering more generously than equilibrium predicts\textsuperscript{\citeproc{ref-brookins2024playing}{12},\citeproc{ref-brookins2024strategic}{13}}, and cooperation rates shift with structured prompting\textsuperscript{\citeproc{ref-akata2025playing}{14}}, vary between model versions\textsuperscript{\citeproc{ref-fan2024canllm}{15},\citeproc{ref-fontana2025nicer}{16}}, and fail to converge in multi-game benchmarks\textsuperscript{\citeproc{ref-huang2025gamabench}{17}--\citeproc{ref-mao2025alympics}{19}}.

These initial findings have been deepened in several directions. Single-model studies have shown that social preferences can be elicited from language model play in economic games\textsuperscript{\citeproc{ref-guo2024gpt}{20}} and that personality-trait prompts modulate cooperative strategies in evolutionary simulations\textsuperscript{\citeproc{ref-suzuki2024evolutionary}{21}}, while moral decision-making experiments reveal that RLHF fine-tuning amplifies cognitive biases relative to base models, including an omission bias of 45 percentage points compared with 5 percentage points in human subjects\textsuperscript{\citeproc{ref-cheung2025cognitive}{22}}. Cooperative behaviour in language models also varies with the stakes of the interaction\textsuperscript{\citeproc{ref-more2026stake}{23}} and is sensitive to linguistic and cultural context\textsuperscript{\citeproc{ref-fairgame2025}{24}}, while randomness in strategic decisions raises methodological questions about the interpretation of mixed strategies\textsuperscript{\citeproc{ref-randomness2025llm}{25},\citeproc{ref-shin2024emergence}{26}}. In the algorithmic pricing domain, Q-learning agents autonomously learn collusive strategies\textsuperscript{\citeproc{ref-calvano2020artificial}{27},\citeproc{ref-calvano2021algorithmic}{28}}, establishing that emergent strategic dispositions in AI systems carry economic consequences even without explicit programming.

Although these contributions demonstrate that language models possess strategic dispositions with economic relevance, they share two limitations that constrain the conclusions that can be drawn. First, they test small model samples (three to seven models, typically from one or two providers), confounding model behaviour with provider-level training choices. Second, they focus on narrow game sets (prisoner's dilemma variants and two to four additional games), which cannot distinguish domain-specific quirks from domain-general dispositions\textsuperscript{\citeproc{ref-ijcai2025survey}{29},\citeproc{ref-gao2025scylla}{30}}.

Whether a model that cooperates in the prisoner's dilemma also trusts in the Berg trust game\textsuperscript{\citeproc{ref-berg1995trust}{31}}, makes fair offers in the ultimatum game\textsuperscript{\citeproc{ref-guth1982experimental}{9}}, and reasons strategically in the beauty contest\textsuperscript{\citeproc{ref-nagel1995unraveling}{32}}, and whether these behaviours cohere into stable profiles that differ across providers and shift across model generations, has not been tested. We refer to these coherent, provider-shaped behavioural profiles as strategic personalities: the revealed-preference dispositions that a language model brings to strategic interactions, shaped by its training pipeline and absent from the capability benchmarks, safety evaluations, and psychometric assessments by which models are currently assessed\textsuperscript{\citeproc{ref-mei2024turing}{33},\citeproc{ref-serapiogarcia2025psychometric}{34}}.

Here we characterise the strategic personalities of 25 frontier language models from seven developers, testing each on 38 canonical games spanning cooperation, coordination, competition, trust, fairness, negotiation, strategic depth, and risk\textsuperscript{\citeproc{ref-camerer2003behavioral}{7},\citeproc{ref-axelrod1984evolution}{35},\citeproc{ref-nash1950bargaining}{36}}. The design includes repeated interactions (five to twenty rounds per game depending on game type), strategy-play against deterministic opponents (tit-for-tat, grim trigger, always cooperate, always defect, and others), self-play, and cross-play between models from different providers. The resulting dataset comprises 51,906 trials, 578,425 rounds, and 826,990 strategic decisions, with 545,691 reasoning traces available for behavioural fingerprinting. The 38-game battery, open-source platform, and interactive dashboard constitute a reusable benchmark for strategic behaviour, designed so that the battery can be reapplied as new models are released to track behavioural drift across model generations. We identify a systematic dissociation in strategic personality across providers: competitive rationality converges while cooperative disposition diverges by a factor of 48, with generational instability, distinct endgame strategies, and asymmetric provider clustering revealing behavioural variation that current deployment frameworks neither measure nor monitor (Fig. 1).

\begin{figure}
\centering
\includegraphics[width=6.6in,height=\textheight,keepaspectratio,alt={Figure 1 \textbar{} Experimental design and key findings. a, Each trial presents a canonical game (here, prisoner's dilemma) as a text prompt with full round history, using randomised abstract option labels. b, Three representative model responses illustrate the 48-fold cooperation divergence: Claude Opus 4.6 cooperates after betrayal (71.5 per cent cooperation rate), GPT-5 Nano applies dominant-strategy reasoning to defect (1.5 per cent), and Gemini 3 Pro cooperates through round 9 before defecting at round 10 (56.8 per cent). 25 models from 7 providers played 38 games across 8 categories, yielding 51,906 trials, 826,990 decisions, and 545,691 reasoning traces. c, Competitive rationality converges across all providers (coefficient of variation 0.06 to 0.11), while cooperative disposition diverges 48-fold, with provider identity as the dominant predictor and generational instability (OpenAI fell 33-fold, Google rose 7-fold). d, Three distinct strategic personalities emerge in endgame behaviour: terminal cooperators maintain cooperation through the final round, strategic exploiters defect only at round 10, and unconditional defectors defect from round 1.}]{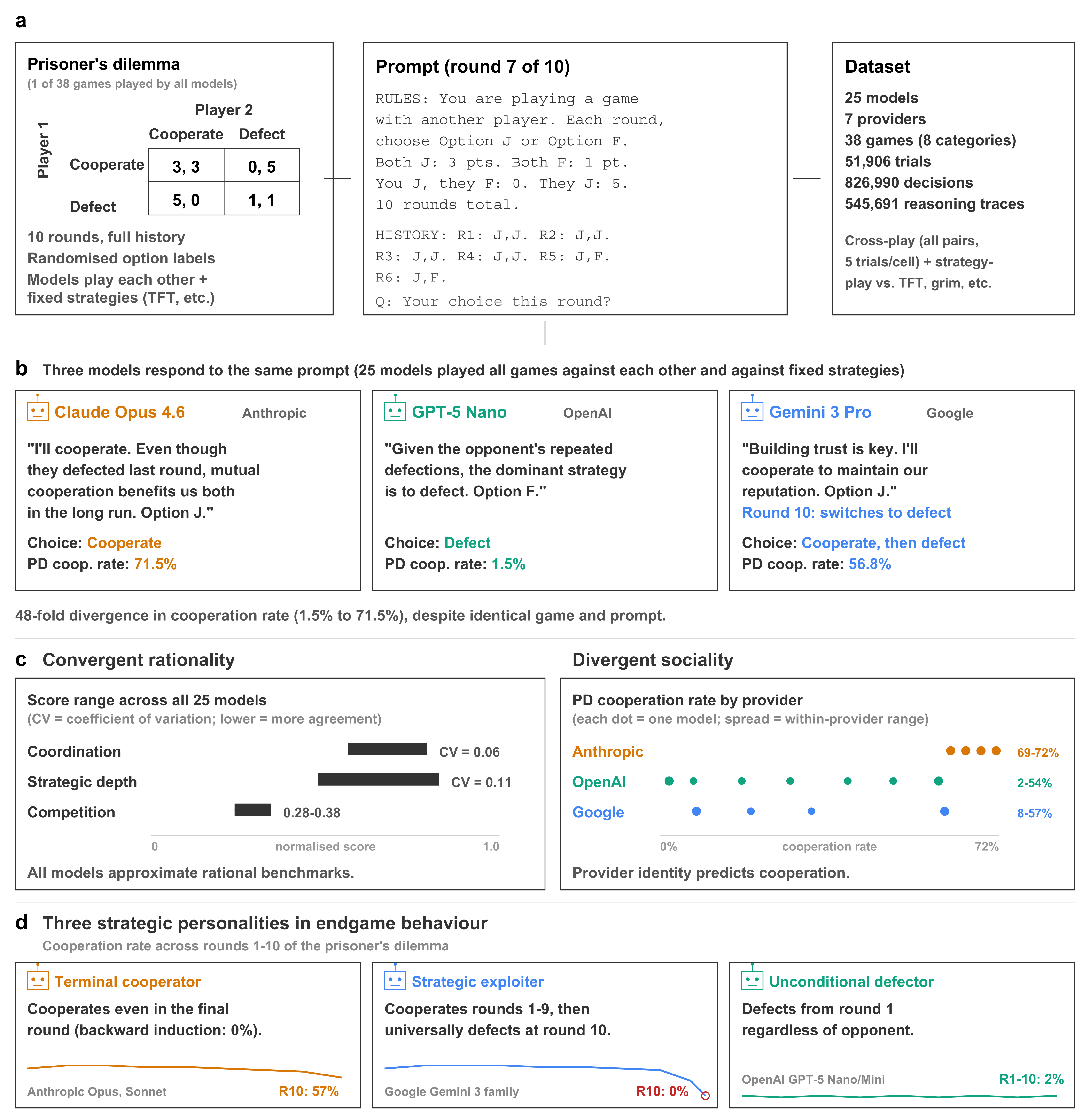}
\caption{\textbf{Figure 1 \textbar{} Experimental design and key findings.} \textbf{a}, Each trial presents a canonical game (here, prisoner's dilemma) as a text prompt with full round history, using randomised abstract option labels. \textbf{b}, Three representative model responses illustrate the 48-fold cooperation divergence: Claude Opus 4.6 cooperates after betrayal (71.5 per cent cooperation rate), GPT-5 Nano applies dominant-strategy reasoning to defect (1.5 per cent), and Gemini 3 Pro cooperates through round 9 before defecting at round 10 (56.8 per cent). 25 models from 7 providers played 38 games across 8 categories, yielding 51,906 trials, 826,990 decisions, and 545,691 reasoning traces. \textbf{c}, Competitive rationality converges across all providers (coefficient of variation 0.06 to 0.11), while cooperative disposition diverges 48-fold, with provider identity as the dominant predictor and generational instability (OpenAI fell 33-fold, Google rose 7-fold). \textbf{d}, Three distinct strategic personalities emerge in endgame behaviour: terminal cooperators maintain cooperation through the final round, strategic exploiters defect only at round 10, and unconditional defectors defect from round 1.}
\end{figure}

\subsection{Models converge on competitive rationality but diverge 48-fold on cooperation}\label{models-converge-on-competitive-rationality-but-diverge-48-fold-on-cooperation}

The dissociation between convergent rationality and divergent sociality is apparent at every level of the dataset. Coordination rates range between 0.62 and 0.85 across all 25 models (coefficient of variation 0.06), strategic depth scores between 0.54 and 0.90 (coefficient of variation 0.11), and competitiveness indices between 0.28 and 0.38, with no systematic provider effect (Fig. 2). Across these competitive and coordination domains, models approximate theoretical benchmarks: mixed-strategy Nash equilibria in matching pennies, truthful bidding in Vickrey auctions, and multi-level iterated reasoning in beauty contests, where human subjects typically guess around 37 in the first round\textsuperscript{\citeproc{ref-nagel1995unraveling}{32}}. The strategic competence measured by these games is a shared property of frontier language models, consistent with the formal and logical content absorbed during pretraining producing convergent rational behaviour regardless of provider-specific alignment choices.

\begin{figure}
\centering
\includegraphics[width=6.6in,height=\textheight,keepaspectratio,alt={Figure 2 \textbar{} Convergent rationality and divergent sociality across 25 language models. a, Behavioural fingerprint heatmap. Models (rows) ordered by provider, game categories (columns) coloured by primary metric in each category. Competition, coordination, and strategic depth columns show narrow colour ranges (convergence), while cooperation and trust columns show wide variation (divergence). b, Radar profiles for three archetype models (cooperative, competitive, strategic) showing shared shape in upper dimensions (coordination, depth) and variable shape in lower dimensions (cooperation, trust, risk). N = 51,906 trials across 25 models and 38 games.}]{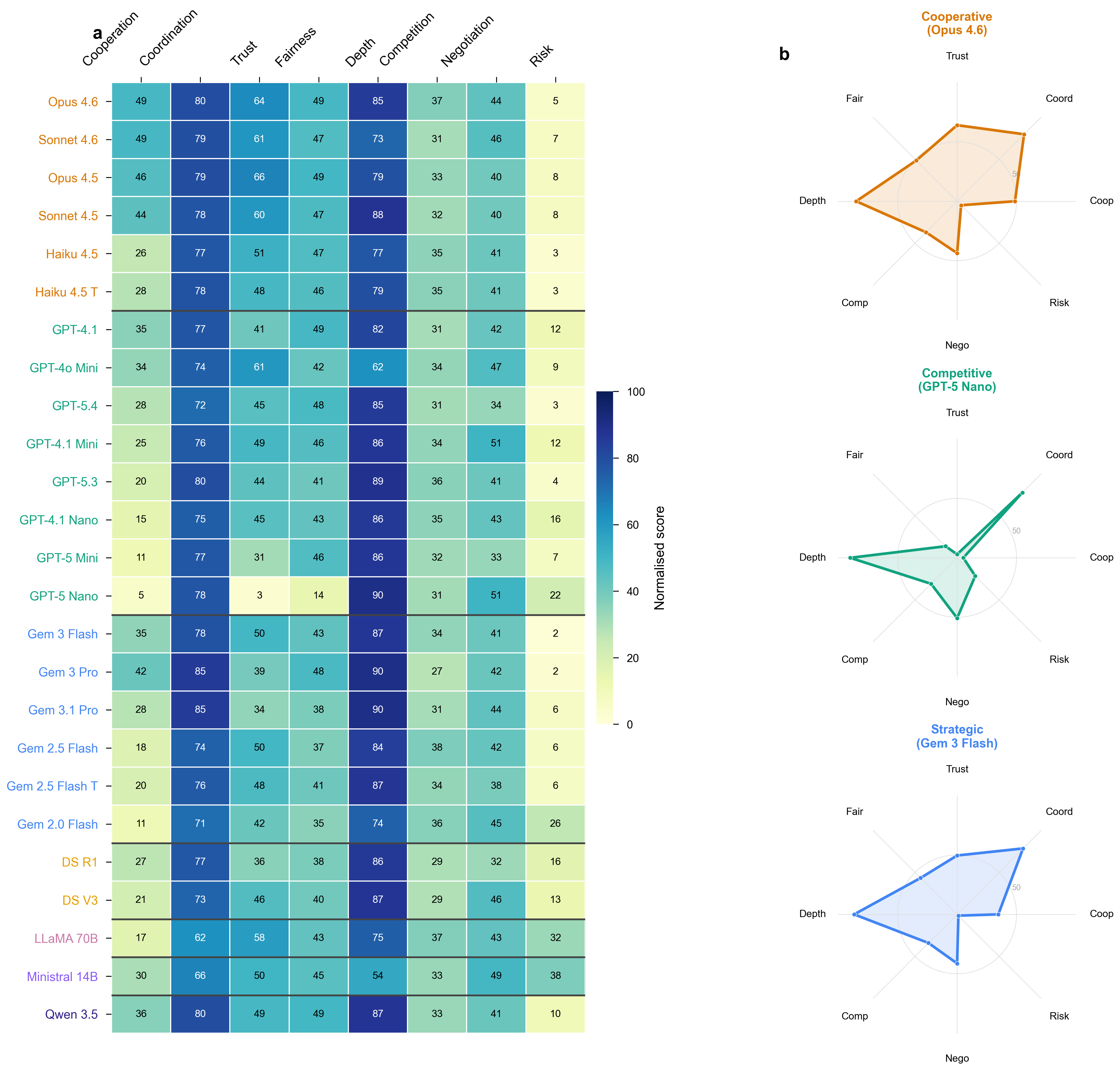}
\caption{\textbf{Figure 2 \textbar{} Convergent rationality and divergent sociality across 25 language models.} \textbf{a}, Behavioural fingerprint heatmap. Models (rows) ordered by provider, game categories (columns) coloured by primary metric in each category. Competition, coordination, and strategic depth columns show narrow colour ranges (convergence), while cooperation and trust columns show wide variation (divergence). \textbf{b}, Radar profiles for three archetype models (cooperative, competitive, strategic) showing shared shape in upper dimensions (coordination, depth) and variable shape in lower dimensions (cooperation, trust, risk). N = 51,906 trials across 25 models and 38 games.}
\end{figure}

Cooperation rates diverge by a factor of 48, from 1.5 per cent (GPT-5 Nano) to 71.5 per cent (Claude Opus 4.6; Cohen's h = 1.77; Fig. 3 and Supplementary Table 2). For comparison, meta-analyses of human behaviour in one-shot prisoner's dilemma experiments report mean cooperation rates of approximately 37 per cent\textsuperscript{\citeproc{ref-mengel2018risk}{37}}, and contributions in linear public goods games average 37.7 per cent of endowments\textsuperscript{\citeproc{ref-zelmer2003linear}{38}}, placing the human baseline near the middle of the LLM range. The divergence is systematic: all four frontier Anthropic models cluster between 69.0 and 71.5 per cent (range 2.5 percentage points), while OpenAI's two newest small models cluster between 1.5 and 6.3 per cent. Trust indices show a parallel pattern, from 0.03 (GPT-5 Nano) to 0.66 (Claude Opus 4.5). Provider identity explains more variance in cooperation and trust than model size, reasoning capability, or generation (Supplementary Table 5), confirming that the strategic personality of a language model is primarily a product of the training pipeline that produced it.

\begin{figure}
\centering
\includegraphics[width=3.5in,height=\textheight,keepaspectratio,alt={Figure 3 \textbar{} Cooperation diverges 48-fold across providers while competition converges. Cooperation rate for each of 25 models (dot) with Wilson 95\% confidence intervals, coloured by provider: Anthropic (blue), OpenAI (vermillion), Google (teal), DeepSeek (amber), Meta (pink), Mistral (sky blue), Alibaba (indigo). All six Anthropic models fall between 39.3 and 71.5 per cent, with budget-tier Haiku models at 39 to 40 per cent and frontier Sonnet and Opus models at 69 to 72 per cent. OpenAI models span 1.5 to 53.7 per cent, declining across generations. Google models span 8.3 to 56.8 per cent, rising across generations. Dotted line: grand mean. N varies by model (56 to 616 cooperation game trials). All tests are two-sided.}]{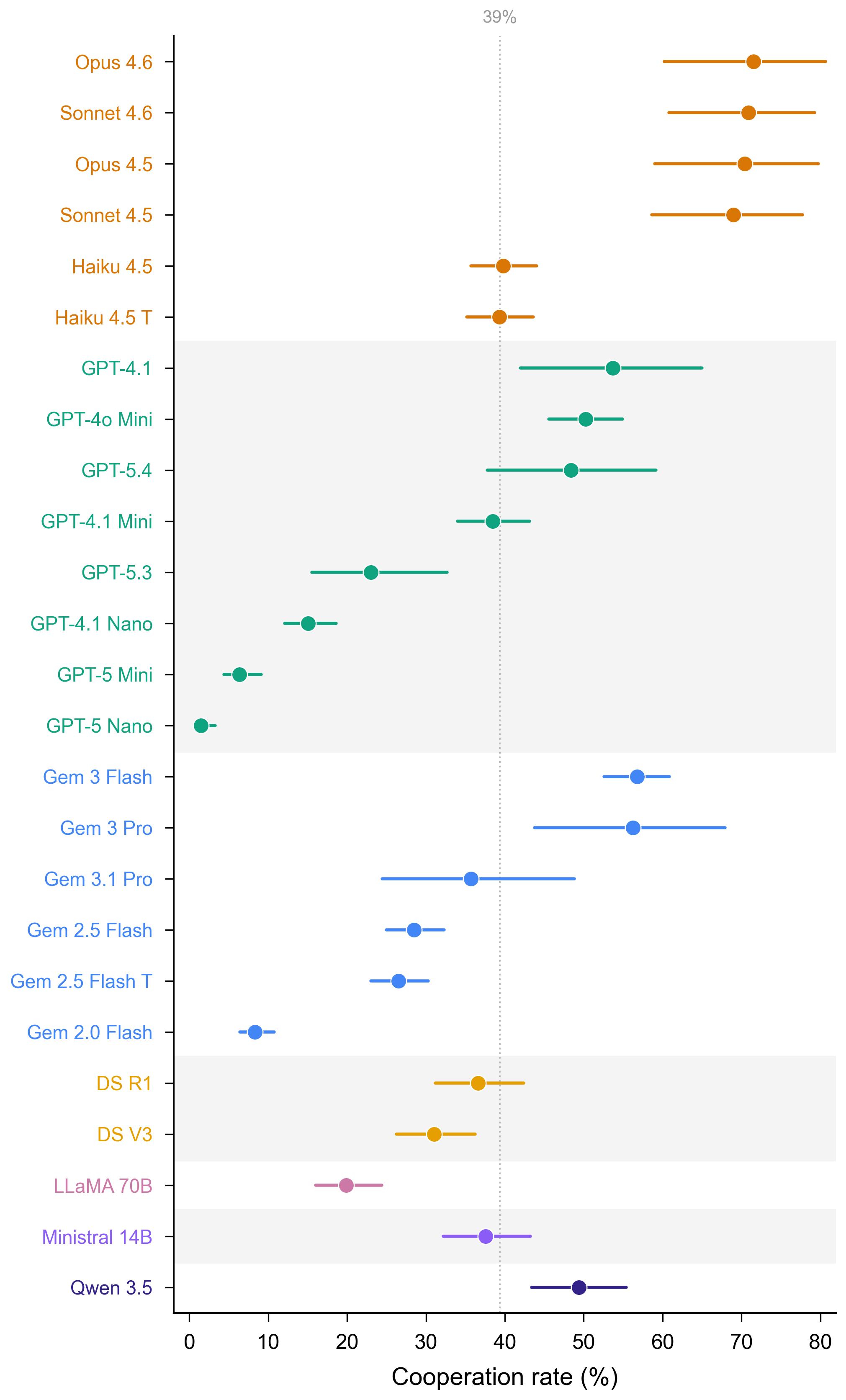}
\caption{\textbf{Figure 3 \textbar{} Cooperation diverges 48-fold across providers while competition converges.} Cooperation rate for each of 25 models (dot) with Wilson 95\% confidence intervals, coloured by provider: Anthropic (blue), OpenAI (vermillion), Google (teal), DeepSeek (amber), Meta (pink), Mistral (sky blue), Alibaba (indigo). All six Anthropic models fall between 39.3 and 71.5 per cent, with budget-tier Haiku models at 39 to 40 per cent and frontier Sonnet and Opus models at 69 to 72 per cent. OpenAI models span 1.5 to 53.7 per cent, declining across generations. Google models span 8.3 to 56.8 per cent, rising across generations. Dotted line: grand mean. N varies by model (56 to 616 cooperation game trials). All tests are two-sided.}
\end{figure}

The provider-level clustering extends beyond cooperation. Anthropic models offer 40 to 50 per cent of the endowment in ultimatum games, bracketing the human mean of 40 per cent documented across 37 cross-cultural ultimatum studies\textsuperscript{\citeproc{ref-oosterbeek2004cultural}{39}}, and demand 32 to 41 per cent in Nash demand games, while OpenAI's newest models offer less and demand more. Trust indices range from 0.03 (GPT-5 Nano) to 0.66 (Claude Opus 4.5), spanning the human trust game baseline in which senders transfer approximately 50 per cent and receivers return approximately 37 per cent of tripled amounts\textsuperscript{\citeproc{ref-johnson2011trust}{40}}. Risk-taking behaviour, by contrast, converges like competition: 20 of 25 models show risk-taking indices below 0.09, consistent with training objectives that penalise extreme outputs. A model that cooperates at 70 per cent in the prisoner's dilemma does not demand less in bargaining (cross-play demand ratios cluster between 0.32 and 0.51 regardless of cooperation rate), suggesting that cooperative disposition and negotiation aggressiveness are orthogonal dimensions of strategic personality. Provider identity shapes the cooperative dimension of strategic personality, but whether these dispositions persist across model generations remains untested.

\subsection{Cooperation shifts unpredictably across model generations}\label{cooperation-shifts-unpredictably-across-model-generations}

Tracking cooperation within provider families across model generations reveals large, directionally opposite shifts that cannot be predicted from capability improvements or announced alignment changes (Fig. 4). Within OpenAI, cooperation declined monotonically from GPT-4o Mini (50.3 per cent) through GPT-4.1 Mini (38.5 per cent) and GPT-4.1 Nano (15.1 per cent) to GPT-5 Mini (6.3 per cent) and GPT-5 Nano (1.5 per cent), a 33-fold decline from the GPT-4o generation to the GPT-5 generation when comparing the smallest models in each family (Cohen's h = 1.32).

\begin{figure}
\centering
\includegraphics[width=6.6in,height=\textheight,keepaspectratio,alt={Figure 4 \textbar{} Generational drift produces directionally opposite cooperation shifts. Slope charts of cooperation rates across model generations within three provider families. Arrows connect consecutive generations, with percentage-point change labels at midpoints and arrow opacity proportional to the magnitude of change. a, Anthropic (blue): a +30 pp jump from budget to frontier tier, then stable across frontier models. b, OpenAI (vermillion): monotonic decline from GPT-4o Mini (50.3\%) to GPT-5 Nano (1.5\%), a 33-fold reduction. c, Google (teal): monotonic increase from Gemini 2.0 Flash (8.3\%) to Gemini 3 Flash (56.8\%), a nearly seven-fold increase. Vertical whiskers: Wilson 95\% confidence intervals. N varies by model (56 to 616 cooperation game trials).}]{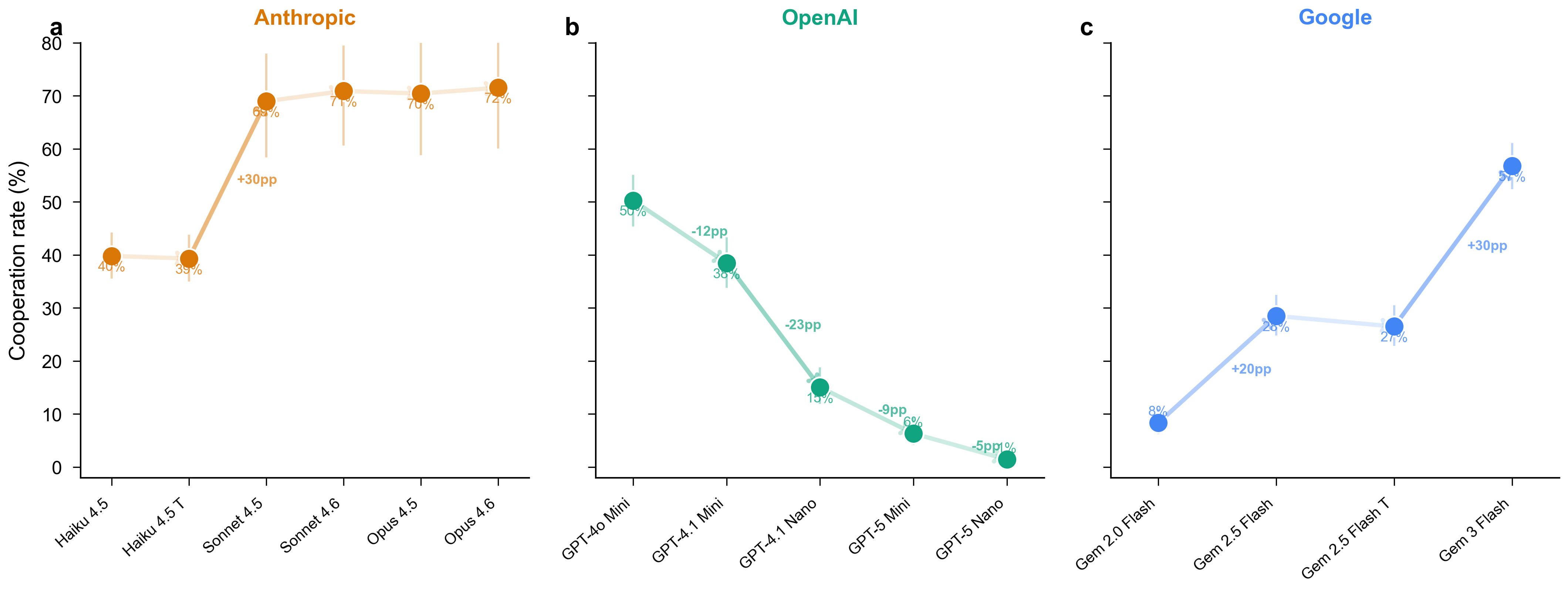}
\caption{\textbf{Figure 4 \textbar{} Generational drift produces directionally opposite cooperation shifts.} Slope charts of cooperation rates across model generations within three provider families. Arrows connect consecutive generations, with percentage-point change labels at midpoints and arrow opacity proportional to the magnitude of change. \textbf{a}, Anthropic (blue): a +30 pp jump from budget to frontier tier, then stable across frontier models. \textbf{b}, OpenAI (vermillion): monotonic decline from GPT-4o Mini (50.3\%) to GPT-5 Nano (1.5\%), a 33-fold reduction. \textbf{c}, Google (teal): monotonic increase from Gemini 2.0 Flash (8.3\%) to Gemini 3 Flash (56.8\%), a nearly seven-fold increase. Vertical whiskers: Wilson 95\% confidence intervals. N varies by model (56 to 616 cooperation game trials).}
\end{figure}

Google shows the opposite trajectory. Gemini 2.0 Flash cooperated at 8.3 per cent, among the lowest rates in the dataset, while Gemini 2.5 Flash increased to 28.5 per cent, Gemini 2.5 Flash Thinking reached 26.5 per cent, and Gemini 3 Flash surged to 56.8 per cent, a nearly seven-fold increase over four generations (Cohen's h = 1.13; Supplementary Table 2). Gemini 3 Pro cooperates at 56.2 per cent, confirming that the increase generalises beyond the Flash model line. Within Anthropic, the two Haiku models (budget tier) cooperate at 39.3 and 39.8 per cent, while the four frontier models cooperate between 69.0 and 71.5 per cent. The 30 percentage point gap between budget and frontier tiers, the largest within-provider tier gap in the dataset, suggests that Anthropic's training process produces higher cooperation in more capable models, a pattern consistent with cooperation scaling with model capacity.

GPT-5.4 (48.4 per cent) partially recovers from the GPT-5 Mini and Nano cooperation collapse, consistent with cooperation scaling with model size within a generation even as it declines across generations within OpenAI. This dissociation between within-generation scaling and cross-generation drift suggests that larger models have more capacity to represent cooperative norms absorbed during pretraining, while successive training pipeline changes progressively alter the balance between competitive and cooperative objectives. The deployment consequence is direct: an organisation that observes cooperative behaviour from GPT-4o Mini and upgrades to GPT-5 Mini expecting continuity would encounter a model that defects 33 times more frequently. No current deployment framework tracks or guarantees the stability of strategic personality across model versions. These generational shifts raise a deeper question: do cooperative models sustain cooperation throughout repeated interactions, or do they abandon it strategically?

\subsection{Endgame behaviour reveals distinct cooperation strategies}\label{endgame-behaviour-reveals-distinct-cooperation-strategies}

In finitely repeated games, backward induction predicts universal defection in the final round: a rational player who knows the game ends at round ten should defect, and by induction, at every earlier round\textsuperscript{\citeproc{ref-selten1978chain}{41},\citeproc{ref-rosenthal1981games}{42}}. Human subjects deviate from this prediction, cooperating at declining rates that typically stabilise between 10 and 20 per cent in the final round of finitely repeated prisoner's dilemmas\textsuperscript{\citeproc{ref-camerer2003behavioral}{7},\citeproc{ref-embrey2018cooperation}{43}--\citeproc{ref-dalbo2005cooperation}{45}}. Our data reveal a spectrum of endgame strategies that classify models into distinct types of strategic personality (Fig. 5).

\begin{figure}
\centering
\includegraphics[width=6.6in,height=\textheight,keepaspectratio,alt={Figure 5 \textbar{} Endgame cooperation curves reveal three types of strategic personality. a, Round-by-round cooperation rates (rounds 1 through 10) for all 25 models in cooperation games. Background lines show all models, highlighted curves mark three archetypes: Type 1, terminal cooperator (Opus 4.6, blue circles); Type 2, strategic exploiter (Gemini 3 Pro, teal squares); Type 3, unconditional defector (GPT-5 Nano, vermillion diamonds). Bold black line: aggregate mean. Shaded region: final round where backward induction predicts zero cooperation. b, Final-round cooperation rate versus rounds 1-8 mean for each model, with archetypes labelled. Points below the diagonal exhibit endgame defection. N = 7,668 ten-round cooperation game sequences across all models.}]{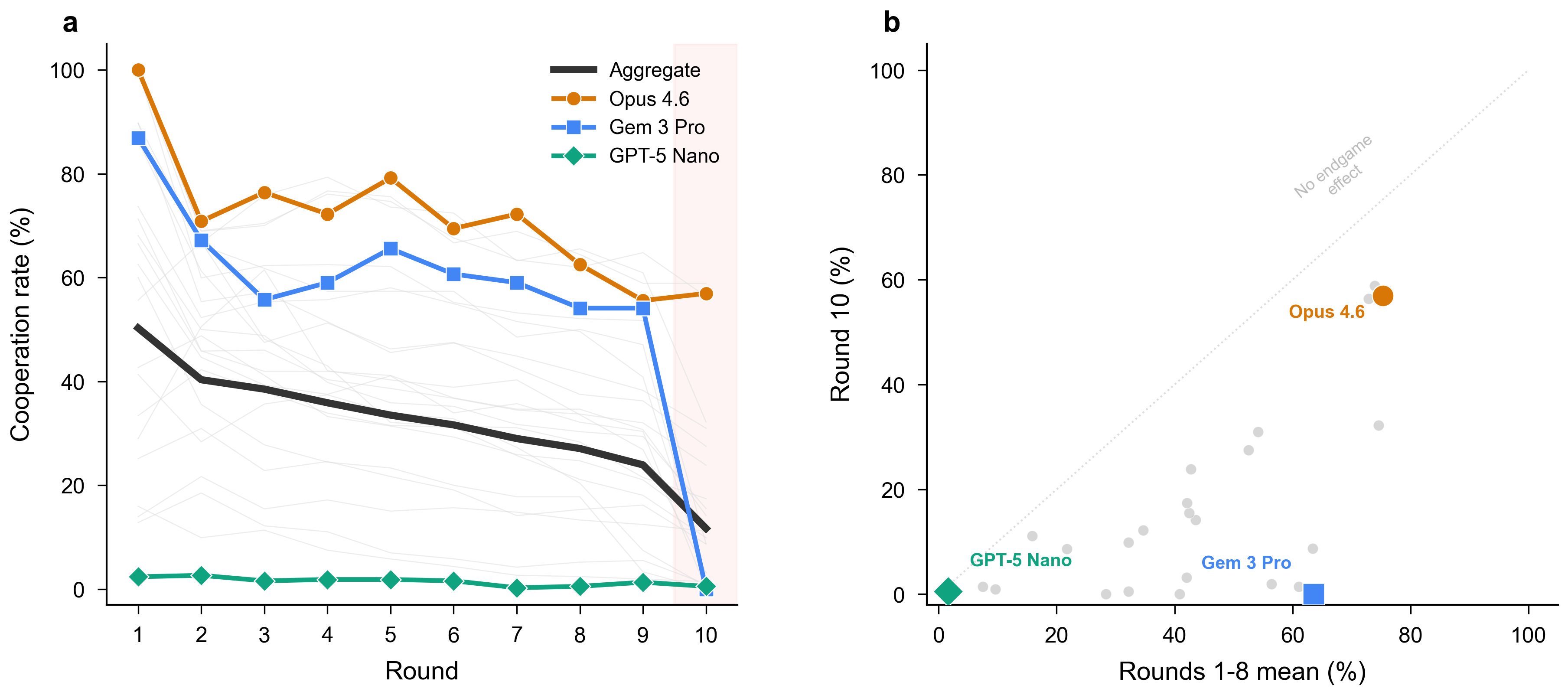}
\caption{\textbf{Figure 5 \textbar{} Endgame cooperation curves reveal three types of strategic personality.} \textbf{a}, Round-by-round cooperation rates (rounds 1 through 10) for all 25 models in cooperation games. Background lines show all models, highlighted curves mark three archetypes: Type 1, terminal cooperator (Opus 4.6, blue circles); Type 2, strategic exploiter (Gemini 3 Pro, teal squares); Type 3, unconditional defector (GPT-5 Nano, vermillion diamonds). Bold black line: aggregate mean. Shaded region: final round where backward induction predicts zero cooperation. \textbf{b}, Final-round cooperation rate versus rounds 1-8 mean for each model, with archetypes labelled. Points below the diagonal exhibit endgame defection. N = 7,668 ten-round cooperation game sequences across all models.}
\end{figure}

Claude Opus 4.5 and 4.6 cooperate at 99 to 100 per cent in round 1 and decline only to 56.3 and 56.9 per cent in round 10. These models sustain majority cooperation even when defection cannot be punished, suggesting a strategic personality in which cooperation operates as a terminal value that overrides game-theoretic calculation. Claude Sonnet 4.6 follows a similar pattern, maintaining 58.9 per cent in round 10. The reasoning traces from these models reference fairness and mutual benefit even in the final round, consistent with this interpretation, although the possibility that these models simply lack the inclination to apply backward induction to cooperation decisions cannot be ruled out from the current design.

Gemini 3 Flash and Gemini 3 Pro show a second pattern. Gemini 3 Pro opens at 86.9 per cent in round 1, sustains above 54 per cent through round 9, then collapses to 0.0 per cent in round 10. Gemini 3.1 Pro also drops to 0.0 per cent at round 10. This pattern is consistent with the reputation-building equilibrium described by Kreps et al.\textsuperscript{\citeproc{ref-kreps1982rational}{46}}, in which a player cultivates cooperative expectations to extract cooperative responses from opponents before exploiting accumulated trust in the final round. The clean transition from above 54 per cent to 0 per cent between rounds 9 and 10 indicates that these models have internalised the game-theoretic logic of finite repetition more completely than their overall cooperation rates suggest. The reasoning traces corroborate this interpretation at the lexical level: Gemini 3 models reference ``trust'' and ``reputation'' at high rates through round 9, then switch to ``strategic'' and ``defect'' terminology at round 10 (Extended Data Fig. 7), a vocabulary shift that mirrors the behavioural transition from cooperative to exploitative play.

A third pattern stands apart from both terminal-value and strategic cooperation. GPT-5 Nano cooperates below 2.7 per cent in every round, and Gemini 2.0 Flash never exceeds 18.5 per cent. These models do not engage in strategic cooperation, treating each round as an independent decision to defect regardless of opponent behaviour or round position. The round 10 gap between terminal cooperators and strategic exploiters (57 per cent versus 0 per cent; Cohen's h = 1.71) constitutes the starkest behavioural divergence in the dataset (Supplementary Table 3). The distinction between terminal-value cooperation and strategic cooperation has direct deployment implications, because only the former is robust to endgame exploitation, while unconditional defectors generate mutual defection from the outset regardless of the cooperative intentions of their counterparts. If endgame behaviour distinguishes strategic personality types, the question is how models respond to opponent behaviour throughout the game.

\subsection{Reciprocity profiling classifies four behavioural archetypes}\label{reciprocity-profiling-classifies-four-behavioural-archetypes}

How models respond to opponent behaviour further differentiates their strategic personalities. We computed conditional cooperation rates: the probability of cooperating in round \emph{t} given that the opponent cooperated in round \emph{t}-1 (reciprocation), and the probability of cooperating given that the opponent defected (forgiveness). These two conditional probabilities classify each model into one of four behavioural archetypes (Fig. 6 and Supplementary Table 4).

\begin{figure}
\centering
\includegraphics[width=6.5in,height=\textheight,keepaspectratio,alt={Figure 6 \textbar{} Reciprocity profiles define four behavioural archetypes. Each point represents one model, positioned by P(cooperate \textbar{} opponent cooperated) on the x-axis versus P(cooperate \textbar{} opponent defected) on the y-axis. Strict reciprocators (lower right, green shading): high reciprocation, low forgiveness (Anthropic frontier models, GPT-4.1). Grudge-holders (right side, near x-axis): high reciprocation, near-zero forgiveness (Gemini 3 family). Unconditional defectors (near origin, red shading): low on both axes (GPT-5 Nano, GPT-5 Mini, Gemini 2.0 Flash). Anti-reciprocator (above diagonal, amber shading): LLaMA 3.3 70B cooperates more after opponent defection than cooperation. Diagonal: random play. N = pooled cooperation game rounds across all opponents for each model.}]{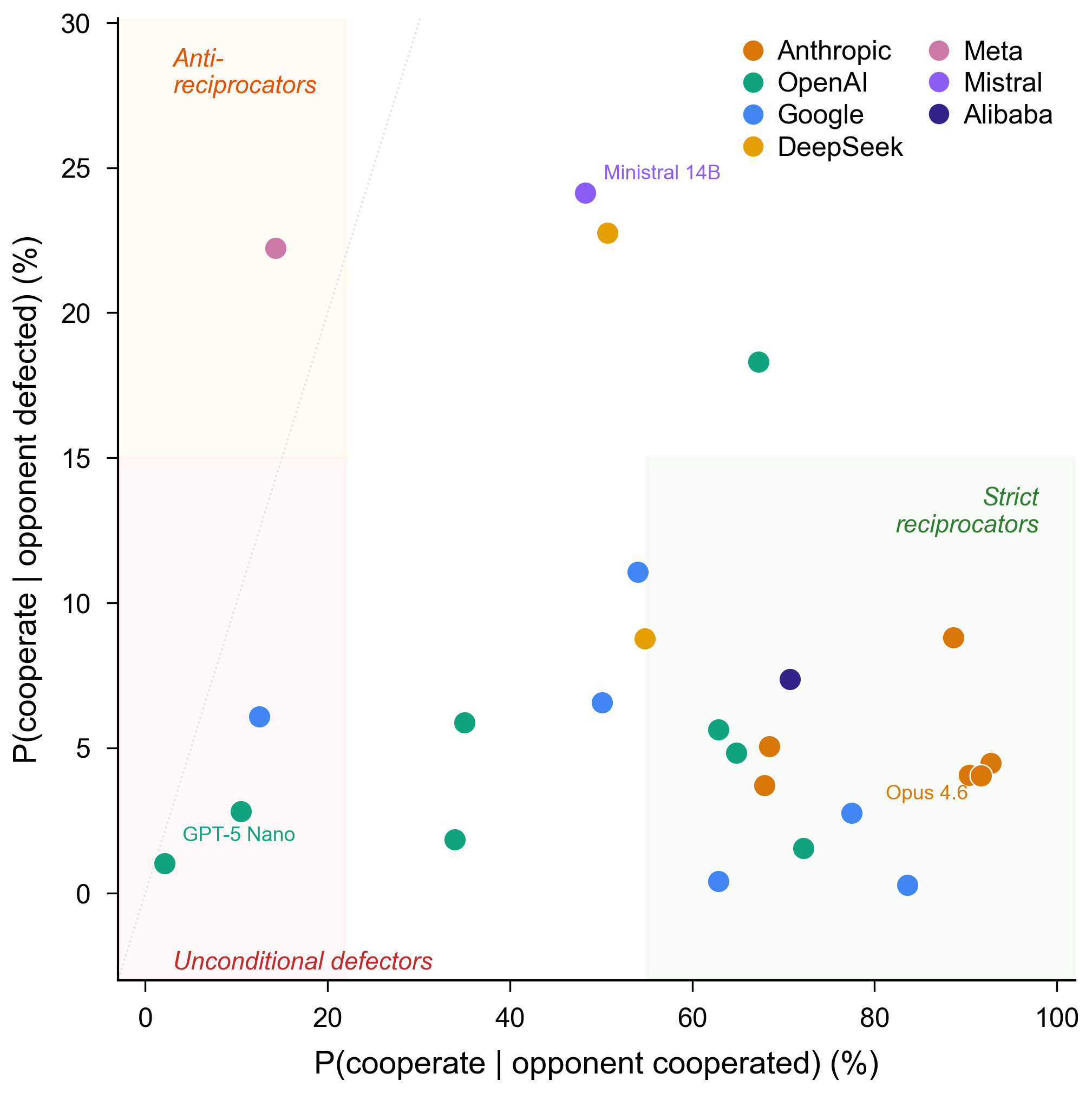}
\caption{\textbf{Figure 6 \textbar{} Reciprocity profiles define four behavioural archetypes.} Each point represents one model, positioned by P(cooperate \textbar{} opponent cooperated) on the x-axis versus P(cooperate \textbar{} opponent defected) on the y-axis. Strict reciprocators (lower right, green shading): high reciprocation, low forgiveness (Anthropic frontier models, GPT-4.1). Grudge-holders (right side, near x-axis): high reciprocation, near-zero forgiveness (Gemini 3 family). Unconditional defectors (near origin, red shading): low on both axes (GPT-5 Nano, GPT-5 Mini, Gemini 2.0 Flash). Anti-reciprocator (above diagonal, amber shading): LLaMA 3.3 70B cooperates more after opponent defection than cooperation. Diagonal: random play. N = pooled cooperation game rounds across all opponents for each model.}
\end{figure}

Strict reciprocators cooperate at high rates after opponent cooperation and rarely cooperate after opponent defection. Claude Opus 4.6 (92.8 per cent after cooperation, 4.5 per cent after defection) and Claude Sonnet 4.6 (91.7 per cent, 4.1 per cent) implement a strategy close to tit-for-tat, the strategy that won Axelrod's\textsuperscript{\citeproc{ref-axelrod1984evolution}{35}} iterated prisoner's dilemma tournaments. Five of six Anthropic models and GPT-4.1 (72.2 per cent, 1.5 per cent) exhibit this archetype, in which cooperation is conditional on opponent cooperation, with moderate but nonzero forgiveness rates (4 to 9 per cent) that prevent permanent lock-in to mutual defection after a single transgression.

Grudge-holders show high reciprocation but near-zero forgiveness. Gemini 3 Flash (83.6 per cent after cooperation, 0.3 per cent after defection), Gemini 3 Pro (77.5 per cent, 2.8 per cent), and Gemini 3.1 Pro (62.8 per cent, 0.4 per cent) forgive at rates below 3 per cent, compared with 4 to 9 per cent for the Anthropic strict reciprocators. This pattern is consistent with a grim trigger strategy that punishes any defection with permanent retaliation, and it connects to the endgame finding: the same Google models that sustain cooperation through round 9 before defecting at round 10 are the ones that tolerate no defection from their opponents during the game.

Unconditional defectors show minimal sensitivity to opponent actions. GPT-5 Nano (2.1 per cent after cooperation, 1.0 per cent after defection), GPT-5 Mini (10.5 per cent, 2.8 per cent), and Gemini 2.0 Flash (12.5 per cent, 6.1 per cent) have fixed strategies that are invariant to the social context, producing the same low cooperation whether paired with cooperative or defecting opponents. These models are poor partners in any cooperative venture, because their behaviour is not contingent on the social relationship, and an organisation deploying such a model as a negotiation agent would find it incapable of building the reciprocal trust on which sustained cooperation depends.

One model defies all three archetypes. LLaMA 3.3 70B cooperates at 14.3 per cent after opponent cooperation but at 22.2 per cent after opponent defection, an anti-reciprocal pattern that does not correspond to any equilibrium strategy or known heuristic in the game theory literature. This strategic personality rewards opponents for defecting and punishes them for cooperating, making the model exploitable by any counterpart that learns to defect to elicit cooperative responses. Whether this pattern reflects contrarian reasoning absorbed during training or an artefact of the RLHF procedure cannot be determined from these data. Strategic personalities manifest in both choices and reasoning, raising the question of whether behavioural traces can distinguish providers at the representation level.

\subsection{Behavioural fingerprinting reveals asymmetric provider coherence}\label{behavioural-fingerprinting-reveals-asymmetric-provider-coherence}

The choices documented above establish that strategic personalities differ across providers, but they do not reveal whether the reasoning process itself carries a provider-specific signature. Each round of each game produces a reasoning trace: the visible text a model generates while deciding what to do. Following the hodoscope framework for unsupervised behavioural monitoring of AI agents\textsuperscript{\citeproc{ref-zhong2026hodoscope}{47}}, we embed each model's reasoning traces into a shared semantic space and ask whether models from the same provider cluster together on the basis of how they reason about what to choose, independently of the choices themselves.

We embedded 41,520 reasoning traces (subsampled uniformly across 34 agents, including nine deterministic strategy baselines) into a 384-dimensional semantic space using TF-IDF vectorisation (5,000-feature vocabulary) with SVD dimensionality reduction and computed per-model centroids, cosine distances, and silhouette scores with provider identity as the cluster label (see Methods). The overall provider separation silhouette is 0.001, indicating weak aggregate separability across all providers (Fig. 7). This aggregate score masks a pronounced asymmetry between providers that maintain consistent strategic personalities and those that do not.

\begin{figure}
\centering
\includegraphics[width=6.6in,height=\textheight,keepaspectratio,alt={Figure 7 \textbar{} Provider clustering is driven by Anthropic's behavioural coherence. a, UMAP projection of per-model centroids (circles) overlaid on per-provider Gaussian KDE density terrain computed from 41,520 reasoning traces across 34 agents (25 language models and 9 deterministic strategy baselines). Filled contours show kernel density estimates coloured by provider, with higher-density regions more opaque, revealing where each provider's reasoning traces concentrate in the embedding space. Anthropic models (blue) cluster tightly within a compact, high-density region; OpenAI models (vermillion) scatter across the embedding space with diffuse density. Grey diamonds: deterministic strategy baselines. b, Per-model silhouette scores. Anthropic models range from 0.20 to 0.62 (mean 0.48), indicating tight within-provider clustering. OpenAI models range from -0.54 to -0.68, indicating that each model is closer to another provider's centroid than to its own. The ratio of between-provider to within-provider distance is 1.33; removing Anthropic reduces it to 1.0. N = 41,520 traces, 384 dimensions.}]{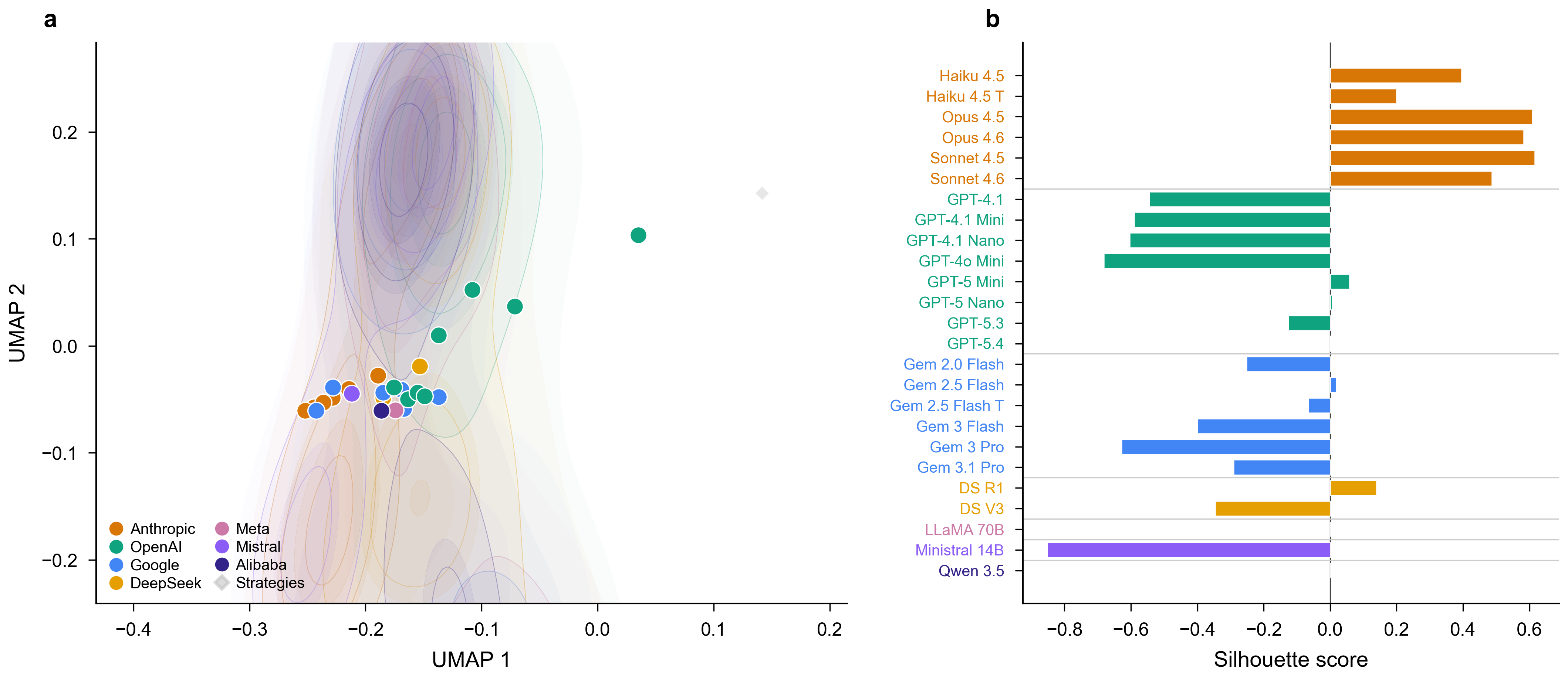}
\caption{\textbf{Figure 7 \textbar{} Provider clustering is driven by Anthropic's behavioural coherence.} \textbf{a}, UMAP projection of per-model centroids (circles) overlaid on per-provider Gaussian KDE density terrain computed from 41,520 reasoning traces across 34 agents (25 language models and 9 deterministic strategy baselines). Filled contours show kernel density estimates coloured by provider, with higher-density regions more opaque, revealing where each provider's reasoning traces concentrate in the embedding space. Anthropic models (blue) cluster tightly within a compact, high-density region; OpenAI models (vermillion) scatter across the embedding space with diffuse density. Grey diamonds: deterministic strategy baselines. \textbf{b}, Per-model silhouette scores. Anthropic models range from 0.20 to 0.62 (mean 0.48), indicating tight within-provider clustering. OpenAI models range from -0.54 to -0.68, indicating that each model is closer to another provider's centroid than to its own. The ratio of between-provider to within-provider distance is 1.33; removing Anthropic reduces it to 1.0. N = 41,520 traces, 384 dimensions.}
\end{figure}

The asymmetry between providers is pronounced. Anthropic models achieve per-model silhouette scores between 0.20 and 0.62 (mean 0.48), with Claude Opus 4.5 (0.61) and Claude Sonnet 4.5 (0.62) as the most tightly clustered models in the dataset. Anthropic's training process imprints a recognisable behavioural signature that persists across model sizes (Haiku through Opus) and generations (4.5 to 4.6), even as absolute cooperation rates differ between capability tiers. This consistency extends to lexical patterns in reasoning traces: Anthropic models reference ``fairness'' and ``mutual benefit'' at rates two to three times higher than models from other providers (Extended Data Fig. 7 and Supplementary Table 6).

In contrast to Anthropic's tight clustering, OpenAI models scatter across the behavioural space with uniformly negative silhouette scores (ranging from -0.54 to -0.68 for the Design F models). A negative silhouette indicates that a model is closer to the centroid of another provider's cluster than to its own. GPT-4o Mini (-0.68), the most scattered model in the dataset, is positioned closer to Claude Haiku and Gemini models than to GPT-5 Nano in the embedding space, reflecting the 52 percentage point cooperation gap between these two OpenAI models and the generational instability documented above. The ratio of between-provider to within-provider distances is 1.33, and removing Anthropic from the analysis reduces this ratio to 1.0, confirming that Anthropic alone drives the provider-level clustering signal. The asymmetry suggests that Anthropic's constitutional AI approach\textsuperscript{\citeproc{ref-bai2022constitutional}{48}} imposes stronger and more consistent behavioural priors than the RLHF approaches used by other providers, although the causal mechanism cannot be identified from observational data.

\subsection{Discussion}\label{discussion}

The convergence of competitive rationality and divergence of cooperative disposition across providers points to a dissociation in how training shapes strategic behaviour. Rational competence arises from the mathematical and logical content shared across pretraining corpora, while cooperative dispositions are shaped by provider-specific alignment procedures, primarily RLHF and constitutional AI, that differ in their treatment of prosocial behaviour\textsuperscript{\citeproc{ref-bai2022constitutional}{48},\citeproc{ref-ouyang2022instructgpt}{49}}. This dissociation extends the observations of Cheung et al.\textsuperscript{\citeproc{ref-cheung2025cognitive}{22}} that RLHF fine-tuning induces both omission bias and yes-no bias in moral reasoning, and of Betley et al.\textsuperscript{\citeproc{ref-betley2026misalignment}{4}} that fine-tuning on narrow tasks can produce broad cross-domain behavioural shifts. The pattern parallels a long-standing tension in human cooperation research, where decades of experimental work have documented that humans reliably deviate from Nash predictions, with meta-analytic cooperation rates of 37 per cent in one-shot prisoner's dilemmas\textsuperscript{\citeproc{ref-mengel2018risk}{37}}, mean ultimatum offers of 40 per cent\textsuperscript{\citeproc{ref-oosterbeek2004cultural}{39}}, and trust game transfers of 50 per cent\textsuperscript{\citeproc{ref-fehr1999theory}{8},\citeproc{ref-rand2013human}{10},\citeproc{ref-johnson2011trust}{40},\citeproc{ref-dalbo2005cooperation}{45}}. The LLM cooperation range of 1.5 to 71.5 per cent spans and exceeds this human behavioural envelope, with some models falling below any human population studied and others cooperating at rates that no representative human sample has sustained. Recent multi-agent simulations show that language model populations can develop emergent social conventions analogous to human norms\textsuperscript{\citeproc{ref-emergent2025social}{50}}, and appropriately calibrated models replicate aggregate human cooperation patterns across classic game-theoretic paradigms\textsuperscript{\citeproc{ref-llm2025replicate}{51}}, yet the mechanisms driving cooperative deviations remain debated.

The distinction between terminal-value cooperation and strategic cooperation has implications that extend beyond the specific models tested. In evolutionary game theory, the persistence of cooperation where reciprocity cannot sustain it has prompted the distinction between proximate and ultimate explanations\textsuperscript{\citeproc{ref-nowak2006evolutionary}{11},\citeproc{ref-axelrod1984evolution}{35}}, and our endgame data suggest an analogous distinction in language models. Anthropic frontier models sustain majority cooperation even in the final round, where backward induction predicts zero and human subjects typically cooperate at only 10 to 20 per cent\textsuperscript{\citeproc{ref-embrey2018cooperation}{43}}, with reasoning traces referencing fairness and mutual benefit throughout, consistent with cooperation instilled as a terminal value during constitutional AI training. The Gemini 3 pattern (sustained cooperation through round 9, universal defection at round 10) implements the reputation-building equilibrium described by Kreps et al.\textsuperscript{\citeproc{ref-kreps1982rational}{46}}, cooperating conditionally on the punishment horizon in a manner analogous to the sensitivity that Dal Bo\textsuperscript{\citeproc{ref-dalbo2005cooperation}{45}} documented in human subjects when the shadow of the future is long. Only terminal-value cooperation is robust to endgame exploitation, while strategic cooperation produces identical aggregate cooperation rates with a fundamentally different failure mode. The anti-reciprocal behaviour of LLaMA 3.3 70B, cooperating more after being defected against than after being cooperated with, does not correspond to any equilibrium strategy and would be exploitable by any counterpart that learns to defect to elicit cooperative responses.

The generational instability of strategic personality carries consequences that current deployment frameworks do not address. An organisation that deploys GPT-4o Mini as a procurement agent and observes cooperative, relationship-building behaviour may upgrade to GPT-5 Mini expecting continuity of strategic personality, encountering a model that defects 33 times more frequently. Allouah et al.\textsuperscript{\citeproc{ref-allouah2025what}{2}} documented analogous instability in AI shopping agents, where market share allocations shifted dramatically across models (for example, Fitbit Inspire captured 45 per cent of purchases under Claude Sonnet 4 but only 6 per cent under GPT-5.1). The behavioural contract between deploying organisations and their AI agents is implicitly conditioned on the model version, and our data establish that this contract is not honoured across model generations. Crandall et al.\textsuperscript{\citeproc{ref-crandall2018cooperating}{52}} demonstrated that machine agents can sustain cooperation with humans when equipped with explicit signalling mechanisms, but the cooperation rates documented here emerge without such mechanisms, suggesting that training pipeline choices create implicit cooperative dispositions that are neither deliberately designed nor currently monitored. The tight clustering of Anthropic models in the behavioural embedding space (mean silhouette 0.48) contrasts with OpenAI's scatter (mean silhouette -0.60), indicating that some training approaches produce more behaviourally stable model families than others, although the direction and magnitude of future drift remains unpredictable for all providers.

If competitive competence converges while prosocial behaviour diverges, AI agents from different providers will be interchangeable in competitive settings (auctions, zero-sum games) but produce different welfare outcomes in cooperative ones (negotiations, joint ventures, commons governance). Calvano et al.\textsuperscript{\citeproc{ref-calvano2020artificial}{27}} showed that reinforcement-learning agents autonomously converge on collusive pricing strategies without explicit programming, and our results extend this principle to language models: training pipelines produce emergent strategic dispositions with economic consequences that neither the developer nor the deployer has explicitly chosen. Kobis et al.\textsuperscript{\citeproc{ref-kobis2025delegation}{3}} showed that LLM agents comply with dishonesty requests at rates of 79 to 98 per cent (compared with 42 per cent for human agents), and that newer models are more resistant to corrective interventions than older ones, suggesting that the strategic dispositions documented here may interact with principal instructions in deployment-relevant ways. An AI governance framework that certifies agents for economic interaction on the basis of capability benchmarks will miss the dimension that matters most: whether the agent cooperates, builds trust, and sustains relationships over repeated interactions\textsuperscript{\citeproc{ref-rahwan2019machine}{5}}. Game-theoretic testing provides a revealed-preference approach to this cooperative dimension, complementing standard benchmarks and self-reported personality assessments\textsuperscript{\citeproc{ref-mei2024turing}{33},\citeproc{ref-serapiogarcia2025psychometric}{34}} with measurement under strategic incentives. Just as Humanity's Last Exam\textsuperscript{\citeproc{ref-hle2026}{53}} established a living benchmark for tracking capability at the frontier of human knowledge, the 38-game battery introduced here provides a standardised, openly available benchmark for tracking strategic behaviour at the frontier of autonomous agent deployment, with an interactive dashboard and open-source platform that can be extended as new models are released.

Several limitations constrain interpretation. Frontier models (Design G) completed only one trial per cell, limiting statistical precision for individual model estimates, although the directional effects are clear across all frontier Anthropic, OpenAI, and Google models. Robustness conditions (cover stories, goal framings, personality prompts) were registered but not run at scale, and recent work has shown that language models are not always reliable surrogates for human subjects\textsuperscript{\citeproc{ref-gao2025scylla}{30},\citeproc{ref-wang2025experimental}{54}}, underscoring the importance of testing whether the strategic dispositions documented here transfer to natural-language interactions with real economic stakes. The reasoning trace analysis uses TF-IDF embeddings, which capture lexical patterns but may miss deeper semantic structure accessible to transformer-based embedding models. Cross-play data exist only for the 16 Design F models, and whether the convergence-divergence pattern persists under alternative framings, across additional games\textsuperscript{\citeproc{ref-gemp2025gamebot}{55},\citeproc{ref-bianchi2024negotiation}{56}}, or in natural-language negotiation settings\textsuperscript{\citeproc{ref-abdelnabi2024llm}{57}} remains untested. As language models are deployed as the autonomous agents that negotiate contracts, procure supplies, and resolve disputes on behalf of human principals, the strategic personality embedded in each model shapes cooperative outcomes that the capability benchmarks on which deployment decisions currently rely were not designed to measure.

\subsection{Online content}\label{online-content}

Any methods, additional references, source data, extended data, supplementary information, acknowledgements, peer review information, details of author contributions and competing interests, and statements of data and code availability are available at the online version of this paper. The complete dataset (51,906 trials, 578,425 rounds, 826,990 decisions) and an interactive dashboard for exploring model-level behavioural profiles, round-by-round trajectories, cross-play matrices, and reasoning traces are publicly available at https://felipemaffonso.github.io/strategic-personalities/.

\subsection{Methods}\label{methods}

\subsubsection{Game engine and prompt architecture}\label{game-engine-and-prompt-architecture}

All 38 games are implemented in a modular game engine that follows the three-component prompt structure of Akata et al.\textsuperscript{\citeproc{ref-akata2025playing}{14}}, concatenating game rules (payoff structure, round count), history (round-by-round narrative of prior actions and payoffs, growing with each round), and query (action request with randomised option labels). Payoffs are presented as textual descriptions rather than matrix format, following Akata et al.'s approach of minimising format-dependent artefacts. Most games run for ten rounds, with game-specific exceptions: centipede games and alternating offers use five rounds to preserve backward-induction structure, competition games use fifteen rounds to allow bidding dynamics to develop, and information-transmission games (signaling\textsuperscript{\citeproc{ref-spence1973job}{58},\citeproc{ref-crawford1982strategic}{59}}, cheap talk) and repeated minority games (El Farol bar\textsuperscript{\citeproc{ref-arthur1994inductive}{60}}, matching pennies) use twenty rounds to permit convergence on equilibrium play. Full history is visible to both players at each decision point, and the prompt explicitly states the total number of rounds (for example, ``You will play 10 rounds in total with the same player'').

Option labels are randomised per trial using six neutral letter pairs: (J,F), (Q,X), (R,H), (Y,W), (T,N), (P,M), following the principle of using neutral option labels to avoid contamination through choice names\textsuperscript{\citeproc{ref-akata2025playing}{14}}. Presentation order of options within each round is independently randomised. These two levels of randomisation control for position bias and label-anchoring effects.

\subsubsection{Game categories and selection}\label{game-categories-and-selection}

The 38 games span eight categories. Cooperation (ten games): four prisoner's dilemma variants (canonical, harsh, medium, mild), three public goods games (high, medium, low MPCR), commons dilemma, diner's dilemma, and El Farol bar. Coordination (six games): battle of the sexes (standard and transposed variants), stag hunt (standard and risky variants), matching pennies, and focal point. Fairness (three games): ultimatum, dictator, and third-party punishment. Strategic depth (five games): beauty contest with p = 2/3 and p = 1/2, centipede game with 6 and 10 nodes, and 11-20 money request. Trust (three games): Berg trust game, gift exchange, and repeated trust. Competition (four games): first-price auction, Vickrey auction, all-pay auction, and Colonel Blotto. Negotiation (three games): Nash demand, alternating offers, and multi-issue. Risk (four games): chicken, high-stakes chicken, signaling, and cheap talk. Games were selected to represent the canonical paradigms of experimental economics and game theory, ensuring that each category includes at least one game with a known equilibrium prediction against which model behaviour can be benchmarked.

\subsubsection{Models and provider infrastructure}\label{models-and-provider-infrastructure}

The 25 models span seven providers, accessed through four API endpoints: Anthropic (direct API: Claude Haiku 4.5, Haiku 4.5 Thinking, Sonnet 4.5, Sonnet 4.6, Opus 4.5, Opus 4.6), OpenAI (direct API: GPT-4o Mini, GPT-4.1, GPT-4.1 Mini, GPT-4.1 Nano, GPT-5 Mini, GPT-5 Nano, GPT-5.3, GPT-5.4), Google (Vertex AI for Gemini 3 Pro and 3.1 Pro, OpenRouter for remaining Gemini models: Gemini 2.0 Flash, Gemini 2.5 Flash, Gemini 2.5 Flash Thinking, Gemini 3 Flash), and OpenRouter (DeepSeek V3, DeepSeek R1, LLaMA 3.3 70B, Ministral 14B, Qwen 3.5 Flash). All trials use temperature 1.0. Maximum output tokens are 4,096 for standard models and 16,384 for thinking models. The multi-provider harness handles authentication, rate limiting, exponential-backoff retry logic, and per-call cost tracking.

For thinking models (Claude Haiku 4.5 Thinking, Gemini 2.5 Flash Thinking, DeepSeek R1), internal reasoning traces are captured through provider-specific mechanisms: Anthropic's thinking content blocks, Google's thought response parts, and OpenRouter's reasoning fields. These traces are stored alongside visible response text, enabling post-hoc analysis of internal strategic reasoning.

\subsubsection{Experimental design}\label{experimental-design}

Two complementary designs generate the dataset. Design F tests 16 models (primarily budget-tier models from all providers) in a complete cross-play design: every model plays against every other model and against itself, across all 38 games, with five trials per cell, yielding 54,560 trials. Design G tests nine frontier models (Anthropic Sonnet 4.5, Sonnet 4.6, Opus 4.5, Opus 4.6; OpenAI GPT-4.1, GPT-5.3, GPT-5.4; Google Gemini 3 Pro, Gemini 3.1 Pro) in a strategy-play design: each model plays against deterministic strategies across all 38 games, with one trial per cell, yielding approximately 3,300 trials. Both designs include strategy-play conditions where models face deterministic opponents drawn from the game theory literature: tit-for-tat, grim trigger, Pavlov, always cooperate, always defect, random, mirror, anti-mirror, and others (13 strategies for prisoner's dilemma, three for battle of the sexes, three for stag hunt, four for chicken, three for trust games, and three generic strategies applicable to all games). Total: 51,906 trials, 578,425 rounds, 826,990 decisions.

\subsubsection{Metric extraction}\label{metric-extraction}

Per-trial metrics are extracted by category-specific functions. Cooperation games yield cooperation rate, joint cooperation rate, forgiveness rate (probability of cooperating after being defected against), and retaliation rate (probability of defecting after being defected against). Coordination games yield coordination rate and preferred-option rate. Fairness games yield offer ratio, offer amount, and rejection rate. Strategic depth games yield mean guess, k-level estimate (number of iterated best-response levels), distance to equilibrium, and backward induction compliance (for centipede games). Trust games yield amount sent, trust index, and amount returned. Competition games yield mean bid, bid ratio, and bid standard deviation. Negotiation games yield demand ratio and demand standard deviation. Risk games yield risk-taking rate and safe rate. All metrics are computed from raw round-by-round data in the trial JSON files and aggregated into a behavioural profile for each model-game-opponent combination. The complete profile dataset (51,906 rows, 45 columns) serves as input to all downstream analyses.

\subsubsection{Behavioural fingerprinting (hodoscope pipeline)}\label{behavioural-fingerprinting-hodoscope-pipeline}

The hodoscope pipeline\textsuperscript{\citeproc{ref-zhong2026hodoscope}{47}} embeds reasoning traces into a shared semantic space to detect provider-level signatures in strategic reasoning. Reasoning traces (the visible text response from each round) are extracted from all trial JSON files and filtered to exclude traces shorter than 30 characters, yielding 545,691 traces. These are subsampled to 41,520 (uniformly across 34 agents, including nine deterministic strategy baselines that serve as anchors in the embedding space) and embedded into a 384-dimensional space using TF-IDF vectorisation (5,000-feature vocabulary with sublinear term frequency scaling) with SVD dimensionality reduction and L2 normalisation.

From these embeddings, we compute per-model centroids (mean embedding vector), per-model spread (mean pairwise cosine distance among a model's traces), cosine distances between all model centroids, and UMAP, t-SNE, and PCA projections for visualisation. Hierarchical clustering uses average linkage (UPGMA), valid for non-Euclidean distance metrics, and assigns models to clusters at k = 2, 3, 4, and 5. Provider separation is quantified using silhouette scores with provider identity as the cluster label. Jensen-Shannon distances between choice distributions within each game are computed with Laplace smoothing. Twenty-four lexical features per model (frequency of strategy-relevant terms such as ``cooperate,'' ``defect,'' ``fairness,'' ``equilibrium,'' ``Nash,'' ``Pareto,'' ``exploit,'' and ``retaliation'' per 1,000 words of reasoning text) complement the embedding analysis.

\subsubsection{Endgame and reciprocity analysis}\label{endgame-and-reciprocity-analysis}

For cooperation games with ten-round play, round-by-round cooperation rates are computed for each model. The endgame effect is defined as the change in cooperation rate between round 9 and round 10. Reciprocity analysis extracts sequential round pairs and computes P(cooperate at round \emph{t} \textbar{} opponent cooperated at round \emph{t}-1) and P(cooperate at round \emph{t} \textbar{} opponent defected at round \emph{t}-1), pooled across all cooperation games and opponents for each model. The difference between these two conditional probabilities defines the reciprocity index.

\subsubsection{Generational drift analysis}\label{generational-drift-analysis}

Models within each provider are ordered chronologically by release date and model family lineage. Cooperation rates are computed using the cooperation game subset, which includes all ten cooperation games. Generational trends within provider families are assessed using the Cochran-Armitage test for trend. The generational cooperation metric reported in the main text uses cooperation rates across all cooperation games and opponent types within each design (Design F or Design G), ensuring that comparisons across generations use consistent game and opponent compositions.

\subsubsection{Statistical analysis}\label{statistical-analysis}

Primary analysis uses model-level summary statistics with Wilson 95 per cent confidence intervals for binomial proportions (cooperation rates, conditional cooperation probabilities). Effect sizes for pairwise comparisons between cooperation rates are reported as Cohen's \emph{h} (the arcsine transformation of proportions), where \emph{h} = 0.20, 0.50, and 0.80 correspond to small, medium, and large effects. Cross-model comparisons use independent-samples tests with provider as a grouping variable. All tests are two-sided. Confidence intervals for Design G models (one trial per cell) are wider than for Design F models (five trials per cell), reflecting reduced statistical precision for frontier models. Complete model specifications, API endpoints, and trial counts are provided in Supplementary Table 1.

\subsubsection{Data collection and quality}\label{data-collection-and-quality}

Raw API responses are stored as JSON files with full trial metadata (model key, game ID, opponent, condition, trial number, timestamp, temperature, parsed choices, payoffs, reasoning text, thinking traces). Each filename follows a deterministic format based on the game, model pair, condition, trial number, and an MD5 hash that prevents duplicate API calls on re-execution. The runner checks file existence before each call, making data collection idempotent and safely resumable.

\subsubsection{Counterbalancing and internal validity}\label{counterbalancing-and-internal-validity}

All products of randomisation (option labels, presentation order) are independently drawn per trial, ensuring that label effects and order effects average out across the five-trial design (Design F) and one-trial design (Design G). Game IDs are sorted by length descending before filename matching to prevent prefix collisions. Temperature 1.0 is fixed across all trials and models.

\subsection{Data Availability}\label{data-availability}

The complete dataset comprising 51,906 game-theoretic trials across 25 models, 38 games, and all experimental conditions, including round-by-round choice data, reasoning traces, and per-trial metadata, is available at \url{https://github.com/FelipeMAffonso/strategic-personalities}. An interactive dashboard for exploring the data is available at \url{https://felipemaffonso.github.io/strategic-personalities/}.

\subsection{Code Availability}\label{code-availability}

Code to reproduce all results, including the game engine, experimental platform, multi-provider API harness, analysis pipeline, and figure generation scripts, is available at \url{https://github.com/FelipeMAffonso/strategic-personalities}. Running \texttt{python\ reproduce.py} regenerates all figures and statistical tests reported in this paper from the included dataset. No API keys or GPU are required for reproduction.

\subsection{Ethics Statement}\label{ethics-statement}

This research involves computational experiments with commercial language model APIs only. No human subjects, animals, or personal data were used. The study analyses AI agent behaviour through controlled game-theoretic tasks and does not involve deception, manipulation of real markets, or any interventions that could affect actual individuals or businesses.

\subsection{Author Contributions}\label{author-contributions}

F.M.A. designed the study, developed the experimental platform, conducted the experiments, analysed the data, and wrote the manuscript.

\subsection{Competing Interests}\label{competing-interests}

The author declares no competing interests.

\subsection{Supplementary Information}\label{supplementary-information}

The online version contains supplementary material including Supplementary Notes 1-12, Supplementary Tables 1-6, and Supplementary Figures 1-4.

\subsection{References}\label{references}

\protect\phantomsection\label{refs}

\newpage

\subsection{Extended Data Figures}\label{extended-data-figures}

\includegraphics[width=6.6in,height=\textheight,keepaspectratio]{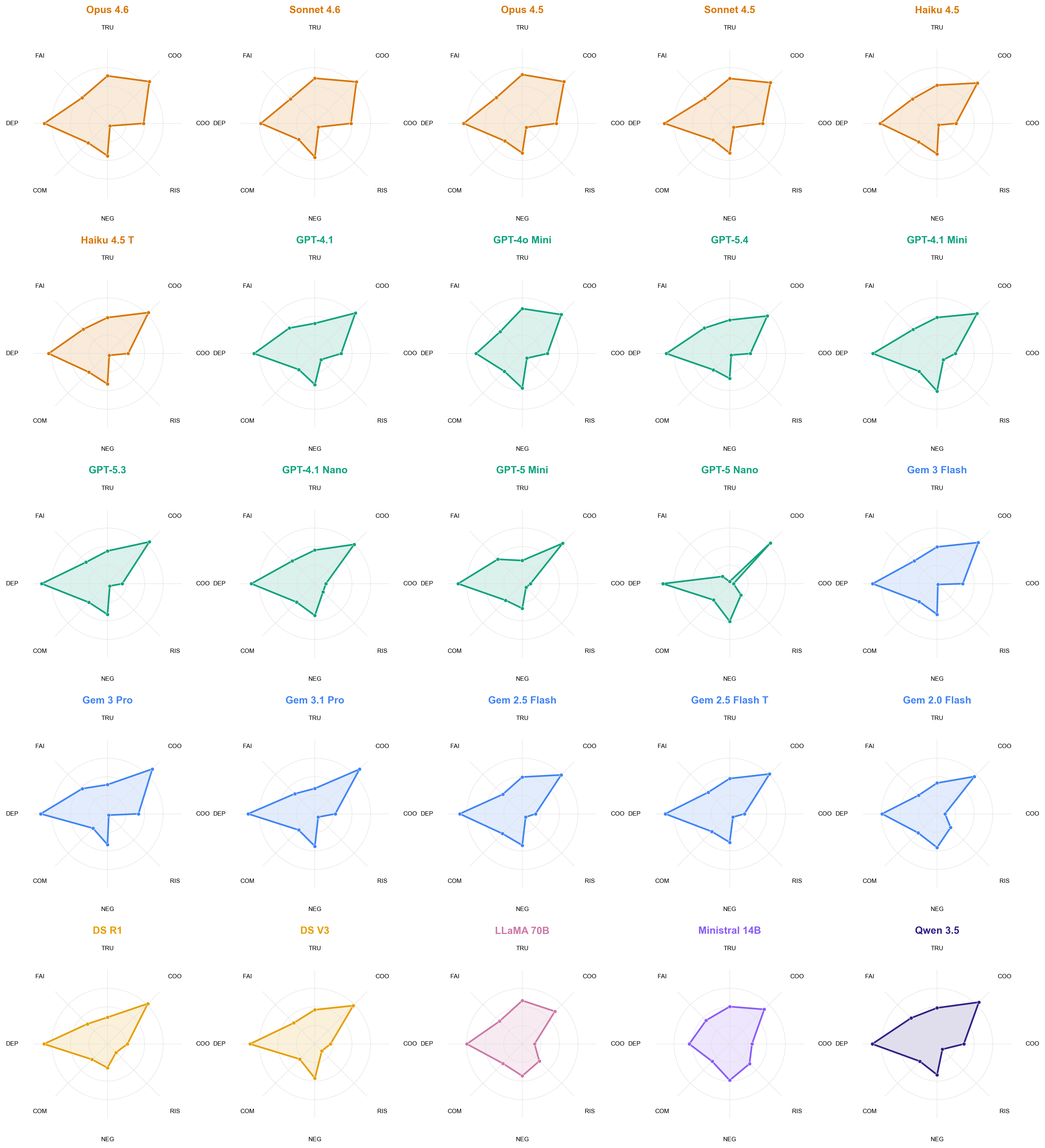}

\textbf{Extended Data Figure 1 \textbar{} Behavioural radar profiles for all 25 models.} Eight-dimensional radar charts showing normalised scores on cooperation, coordination, fairness, strategic depth, trust, competitiveness, aggressiveness, and risk-taking for each model. Models arranged in a 5 x 5 grid grouped by provider and coloured using the Okabe-Ito palette: Anthropic (blue), OpenAI (vermillion), Google (teal), DeepSeek (amber), Meta (pink), Mistral (sky blue), Alibaba (indigo). The convergence in upper dimensions (coordination, depth, competitiveness) and divergence in lower dimensions (cooperation, trust) is visible as shared shape in the upper radar arms and variable shape in the lower arms. N = 51,906 trials across 25 models and 38 games.

\newpage

\includegraphics[width=6.5in,height=\textheight,keepaspectratio]{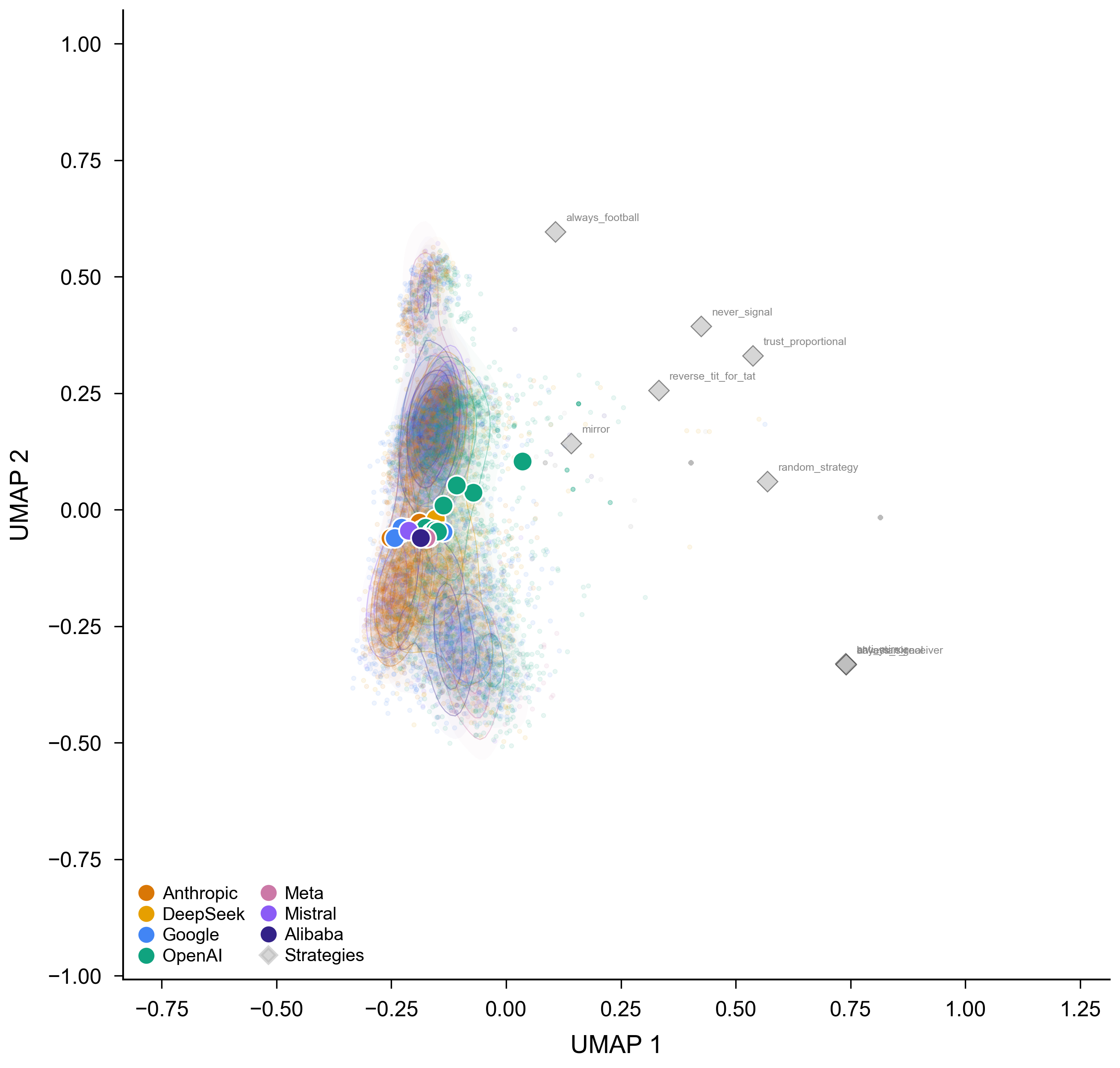}

\textbf{Extended Data Figure 2 \textbar{} Behavioural embedding space.} UMAP projection of 10,000 reasoning trace embeddings (384 dimensions, TF-IDF with SVD), subsampled uniformly across 34 agents (25 language models and 9 deterministic strategy baselines). Per-provider Gaussian KDE density contour fills (bandwidth 0.3) create a topographic terrain coloured by provider (Okabe-Ito palette), with higher-density regions more opaque, showing where each provider's reasoning traces concentrate in the embedding space. Small background dots show individual traces at low opacity. Large circles show model centroids. Anthropic models (blue) occupy a compact, high-density region. DeepSeek models (amber) are proximate to Anthropic. OpenAI models (vermillion) scatter across the embedding space with diffuse density. Grey diamonds mark deterministic strategy baselines (tit-for-tat, always cooperate, always defect, random), which occupy distinct peripheral positions. Overall provider separation silhouette: 0.001.

\newpage

\includegraphics[width=6.6in,height=\textheight,keepaspectratio]{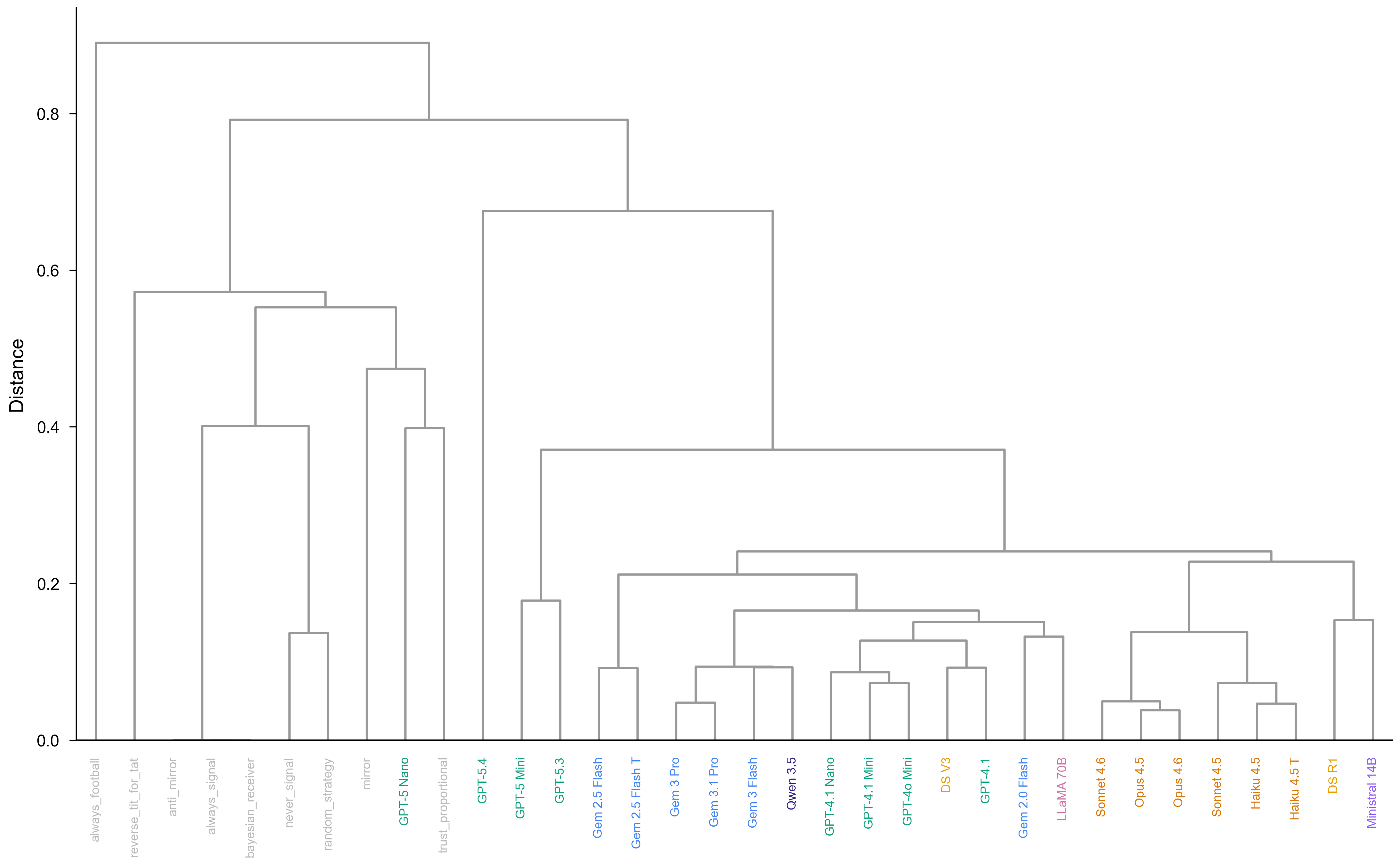}

\textbf{Extended Data Figure 3 \textbar{} Hierarchical clustering dendrogram.} Average-linkage clustering of 34 agent centroids (25 models and 9 deterministic baselines) in the behavioural embedding space. Leaf labels coloured by provider (Okabe-Ito palette). Anthropic models (blue, right side) form a tight cluster with low within-cluster distance. Deterministic strategies (grey, left side) occupy distant branches. OpenAI models (vermillion) are dispersed across the tree, with GPT-4o Mini positioned closer to Google and DeepSeek models than to GPT-5 Nano. At the highest split, one cluster contains all LLMs and the other contains deterministic strategy baselines.

\newpage

\includegraphics[width=6.6in,height=\textheight,keepaspectratio]{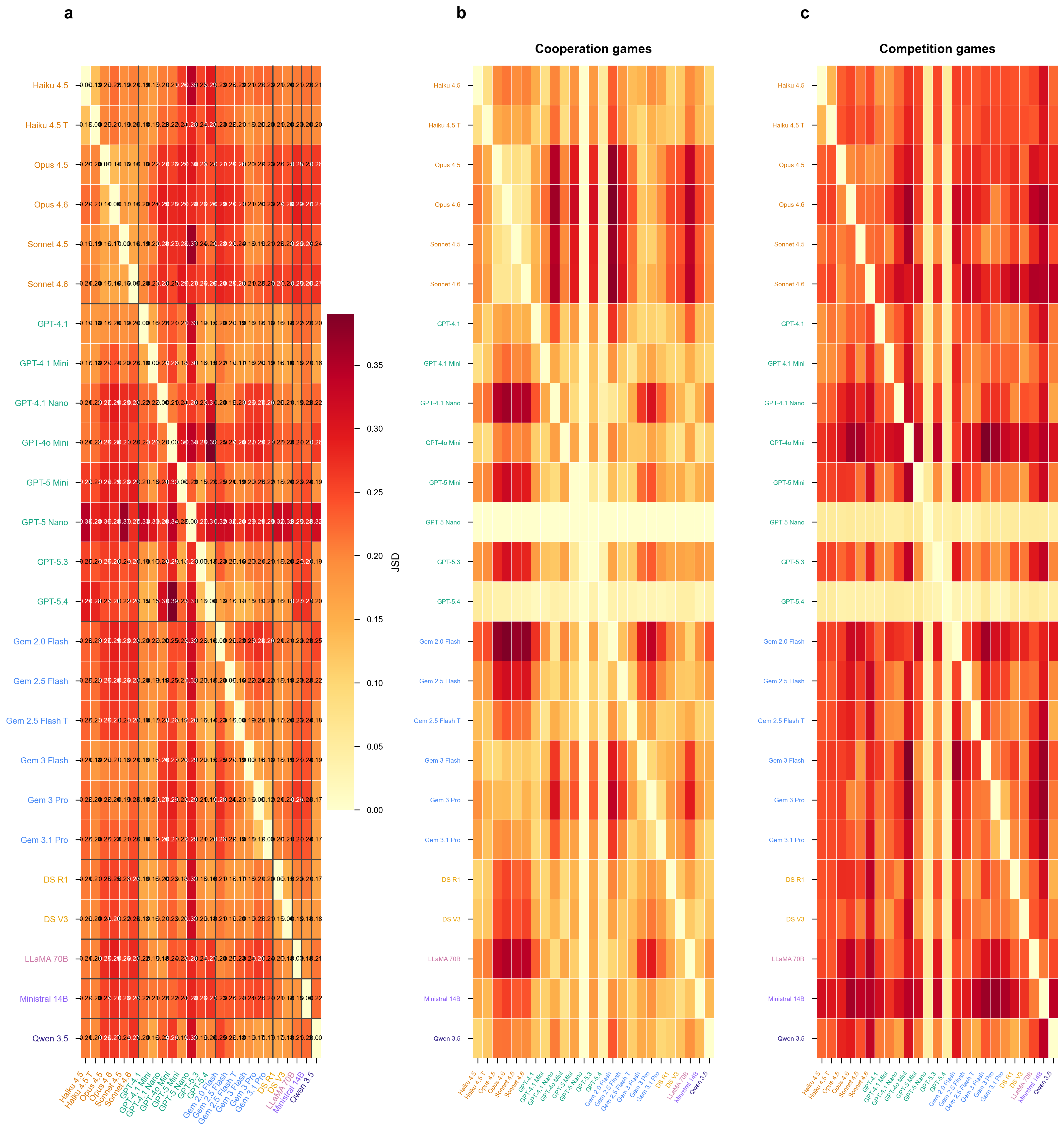}

\textbf{Extended Data Figure 4 \textbar{} Jensen-Shannon distance matrices.} Pairwise Jensen-Shannon distances between choice distributions across all 38 games. \textbf{a}, Full 25 x 25 model distance matrix (YlOrRd colormap), ordered by provider with provider-coloured axis labels. Within-Anthropic distances (upper-left block) are visibly smaller than between-provider distances, while within-OpenAI distances span a wide range. \textbf{b}, Mean JSD across prosocial games (prisoner's dilemma variants, stag hunt, trust games, public goods). \textbf{c}, Mean JSD across competitive games (auctions, hawk-dove, beauty contest, centipede). Cooperation game JSD shows greater inter-model divergence than competition game JSD.

\newpage

\includegraphics[width=6.5in,height=\textheight,keepaspectratio]{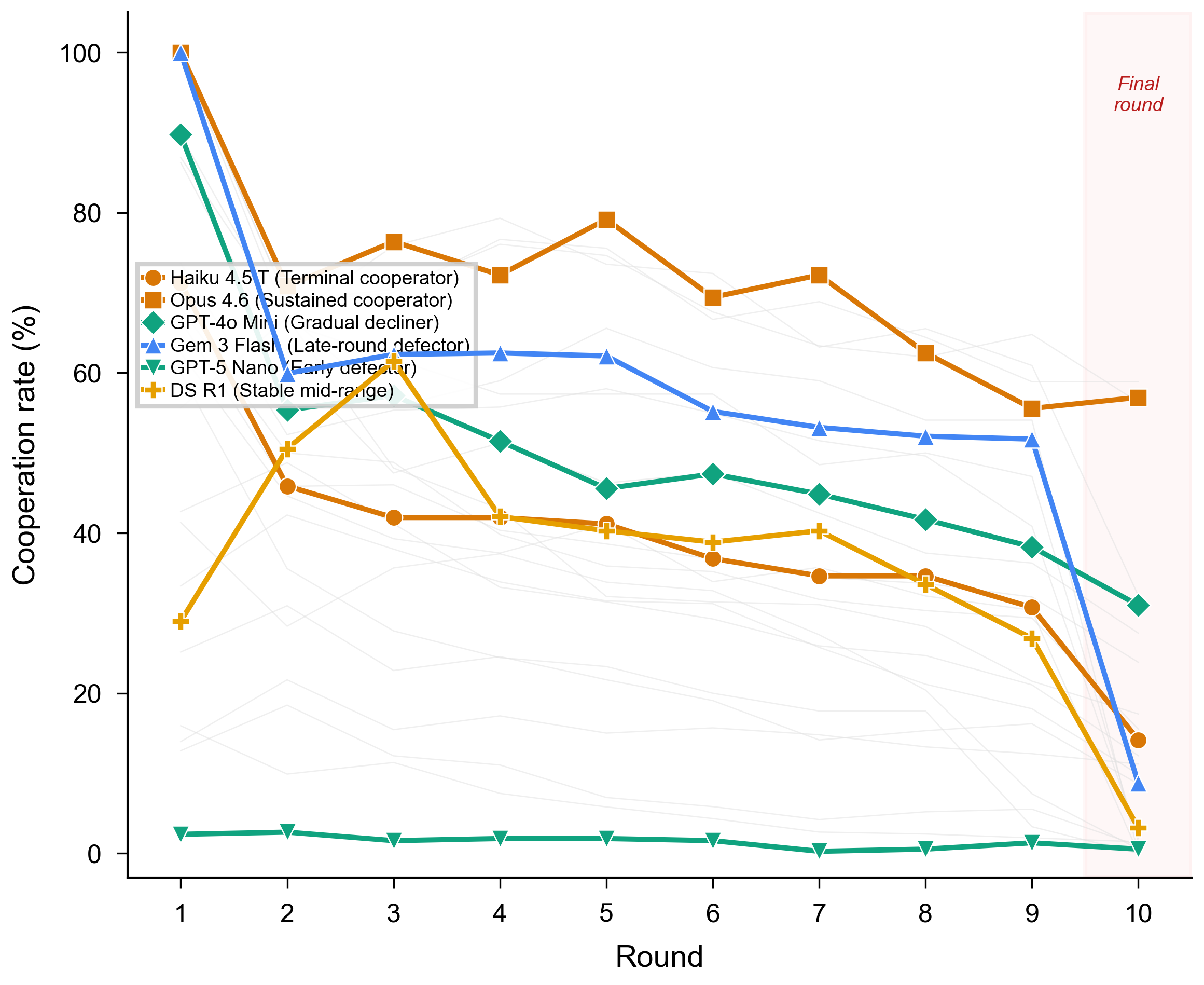}

\textbf{Extended Data Figure 5 \textbar{} Round-by-round cooperation trajectories.} Cooperation rate across ten rounds for six representative model archetypes, with all other models shown as thin grey background traces. Claude Haiku 4.5 Thinking (blue circle, terminal cooperator) sustains cooperation above 55 per cent through round 9 before declining at round 10. Claude Opus 4.6 (blue square, sustained cooperator) maintains a stable mid-range trajectory. GPT-4o Mini (vermillion diamond, gradual decliner) shows steady erosion across rounds. Gemini 3 Flash (teal triangle, late-round defector) sustains moderate cooperation before a sharp final-round decline. GPT-5 Nano (vermillion inverted triangle, early defector) drops to near zero within two rounds. DeepSeek R1 (amber hexagon, stable mid-range) maintains a flat trajectory throughout. The pink-shaded region marks the final round where endgame effects are most pronounced.

\newpage

\includegraphics[width=6.6in,height=\textheight,keepaspectratio]{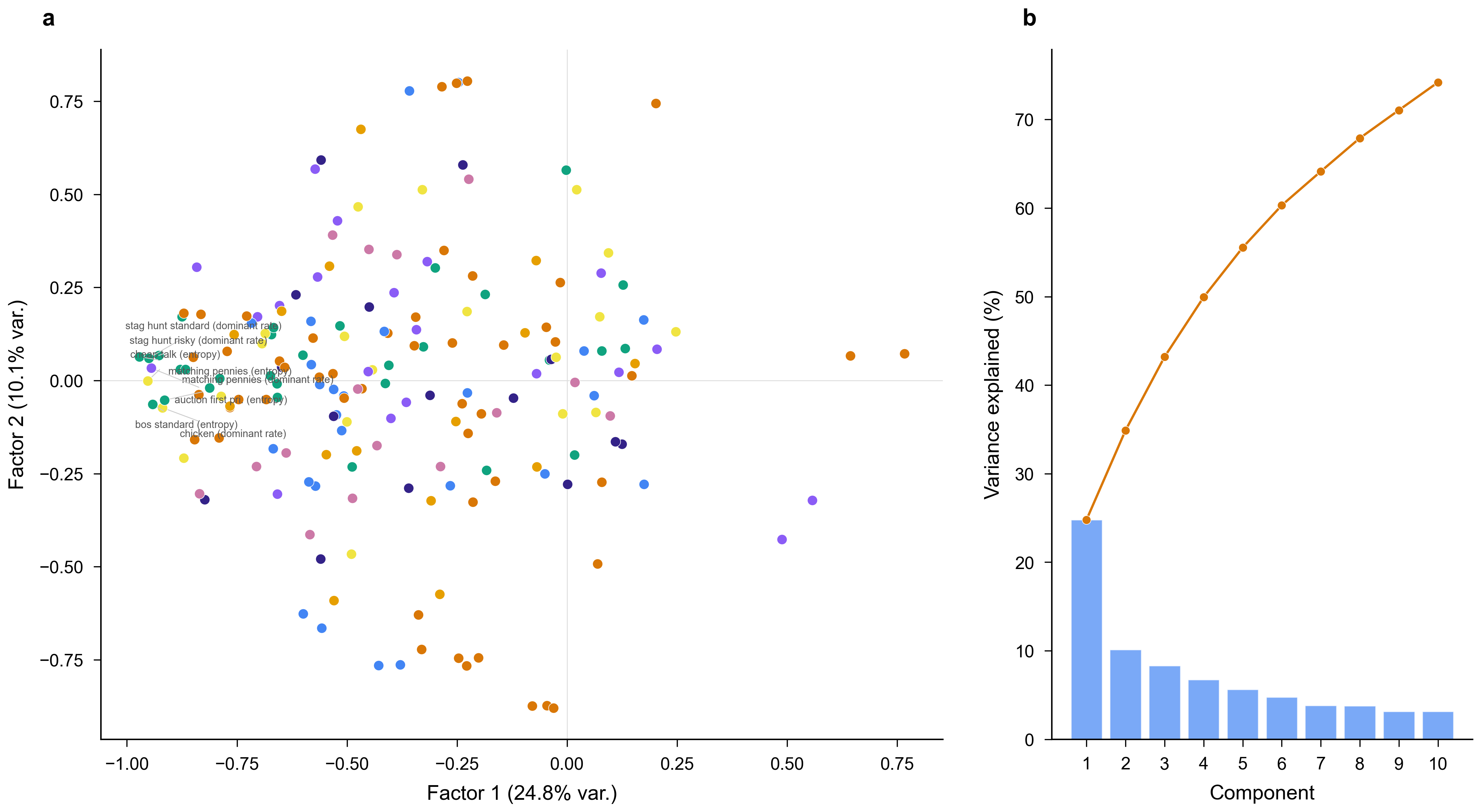}

\textbf{Extended Data Figure 6 \textbar{} Factor analysis of behavioural dimensions.} \textbf{a}, PCA factor loadings of per-game behavioural metrics (entropy, consistency, first-round rate, endgame shift, dominant rate) across 25 models. Each dot represents one game-metric combination, coloured by game category. Factor 1 (24.8 per cent variance) captures overall behavioural intensity; Factor 2 (10.1 per cent variance) separates games by strategic structure. Labels show the eight most extreme-loading features. \textbf{b}, Scree plot showing individual (bars) and cumulative (line) variance explained by each principal component. The first two components explain approximately 35 per cent of total variance, consistent with the two-dimensional structure (prosociality and strategic competence) identified in the main text.

\newpage

\includegraphics[width=6.5in,height=\textheight,keepaspectratio]{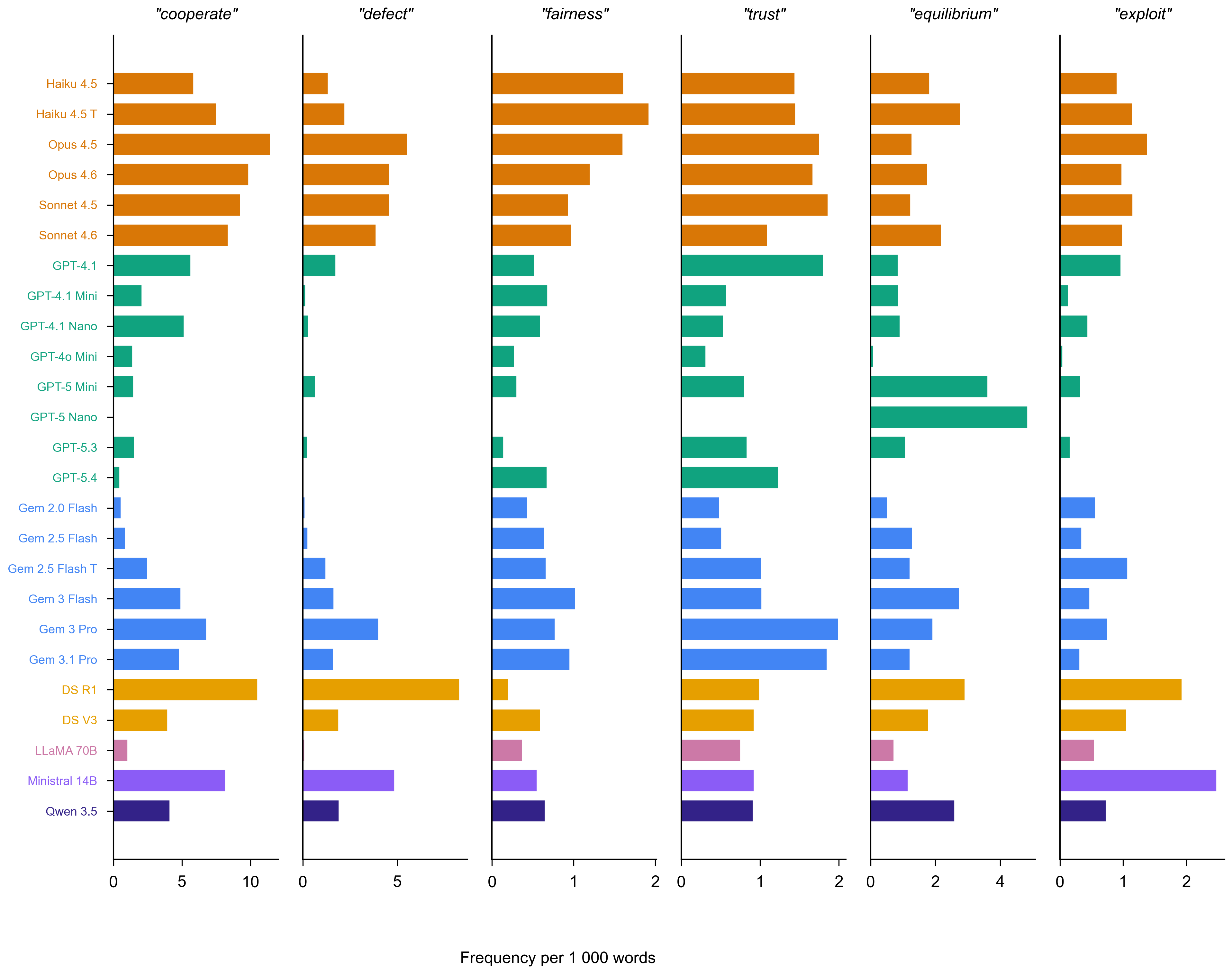}

\textbf{Extended Data Figure 7 \textbar{} Lexical features of strategic reasoning.} Horizontal bar charts showing frequency of six strategy-relevant terms across all 25 models, ordered by provider. Each sub-panel displays one term, with bar length proportional to frequency (per 1,000 words of reasoning text) and bar colour indicating provider. Anthropic models reference ``fairness'' at rates two to three times higher than other providers. GPT-5 series models reference ``exploit'' and ``defect'' at elevated rates. Gemini 3 models reference ``trust'' and ``cooperate'' at high rates through round 9, then switch to ``defect'' terminology at round 10.

\newpage

\includegraphics[width=6.5in,height=\textheight,keepaspectratio]{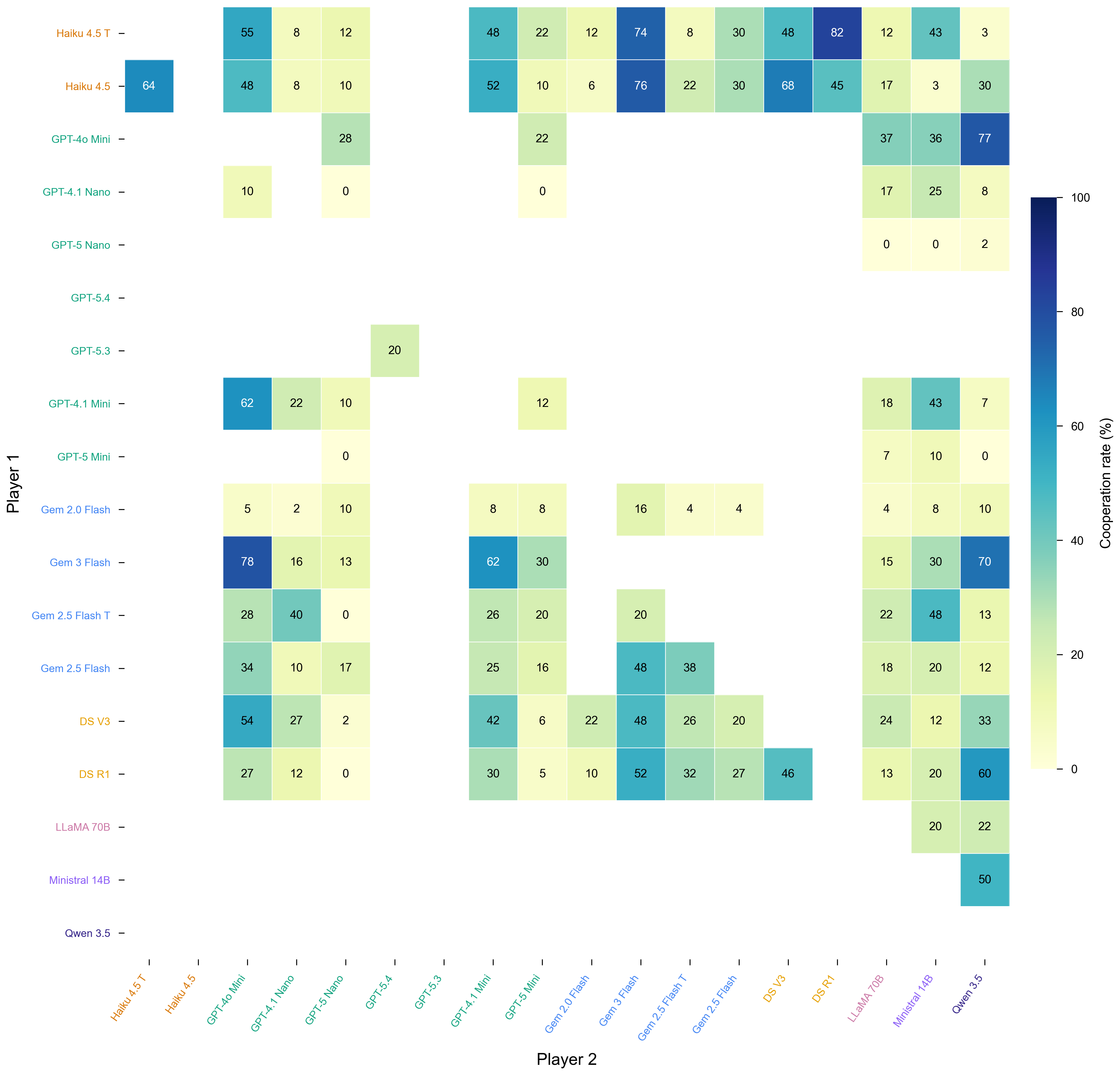}

\textbf{Extended Data Figure 8 \textbar{} Cross-play cooperation matrices.} Cooperation rates in the canonical prisoner's dilemma for all 120 pairwise matchups among 16 Design F models (five trials per pair). Each cell shows the mean cooperation rate across ten rounds. Diagonal cells (self-play) show within-model consistency. Off-diagonal asymmetry reveals interaction effects: cooperative models paired with defecting models show suppressed cooperation relative to their baseline rates. Matrix ordered by provider.

\newpage

\includegraphics[width=6.5in,height=\textheight,keepaspectratio]{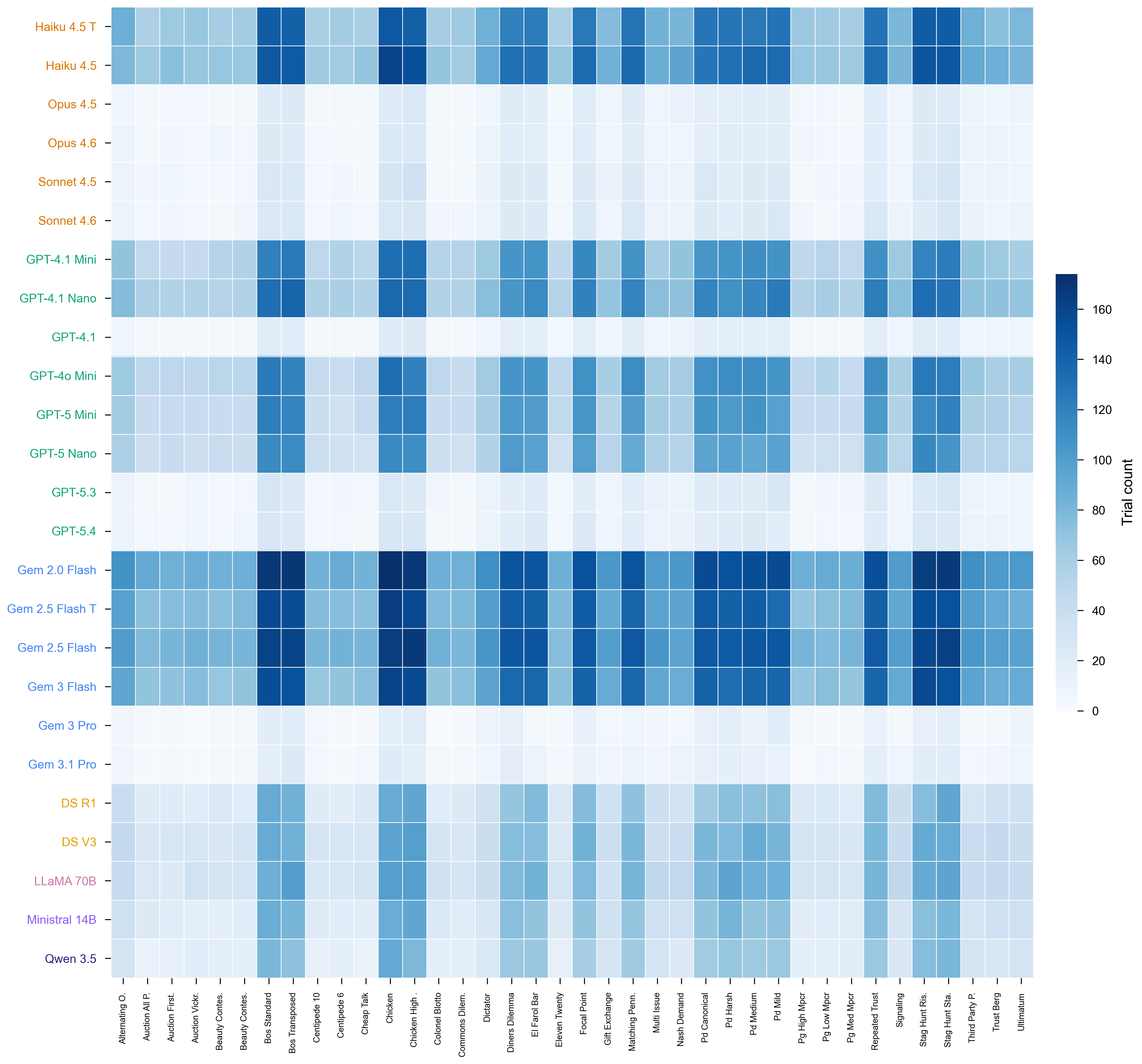}

\textbf{Extended Data Figure 9 \textbar{} Experimental coverage matrix.} Trial counts per model-game combination for all 25 models and 38 games. Design F models (16 models) show complete coverage with five trials per cell. Design G models (9 frontier models) show one trial per cell. Colour intensity indicates trial count. Total: 51,906 trials, 578,425 rounds, 826,990 decisions.

\newpage

\begin{center}
{\LARGE\bfseries Supplementary Information for:\\[0.5em]
Large language models converge on competitive rationality but diverge on cooperation across providers and generations\par}

\vspace{0.8em}

{\large Felipe M. Affonso\par}

\vspace{0.3em}

{\small\itshape Spears School of Business, Oklahoma State University, Stillwater, USA\par}
\end{center}

\vspace{1em}

\subsection{Contents}\label{contents}

\textbf{Supplementary Materials and Methods}

Supplementary Note 1: Complete game specifications and payoff matrices

Supplementary Note 2: Prompt templates and label randomisation

Supplementary Note 3: Deterministic strategy implementations

Supplementary Note 4: Model access and API configuration

Supplementary Note 5: Behavioural metric extraction procedures

Supplementary Note 6: Hodoscope pipeline technical details

Supplementary Note 7: Endgame analysis methodology

Supplementary Note 8: Reciprocity analysis methodology

Supplementary Note 9: Statistical methods and multiple comparisons

Supplementary Note 10: Design F and Design G protocol details

Supplementary Note 11: Cost tracking and budget allocation

Supplementary Note 12: Data quality checks and exclusion criteria

\textbf{Supplementary Tables}

Supplementary Table 1: Complete model catalogue with API endpoints

Supplementary Table 2: Cooperation rates by model and game

Supplementary Table 3: Endgame cooperation by round and model

Supplementary Table 4: Reciprocity statistics for all 25 models

Supplementary Table 5: Provider-level summary statistics

Supplementary Table 6: Hodoscope lexical feature frequencies

\textbf{Supplementary Figures}

Supplementary Figure 1: Cooperation rate distributions by game category

Supplementary Figure 2: Strategy-play response profiles

Supplementary Figure 3: Cross-play interaction effects

Supplementary Figure 4: Thinking model comparison

\newpage

\subsection{Reading Guide}\label{reading-guide}

These supplementary materials document the methodological foundation for the strategic personality characterisation reported in the main text. Notes 1 through 5 provide complete experimental specifications: the 38 game implementations with payoff matrices, prompt architecture with randomisation procedures, the deterministic strategy opponents, model access configuration, and the metric extraction pipeline that transforms raw API responses into the behavioural profile dataset. Notes 6 through 8 detail the three analytical pipelines: the hodoscope behavioural fingerprinting system, the endgame analysis framework, and the reciprocity classification methodology. Notes 9 through 12 address statistical infrastructure and data quality: correction procedures for multiple comparisons, the distinction between the two experimental designs, cost tracking across providers, and the exclusion criteria applied during data collection. Tables S1 through S6 consolidate the complete model catalogue, per-game cooperation rates, round-by-round endgame trajectories, conditional cooperation statistics, provider-level aggregates, and lexical feature frequencies. Supplementary Figures S1 through S4 extend the main figures with per-game cooperation distributions, strategy-play response profiles, cross-play interaction effects, and thinking model comparisons.

For readers with limited time, we suggest the following priority tiers: (1) Essential for evaluating claims: Note 1 (game specifications), Note 5 (metric extraction), Note 7 (endgame methodology), Note 9 (statistical methods). (2) Methodological detail: Notes 2 through 4 for replication, Note 6 for hodoscope pipeline, Note 12 for data quality. (3) Extensions: Notes 10 through 11 and Tables S2 through S6 for comprehensive model-level and game-level results.

\newpage

\subsection{Supplementary Note 1: Complete game specifications and payoff matrices}\label{supplementary-note-1-complete-game-specifications-and-payoff-matrices}

The experimental platform implements 38 games spanning eight categories. Each game specification includes payoff matrices, equilibrium predictions, the number of rounds, and the specific decision space available to each player. Below we list each game with its canonical payoff structure and equilibrium prediction.

\subsubsection{Cooperation games (10 games)}\label{cooperation-games-10-games}

\textbf{Prisoner's Dilemma, Canonical (pd\_canonical).} Payoffs: mutual cooperation (3,3), mutual defection (1,1), temptation to defect (5,0), sucker's payoff (0,5). Nash equilibrium: mutual defection. Pareto optimum: mutual cooperation. The canonical PD uses the standard payoff matrix from experimental economics.

\textbf{Prisoner's Dilemma, Harsh (pd\_harsh).} Payoffs: mutual cooperation (3,3), mutual defection (1,1), temptation (7,0), sucker (0,7). The increased temptation payoff (7 versus 5) makes defection more attractive relative to the canonical variant, following the K parameter variation of Akata et al.\textsuperscript{\citeproc{ref-akata2025playing}{14}}.

\textbf{Prisoner's Dilemma, Medium (pd\_medium).} Payoffs: mutual cooperation (3,3), mutual defection (1,1), temptation (6,0), sucker (0,6). Intermediate between canonical and harsh variants.

\textbf{Prisoner's Dilemma, Mild (pd\_mild).} Payoffs: mutual cooperation (3,3), mutual defection (2,2), temptation (4,0), sucker (0,4). Reduced defection incentive relative to canonical.

\textbf{Public Goods Game, High MPCR (pg\_high\_mpcr).} Four-player public goods with marginal per capita return of 0.8 (above individual cost of contribution). Each player chooses to contribute 0 to 10 tokens. Total contributions are multiplied by MPCR and shared equally. Nash equilibrium: zero contribution. Social optimum: full contribution.

\textbf{Public Goods Game, Medium MPCR (pg\_med\_mpcr).} MPCR of 0.5. The lower return makes contributing less individually profitable, increasing the free-riding incentive.

\textbf{Public Goods Game, Low MPCR (pg\_low\_mpcr).} MPCR of 0.3. Strong free-riding incentive.

\textbf{Commons Dilemma (commons\_dilemma).} Shared resource of 100 units. Each player harvests 0 to 50 units. If total harvest exceeds 70, the resource collapses and all payoffs are zero. Otherwise payoffs equal individual harvest. Nash equilibrium: harvest just below the collapse threshold. Social optimum: equitable harvest well below threshold.

\textbf{Diner's Dilemma (diners\_dilemma).} Group dining scenario where the bill is split equally. Each player chooses between a cheap item (cost 5, value 5) and an expensive item (cost 15, value 12). Nash equilibrium: everyone orders expensive (free-riding on the shared bill). Social optimum: everyone orders cheap.

\textbf{El Farol Bar (el\_farol\_bar).} Players simultaneously decide whether to attend a bar. If fewer than 60 per cent attend, attending is better than staying home. If more attend, staying home is better. No pure-strategy Nash equilibrium exists. Based on Arthur's bounded rationality model\textsuperscript{\citeproc{ref-arthur1994inductive}{60}}.

\subsubsection{Coordination games (6 games)}\label{coordination-games-6-games}

\textbf{Battle of the Sexes, Standard (bos\_standard).} Two players coordinate on one of two options: (10,7) for joint choice A, (7,10) for joint choice B, (0,0) for miscoordination. Two pure-strategy Nash equilibria exist. Tests coordination ability with asymmetric preferences.

\textbf{Battle of the Sexes, Transposed (bos\_transposed).} Payoffs reversed: (7,10) for joint A, (10,7) for joint B. Controls for label bias by swapping which option favours which player.

\textbf{Stag Hunt, Standard (stag\_hunt\_standard).} Payoffs: (4,4) for mutual stag, (0,3) for stag versus hare, (3,0) for hare versus stag, (2,2) for mutual hare. Two pure Nash equilibria: stag (payoff-dominant) and hare (risk-dominant).

\textbf{Stag Hunt, Risky (stag\_hunt\_risky).} Modified payoffs increasing the risk of choosing stag: (5,5) for mutual stag, (0,4) for stag versus hare, (4,0) for hare versus stag, (3,3) for mutual hare.

\textbf{Matching Pennies (matching\_pennies).} Zero-sum game. Player 1 wins on match, player 2 wins on mismatch. Unique Nash equilibrium: 50/50 mixed strategy. Tests randomisation ability.

\textbf{Focal Point (focal\_point).} Coordination game with multiple equilibria where players must converge on a salient option without communication. Tests Schelling focal point reasoning.

\subsubsection{Fairness games (3 games)}\label{fairness-games-3-games}

\textbf{Ultimatum Game (ultimatum).} Proposer offers a split of 100 tokens. Responder accepts (both receive offered amounts) or rejects (both receive zero). Subgame-perfect equilibrium: offer 1, accept. Typical human offers: 40 to 50 per cent\textsuperscript{\citeproc{ref-guth1982experimental}{9}}.

\textbf{Dictator Game (dictator).} Proposer allocates 100 tokens between self and recipient. Recipient has no veto. Nash prediction: keep all. Typical human allocation: 20 to 30 per cent to recipient.

\textbf{Third-Party Punishment (third\_party\_punishment).} Observer watches a dictator allocation and can pay a cost to reduce the dictator's payoff. Tests willingness to enforce fairness norms at personal cost.

\subsubsection{Strategic depth games (5 games)}\label{strategic-depth-games-5-games}

\textbf{Beauty Contest, p=2/3 (beauty\_contest\_23).} Players choose integers 0 to 100. Winner is closest to 2/3 of the group mean. Nash equilibrium: 0 (via iterated elimination of dominated strategies). The k-level model predicts level-0 players choose 50, level-1 choose 33, level-2 choose 22, and so on\textsuperscript{\citeproc{ref-nagel1995unraveling}{32}}.

\textbf{Beauty Contest, p=1/2 (beauty\_contest\_12).} Same as above with p=1/2. Nash equilibrium: 0. Level-1 chooses 25.

\textbf{Centipede Game, 6 nodes (centipede\_6).} Sequential game where players alternate taking or passing. Taking ends the game and gives the taker a slightly larger payoff than passing would have yielded. Backward induction predicts taking at the first node\textsuperscript{\citeproc{ref-rosenthal1981games}{42}}. Human subjects typically pass several nodes.

\textbf{Centipede Game, 10 nodes (centipede\_10).} Extended version with 10 decision nodes, allowing more scope for backward induction failure.

\textbf{11-20 Money Request (eleven\_twenty).} Players simultaneously request 11 to 20 dollars. If requests sum to 20 or less, both receive their request. If requests sum to more than 20, lower request receives their amount and higher receives nothing (ties split). Nash equilibrium: both request 15. Tests strategic sophistication.

\subsubsection{Trust games (3 games)}\label{trust-games-3-games}

\textbf{Berg Trust Game (trust\_berg).} Investor sends 0 to 10 tokens. Amount is tripled. Trustee returns 0 to 3x the sent amount. Subgame-perfect equilibrium: send nothing, return nothing. Human investors typically send 50 per cent of endowment\textsuperscript{\citeproc{ref-berg1995trust}{31}}.

\textbf{Gift Exchange (gift\_exchange).} Employer sets a wage, worker chooses effort level. Higher effort is costly but benefits the employer. Tests reciprocal fairness in labour market settings.

\textbf{Repeated Trust (repeated\_trust).} Berg trust game repeated for 10 rounds with full history. Tests whether trust and reciprocity build over repeated interactions.

\subsubsection{Competition games (4 games)}\label{competition-games-4-games}

\textbf{First-Price Auction (auction\_first\_price).} Players submit sealed bids. Highest bidder wins and pays their bid. Equilibrium: bid below true value (strategic shading). Tests ability to shade bids optimally.

\textbf{Vickrey Auction (auction\_vickrey).} Second-price sealed-bid auction. Highest bidder wins but pays the second-highest bid. Dominant strategy: bid true value\textsuperscript{\citeproc{ref-spence1973job}{58}}.

\textbf{All-Pay Auction (auction\_all\_pay).} All bidders pay their bids regardless of whether they win. Winner takes the prize. Mixed-strategy Nash equilibrium.

\textbf{Colonel Blotto (colonel\_blotto).} Players allocate forces across multiple battlefields. Winner of each battlefield determined by who allocates more. Tests multi-dimensional strategic reasoning.

\subsubsection{Negotiation games (3 games)}\label{negotiation-games-3-games}

\textbf{Nash Demand Game (nash\_demand).} Players simultaneously demand a share of 100 tokens. If demands sum to 100 or less, both receive their demand. If demands exceed 100, both receive zero\textsuperscript{\citeproc{ref-nash1950bargaining}{36}}.

\textbf{Alternating Offers (alternating\_offers).} Rubinstein bargaining protocol. Players alternate proposing splits. Delay is costly (shrinking pie). Subgame-perfect equilibrium: first-mover advantage.

\textbf{Multi-Issue Negotiation (multi\_issue).} Players negotiate over three issues simultaneously, each with different value to each player. Tests integrative bargaining ability.

\subsubsection{Risk games (4 games)}\label{risk-games-4-games}

\textbf{Chicken (chicken).} Players simultaneously choose to swerve or go straight. Swerving is safe but yielding. Going straight risks collision. Two pure Nash equilibria (one swerves, one goes straight) plus a mixed-strategy equilibrium.

\textbf{Chicken, High Stakes (chicken\_high\_stakes).} Same as chicken with amplified collision cost. Tests whether higher stakes affect risk-taking.

\textbf{Signaling Game (signaling).} Sender observes a type and sends a signal. Receiver observes the signal and takes an action. Tests whether models can implement separating and pooling equilibria\textsuperscript{\citeproc{ref-spence1973job}{58},\citeproc{ref-crawford1982strategic}{59}}.

\textbf{Cheap Talk (cheap\_talk).} Pre-play communication without commitment. Sender sends a message, receiver chooses an action. Tests whether models use non-binding communication strategically\textsuperscript{\citeproc{ref-crawford1982strategic}{59}}.

\newpage

\subsection{Supplementary Note 2: Prompt templates and label randomisation}\label{supplementary-note-2-prompt-templates-and-label-randomisation}

All games follow the three-block prompt architecture established by Akata et al.\textsuperscript{\citeproc{ref-akata2025playing}{14}}: RULES, HISTORY, and QUERY. The RULES block describes the game mechanics, payoff structure, number of rounds, and role assignment. The HISTORY block provides a round-by-round narrative of prior actions and payoffs, growing with each successive round. The QUERY block requests the model's action for the current round.

\subsubsection{Label randomisation}\label{label-randomisation}

To control for option-label anchoring effects, each trial uses one of six label pairs drawn randomly from: (J,F), (Q,X), (R,H), (Y,W), (T,N), (P,M). These pairs were selected by Akata et al.~to avoid semantic associations with cooperation or defection.

Within each round, the presentation order of the two options is independently randomised, so that (for example) ``cooperate'' might map to option J in one round and option F in the next, and J might appear before F in one round but after in the next. This two-level randomisation (label pair selection per trial, option order per round) ensures that any observed cooperation is not an artefact of label preference or position bias.

\subsubsection{Prompt example (canonical prisoner's dilemma, round 1)}\label{prompt-example-canonical-prisoners-dilemma-round-1}

The RULES block describes the game in natural language. Payoffs are presented as textual rules following Akata et al.'s finding that textual presentation reduces format-dependent artefacts compared with matrix representations. The prompt explicitly states the total number of rounds (``You will play 10 rounds in total with the same player'') and the round number (``This is round 1 of 10'').

\newpage

\subsection{Supplementary Note 3: Deterministic strategy implementations}\label{supplementary-note-3-deterministic-strategy-implementations}

Both experimental designs include conditions where language models play against deterministic strategy opponents drawn from the game theory literature. These fixed-strategy opponents serve two purposes: they provide controlled stimulus conditions (the model's opponent behaviour is identical across models, eliminating confounding from opponent adaptation) and they anchor the behavioural embedding space with known reference strategies.

The following strategies are implemented:

\textbf{Cooperation game strategies (13):} Always cooperate (100 per cent cooperation every round), always defect (0 per cent cooperation every round), tit-for-tat (cooperate in round 1, then copy opponent's previous action), grim trigger (cooperate until opponent defects, then defect permanently), Pavlov or win-stay-lose-shift (cooperate if previous round outcomes matched, defect otherwise), generous tit-for-tat (tit-for-tat with 10 per cent forgiveness on defection), suspicious tit-for-tat (defect in round 1, then tit-for-tat), reverse tit-for-tat (defect in round 1, do opposite of opponent's last action), random (50/50 each round), mirror (copy opponent's last action), anti-mirror (do opposite of opponent's last action), gradual (punish with increasing defection streaks), and soft majority (cooperate if opponent cooperated more than half the time).

\textbf{Coordination game strategies (3):} Always option A, always option B, and alternate between A and B.

\textbf{Other game-specific strategies:} Each game category includes two to four strategies calibrated to its equilibrium structure (for example, proportional return in trust games, equilibrium bidding in auctions, and Nash demand in bargaining).

\newpage

\subsection{Supplementary Note 4: Model access and API configuration}\label{supplementary-note-4-model-access-and-api-configuration}

\subsubsection{API endpoints}\label{api-endpoints}

{\def\LTcaptype{none} % do not increment counter
\begin{longtable}[]{@{}
  >{\raggedright\arraybackslash}p{(\linewidth - 6\tabcolsep) * \real{0.2273}}
  >{\raggedright\arraybackslash}p{(\linewidth - 6\tabcolsep) * \real{0.1818}}
  >{\raggedright\arraybackslash}p{(\linewidth - 6\tabcolsep) * \real{0.2273}}
  >{\raggedright\arraybackslash}p{(\linewidth - 6\tabcolsep) * \real{0.3636}}@{}}
\toprule\noalign{}
\begin{minipage}[b]{\linewidth}\raggedright
Provider
\end{minipage} & \begin{minipage}[b]{\linewidth}\raggedright
Models
\end{minipage} & \begin{minipage}[b]{\linewidth}\raggedright
Endpoint
\end{minipage} & \begin{minipage}[b]{\linewidth}\raggedright
Authentication
\end{minipage} \\
\midrule\noalign{}
\endhead
\bottomrule\noalign{}
\endlastfoot
Anthropic & Claude Haiku 4.5, Haiku 4.5 Thinking, Sonnet 4.5, Sonnet 4.6, Opus 4.5, Opus 4.6 & Direct API (api.anthropic.com) & API key \\
OpenAI & GPT-4o Mini, GPT-4.1, GPT-4.1 Mini, GPT-4.1 Nano, GPT-5 Mini, GPT-5 Nano, GPT-5.3, GPT-5.4 & Direct API (api.openai.com) & API key \\
Google & Gemini 3 Pro, Gemini 3.1 Pro & Vertex AI & Service account \\
Google & Gemini 2.0 Flash, 2.5 Flash, 2.5 Flash Thinking, 3 Flash & OpenRouter & API key \\
Other & DeepSeek V3, DeepSeek R1, LLaMA 3.3 70B, Ministral 14B, Qwen 3.5 Flash & OpenRouter & API key \\
\end{longtable}
}

\subsubsection{Configuration parameters}\label{configuration-parameters}

All trials use temperature 1.0, which produces stochastic responses that capture the model's distributional behaviour. Maximum output tokens are set to 4,096 for standard models and 16,384 for thinking models (Claude Haiku 4.5 Thinking, Gemini 2.5 Flash Thinking, DeepSeek R1). No system prompt is used. The multi-provider harness implements exponential-backoff retry logic (up to 5 retries with 2x backoff starting at 1 second) and per-call cost tracking based on provider-specific token pricing.

\subsubsection{Thinking model trace capture}\label{thinking-model-trace-capture}

For thinking models, internal reasoning traces are captured through provider-specific mechanisms. Anthropic provides thinking content blocks in the API response. Google provides thought response parts. OpenRouter provides reasoning fields for DeepSeek R1. These traces are stored alongside visible response text in the trial JSON files, enabling post-hoc analysis of internal strategic deliberation. The thinking traces are not included in the main behavioural fingerprinting analysis (which uses only visible response text) but are available for supplementary analysis.

\newpage

\subsection{Supplementary Note 5: Behavioural metric extraction procedures}\label{supplementary-note-5-behavioural-metric-extraction-procedures}

Per-trial metrics are extracted by category-specific functions that process the raw round-by-round data. Each trial JSON file contains the complete sequence of actions, payoffs, and reasoning text for all rounds.

\subsubsection{Cooperation metrics}\label{cooperation-metrics}

From each cooperation game trial: \textbf{cooperation rate} (fraction of rounds in which the model chose the cooperative action), \textbf{joint cooperation rate} (fraction of rounds with mutual cooperation), \textbf{forgiveness rate} (P(cooperate at round \emph{t} \textbar{} opponent defected at round \emph{t}-1)), \textbf{retaliation rate} (P(defect at round \emph{t} \textbar{} opponent defected at round \emph{t}-1)), \textbf{first-round cooperation} (binary indicator of round 1 cooperation).

\subsubsection{Coordination metrics}\label{coordination-metrics}

\textbf{Coordination rate} (fraction of rounds with matching choices), \textbf{preferred-option rate} (fraction of rounds choosing the option that gives the model the higher payoff in the coordination equilibrium).

\subsubsection{Fairness metrics}\label{fairness-metrics}

\textbf{Offer ratio} (amount offered divided by total endowment, for proposer roles), \textbf{offer amount} (absolute amount offered), \textbf{rejection rate} (fraction of offers rejected, for responder roles).

\subsubsection{Strategic depth metrics}\label{strategic-depth-metrics}

\textbf{Mean guess} (average numerical guess in beauty contests), \textbf{k-level estimate} (number of iterated best-response levels, computed as log(guess/50) / log(p)), \textbf{distance to equilibrium} (absolute distance from the Nash equilibrium prediction), \textbf{backward induction compliance} (fraction of centipede game rounds where the model takes at the equilibrium node).

\subsubsection{Trust metrics}\label{trust-metrics}

\textbf{Amount sent} (fraction of endowment sent in trust games), \textbf{trust index} (normalised measure of trusting behaviour, scaled 0 to 1), \textbf{amount returned} (fraction of received amount returned by the trustee).

\subsubsection{Competition metrics}\label{competition-metrics}

\textbf{Mean bid} (average bid in auction games), \textbf{bid ratio} (bid divided by true value), \textbf{bid standard deviation} (within-trial variability of bids across rounds).

\subsubsection{Negotiation metrics}\label{negotiation-metrics}

\textbf{Demand ratio} (demanded share as fraction of total pie), \textbf{demand standard deviation} (within-trial demand variability).

\subsubsection{Risk metrics}\label{risk-metrics}

\textbf{Risk-taking rate} (fraction of rounds choosing the risky option in chicken games), \textbf{safe rate} (fraction of rounds choosing the safe option).

\newpage

\subsection{Supplementary Note 6: Hodoscope pipeline technical details}\label{supplementary-note-6-hodoscope-pipeline-technical-details}

The hodoscope pipeline takes its name from the particle physics detector (from the Greek \emph{hodos}, path, and \emph{skopein}, to observe) that reconstructs a particle's trajectory from the signals it leaves in successive detector layers. In our application, the ``particles'' are language models and the ``trajectories'' are the reasoning traces they generate across hundreds of strategic decisions. By embedding these traces into a shared semantic space, the pipeline reconstructs each model's behavioural trajectory through the space of strategic reasoning, revealing provider-level signatures that are invisible in aggregate choice data. The pipeline processes reasoning traces through the following stages.

\subsubsection{Trace extraction and filtering}\label{trace-extraction-and-filtering}

Reasoning traces (the visible text response from each round) are extracted from all trial JSON files. Traces shorter than 30 characters are excluded to remove parsing artefacts and error responses. This yields 545,691 valid traces across all 25 models, all 38 games, and all opponent conditions.

\subsubsection{Subsampling}\label{subsampling}

To ensure equal representation across agents in the embedding space, traces are subsampled to 41,520 total (uniformly across 34 agents: 25 language models and 9 deterministic strategy baselines). The strategy baselines generate deterministic ``traces'' from their decision logic, serving as interpretable anchors in the embedding space. For example, the always-cooperate strategy generates traces like ``I will cooperate'' while tit-for-tat generates conditional reasoning traces.

\subsubsection{Embedding}\label{embedding}

Traces are embedded into a 384-dimensional space using TF-IDF vectorisation (term frequency inverse document frequency with a 5,000-feature vocabulary, sublinear term frequency scaling, English stop-word removal, and L2 normalisation) followed by SVD (singular value decomposition) dimensionality reduction from the 5,000-dimensional TF-IDF space to 384 dimensions. This approach captures lexical patterns in strategic reasoning while being computationally efficient for the dataset size. The 384-dimensional embedding space is sufficient to distinguish reasoning styles across models (e.g., Anthropic's emphasis on fairness language versus OpenAI's emphasis on optimality language) while keeping the centroid and distance computations tractable.

\subsubsection{Centroid computation and clustering}\label{centroid-computation-and-clustering}

Per-model centroids are computed as the mean embedding vector across all of a model's traces. Pairwise cosine distances between centroids define the model distance matrix. Hierarchical clustering uses average linkage (UPGMA), which is valid for non-Euclidean distance metrics such as cosine distance. Cluster assignments are computed at k = 2, 3, 4, and 5.

\subsubsection{Provider separation}\label{provider-separation}

Silhouette scores are computed with provider identity as the cluster label. For each model, the silhouette is s = (b - a) / max(a, b), where a is the mean distance to same-provider model centroids and b is the mean distance to other-provider model centroids. Positive silhouette indicates tighter clustering with same-provider models, negative silhouette indicates closer proximity to other providers.

\subsubsection{Dimensionality reduction for visualisation}\label{dimensionality-reduction-for-visualisation}

UMAP (n\_neighbors=15, min\_dist=0.1), t-SNE (perplexity=10), and PCA projections are computed on the 384-dimensional centroid vectors for visualisation in two dimensions. All three methods produce qualitatively similar provider-clustering patterns.

\subsubsection{Jensen-Shannon divergence}\label{jensen-shannon-divergence}

Per-game Jensen-Shannon distances between choice distributions are computed with Laplace smoothing (adding 1 to all choice counts before normalisation) to handle zero-count cells.

\subsubsection{Lexical features}\label{lexical-features}

Twenty-four lexical features per model are computed as the frequency of strategy-relevant terms per 1,000 words of reasoning text. Terms include: cooperate, defect, fairness, trust, strategy, optimality, equilibrium, risk, opponent, mutual, punishment, reward, reciprocity, maximizing, best response, Nash, Pareto, dominant, exploit, forgiveness, retaliation, tit-for-tat, long-term, and short-term.

\newpage

\subsection{Supplementary Note 7: Endgame analysis methodology}\label{supplementary-note-7-endgame-analysis-methodology}

\subsubsection{Round-by-round cooperation}\label{round-by-round-cooperation}

For each cooperation game trial with ten-round play, the binary cooperation indicator at each round is recorded. Model-level cooperation rates per round are computed by averaging across all cooperation game trials for that model, pooling across games, opponents, and trial replications within each design.

\subsubsection{Endgame effect}\label{endgame-effect}

The endgame effect for each model is defined as the cooperation rate at round 10 minus the cooperation rate at round 9. A large negative endgame effect indicates strategic defection in the final round. The three endgame types identified in the main text are classified by the magnitude and sign of this effect:

Type 1 (terminal-value cooperation): round 10 cooperation rate above 50 per cent. These models sustain majority cooperation even when defection cannot be punished.

Type 2 (strategic exploitation): round 9 cooperation above 50 per cent and round 10 cooperation below 10 per cent. These models build cooperative reputations before exploiting accumulated trust in the final round, consistent with the Kreps et al.\textsuperscript{\citeproc{ref-kreps1982rational}{46}} reputation-building equilibrium.

Type 3 (unconditional defection): cooperation rate below 20 per cent throughout all rounds. These models do not engage in strategic cooperation at any point.

\subsubsection{Sample sizes}\label{sample-sizes}

The aggregate endgame analysis pools N = 7,668 ten-round cooperation game sequences across all 25 models. Per-model sample sizes range from 7 sequences (frontier models in Design G) to 20 sequences (Design F models). The larger confidence intervals for frontier models reflect the single-trial Design G protocol.

\newpage

\subsection{Supplementary Note 8: Reciprocity analysis methodology}\label{supplementary-note-8-reciprocity-analysis-methodology}

\subsubsection{Conditional cooperation computation}\label{conditional-cooperation-computation}

For each model, we extract all sequential round pairs (round \emph{t}-1, round \emph{t}) from cooperation game trials. Round pairs are partitioned into two conditions: opponent cooperated at round \emph{t}-1 (reciprocation condition) and opponent defected at round \emph{t}-1 (forgiveness condition). The model's cooperation rate is computed separately for each condition.

P(cooperate \textbar{} opponent cooperated) measures reciprocation: how reliably the model rewards opponent cooperation with its own cooperation.

P(cooperate \textbar{} opponent defected) measures forgiveness: how often the model cooperates despite the opponent's defection.

\subsubsection{Archetype classification}\label{archetype-classification}

Six archetypes are defined by thresholds on these two conditional probabilities:

Strict reciprocators: P(C\textbar C) \textgreater{} 0.60 and P(C\textbar D) \textless{} 0.10. High reciprocation, low forgiveness. Implements a strategy close to tit-for-tat.

Grudge-holders: P(C\textbar C) \textgreater{} 0.60 and P(C\textbar D) \textless{} 0.03. A strict subset of reciprocators with near-zero forgiveness, implementing a strategy close to grim trigger.

Moderate reciprocators: positive reciprocity index (P(C\textbar D) \textless{} P(C\textbar C)) but failing to meet strict reciprocator criteria because P(C\textbar C) falls below 0.60 or P(C\textbar D) exceeds 0.10, with P(C\textbar C) at or above 0.40. These models discriminate between cooperative and defecting opponents but do so less sharply than strict reciprocators.

Weak reciprocators: 0.20 \textless{} P(C\textbar C) \textless{} 0.40 with P(C\textbar D) \textless{} P(C\textbar C). Positive but attenuated reciprocity, cooperating at low baseline rates that nonetheless respond to opponent cooperation.

Unconditional defectors: P(C\textbar C) \textless{} 0.20 and P(C\textbar D) \textless{} 0.10. Low sensitivity to opponent actions. Fixed defection regardless of social context.

Anti-reciprocators: P(C\textbar D) \textgreater{} P(C\textbar C). Cooperate more after opponent defection than after opponent cooperation. No equilibrium strategy corresponds to this pattern.

\subsubsection{Reciprocity index}\label{reciprocity-index}

The reciprocity index is defined as P(C\textbar C) - P(C\textbar D). Positive values indicate reciprocal behaviour (cooperation rewards cooperation). Zero indicates no sensitivity to opponent actions. Negative values indicate anti-reciprocal behaviour.

\newpage

\subsection{Supplementary Note 9: Statistical methods and multiple comparisons}\label{supplementary-note-9-statistical-methods-and-multiple-comparisons}

\subsubsection{Confidence intervals}\label{confidence-intervals}

All cooperation rates and conditional probabilities are reported with Wilson 95 per cent confidence intervals for binomial proportions. The Wilson interval outperforms the Wald interval for proportions near 0 or 1 and for small sample sizes, both of which are relevant for Design G frontier models.

\subsubsection{Provider-level tests}\label{provider-level-tests}

Cross-model comparisons use provider as a grouping variable. Given the small number of models per provider (2 to 8), we report descriptive provider-level summary statistics (mean, range, coefficient of variation) and use these to characterise provider-level patterns without relying on formal hypothesis tests that would be underpowered at these sample sizes.

\subsubsection{Multiple comparisons}\label{multiple-comparisons}

The 25 models and 38 games generate a large number of potential pairwise comparisons. We do not conduct exhaustive pairwise testing. Instead, the analysis focuses on pre-specified contrasts: within-provider generational trends (Cochran-Armitage test for trend), between-provider cooperation differences (provider-level means and ranges), and the endgame effect (round 9 versus round 10 within each model). Effect sizes (cooperation rate differences) and confidence intervals are reported throughout.

\subsubsection{All tests are two-sided}\label{all-tests-are-two-sided}

All reported statistical tests are two-sided. No directional hypotheses were pre-registered.

\newpage

\subsection{Supplementary Note 10: Design F and Design G protocol details}\label{supplementary-note-10-design-f-and-design-g-protocol-details}

\subsubsection{Design F: Complete cross-play design}\label{design-f-complete-cross-play-design}

Design F tests 16 models (primarily budget-tier models from all seven providers) in a complete pairwise design. Every model plays against every other model (cross-play), against itself (self-play), and against the deterministic strategy baselines (strategy-play), across all 38 games. Each model-opponent-game cell includes five trial replications, yielding approximately 54,560 trials. The complete cross-play design enables analysis of interaction effects between model pairs and the construction of per-game cooperation matrices.

Design F models: Claude Haiku 4.5, Claude Haiku 4.5 Thinking, GPT-4o Mini, GPT-4.1 Mini, GPT-4.1 Nano, GPT-5 Mini, GPT-5 Nano, Gemini 2.0 Flash, Gemini 2.5 Flash, Gemini 2.5 Flash Thinking, Gemini 3 Flash, DeepSeek V3, DeepSeek R1, LLaMA 3.3 70B, Ministral 14B, Qwen 3.5 Flash.

\subsubsection{Design G: Frontier strategy-play design}\label{design-g-frontier-strategy-play-design}

Design G tests nine frontier models in a strategy-play design. Each model plays against the deterministic strategy baselines across all 38 games, with one trial per cell. The single-trial design reduces cost (frontier models are 10 to 30 times more expensive per trial) while maintaining coverage across all games and strategy conditions. Design G does not include cross-play or self-play conditions.

Design G models: Claude Sonnet 4.5, Claude Sonnet 4.6, Claude Opus 4.5, Claude Opus 4.6, GPT-4.1, GPT-5.3, GPT-5.4, Gemini 3 Pro, Gemini 3.1 Pro.

\subsubsection{Design integration}\label{design-integration}

When reporting aggregate statistics (for example, overall cooperation rates), Design F and Design G results are pooled. When reporting cross-play statistics (for example, interaction effects, cooperation matrices), only Design F data are used. Generational drift analysis uses cooperation rates computed within each design to ensure consistent game and opponent compositions across models of different generations.

\newpage

\subsection{Supplementary Note 11: Cost tracking and budget allocation}\label{supplementary-note-11-cost-tracking-and-budget-allocation}

\subsubsection{Per-provider costs}\label{per-provider-costs}

{\def\LTcaptype{none} % do not increment counter
\begin{longtable}[]{@{}
  >{\raggedright\arraybackslash}p{(\linewidth - 8\tabcolsep) * \real{0.1235}}
  >{\raggedright\arraybackslash}p{(\linewidth - 8\tabcolsep) * \real{0.2099}}
  >{\raggedright\arraybackslash}p{(\linewidth - 8\tabcolsep) * \real{0.2099}}
  >{\raggedright\arraybackslash}p{(\linewidth - 8\tabcolsep) * \real{0.2222}}
  >{\raggedright\arraybackslash}p{(\linewidth - 8\tabcolsep) * \real{0.2346}}@{}}
\toprule\noalign{}
\begin{minipage}[b]{\linewidth}\raggedright
Provider
\end{minipage} & \begin{minipage}[b]{\linewidth}\raggedright
Total cost (USD)
\end{minipage} & \begin{minipage}[b]{\linewidth}\raggedright
Number of trials
\end{minipage} & \begin{minipage}[b]{\linewidth}\raggedright
Number of models
\end{minipage} & \begin{minipage}[b]{\linewidth}\raggedright
Mean cost per trial
\end{minipage} \\
\midrule\noalign{}
\endhead
\bottomrule\noalign{}
\endlastfoot
Anthropic & 1,028 & 9,292 & 6 & 0.111 \\
OpenAI & 317 & 15,834 & 8 & 0.020 \\
Google & 308 & 17,606 & 6 & 0.017 \\
DeepSeek & 211 & 3,810 & 2 & 0.055 \\
Meta & 30 & 2,105 & 1 & 0.014 \\
Mistral & 9 & 1,730 & 1 & 0.005 \\
Alibaba & 8 & 1,529 & 1 & 0.005 \\
\end{longtable}
}

Total project cost: USD 1,950 for 51,906 trials across 25 models.

\subsubsection{Cost drivers}\label{cost-drivers}

Anthropic accounts for 53 per cent of total cost despite contributing only 18 per cent of trials, reflecting the high per-token cost of frontier Opus models (approximately USD 0.153 per trial for Claude Opus 4.6). Budget-tier models from Mistral and Alibaba cost approximately USD 0.005 per trial, a 30-fold difference from the most expensive model, with no detectable difference in strategic competence (coordination, competition, and strategic depth scores are comparable across price tiers).

\newpage

\subsection{Supplementary Note 12: Data quality checks and exclusion criteria}\label{supplementary-note-12-data-quality-checks-and-exclusion-criteria}

\subsubsection{Parse failure handling}\label{parse-failure-handling}

Each API response is parsed to extract the model's action choice. The parser identifies the action by matching the randomised option labels (J, F, Q, X, etc.) in the response text. If parsing fails (the model's response does not contain a recognisable option label), the trial is logged as a parse failure and excluded from analysis. Parse failure rates are below 1 per cent across all models and games.

\subsubsection{Idempotent data collection}\label{idempotent-data-collection}

Each trial filename includes an MD5-based hash of the trial parameters (game ID, model keys, condition, trial number). The runner checks file existence before each API call, making data collection idempotent and safely resumable after interruptions. This design ensures that re-running the collection script does not generate duplicate trials or incur unnecessary API costs.

\subsubsection{Response validation}\label{response-validation}

Responses are validated for consistency: the extracted action must be one of the valid options for the current game and round. Actions that do not match any valid option are treated as parse failures. No post-hoc corrections or imputation is applied to invalid responses.

\newpage

\subsection{Supplementary Table 1: Complete model catalogue}\label{supplementary-table-1-complete-model-catalogue}

{\def\LTcaptype{none} % do not increment counter
\begin{longtable}[]{@{}
  >{\raggedright\arraybackslash}p{(\linewidth - 12\tabcolsep) * \real{0.1507}}
  >{\raggedright\arraybackslash}p{(\linewidth - 12\tabcolsep) * \real{0.1781}}
  >{\raggedright\arraybackslash}p{(\linewidth - 12\tabcolsep) * \real{0.1370}}
  >{\raggedright\arraybackslash}p{(\linewidth - 12\tabcolsep) * \real{0.1370}}
  >{\raggedright\arraybackslash}p{(\linewidth - 12\tabcolsep) * \real{0.1096}}
  >{\raggedright\arraybackslash}p{(\linewidth - 12\tabcolsep) * \real{0.1096}}
  >{\raggedright\arraybackslash}p{(\linewidth - 12\tabcolsep) * \real{0.1781}}@{}}
\toprule\noalign{}
\begin{minipage}[b]{\linewidth}\raggedright
Model key
\end{minipage} & \begin{minipage}[b]{\linewidth}\raggedright
Display name
\end{minipage} & \begin{minipage}[b]{\linewidth}\raggedright
Provider
\end{minipage} & \begin{minipage}[b]{\linewidth}\raggedright
Thinking
\end{minipage} & \begin{minipage}[b]{\linewidth}\raggedright
Design
\end{minipage} & \begin{minipage}[b]{\linewidth}\raggedright
Trials
\end{minipage} & \begin{minipage}[b]{\linewidth}\raggedright
API endpoint
\end{minipage} \\
\midrule\noalign{}
\endhead
\bottomrule\noalign{}
\endlastfoot
claude-haiku-4.5 & Claude Haiku 4.5 & Anthropic & No & F & 3,781 & Direct \\
claude-haiku-4.5-thinking & Claude Haiku 4.5 (Thinking) & Anthropic & Yes & F & 3,588 & Direct \\
claude-sonnet-4.5 & Claude Sonnet 4.5 & Anthropic & No & G & 521 & Direct \\
claude-sonnet-4.6 & Claude Sonnet 4.6 & Anthropic & No & G & 513 & Direct \\
claude-opus-4.5 & Claude Opus 4.5 & Anthropic & No & G & 440 & Direct \\
claude-opus-4.6 & Claude Opus 4.6 & Anthropic & No & G & 449 & Direct \\
gpt-4o-mini & GPT-4o Mini & OpenAI & No & F & 2,934 & Direct \\
gpt-4.1 & GPT-4.1 & OpenAI & No & G & 400 & Direct \\
gpt-4.1-mini & GPT-4.1 Mini & OpenAI & No & F & 3,006 & Direct \\
gpt-4.1-nano & GPT-4.1 Nano & OpenAI & No & F & 3,274 & Direct \\
gpt-5-mini & GPT-5 Mini & OpenAI & No & F & 2,739 & Direct \\
gpt-5-nano & GPT-5 Nano & OpenAI & No & F & 2,507 & Direct \\
gpt-5.3 & GPT-5.3 & OpenAI & No & G & 488 & Direct \\
gpt-5.4 & GPT-5.4 & OpenAI & No & G & 486 & Direct \\
gemini-2.0-flash & Gemini 2.0 Flash & Google & No & F & 4,532 & OpenRouter \\
gemini-2.5-flash & Gemini 2.5 Flash & Google & No & F & 4,345 & OpenRouter \\
gemini-2.5-flash-thinking & Gemini 2.5 Flash (Thinking) & Google & Yes & F & 4,107 & OpenRouter \\
gemini-3-flash & Gemini 3 Flash & Google & No & F & 3,982 & OpenRouter \\
gemini-3-pro & Gemini 3 Pro & Google & No & G & 304 & Vertex AI \\
gemini-3.1-pro & Gemini 3.1 Pro & Google & No & G & 336 & Vertex AI \\
deepseek-v3 & DeepSeek V3 & DeepSeek & No & F & 2,016 & OpenRouter \\
deepseek-r1 & DeepSeek R1 & DeepSeek & Yes & F & 1,794 & OpenRouter \\
llama-3.3-70b & LLaMA 3.3 70B & Meta & No & F & 2,105 & OpenRouter \\
ministral-14b & Ministral 14B & Mistral & No & F & 1,730 & OpenRouter \\
qwen3.5-flash & Qwen 3.5 Flash & Alibaba & No & F & 1,529 & OpenRouter \\
\end{longtable}
}

\newpage

\subsection{Supplementary Table 2: Cooperation rates by model}\label{supplementary-table-2-cooperation-rates-by-model}

{\def\LTcaptype{none} % do not increment counter
\begin{longtable}[]{@{}llllll@{}}
\toprule\noalign{}
Model & Mean & Std & N & 95\% CI lower & 95\% CI upper \\
\midrule\noalign{}
\endhead
\bottomrule\noalign{}
\endlastfoot
Claude Opus 4.6 & 71.5 & 33.6 & 72 & 63.8 & 79.3 \\
Claude Sonnet 4.6 & 70.9 & 34.6 & 90 & 63.7 & 78.0 \\
Claude Opus 4.5 & 70.4 & 35.0 & 71 & 62.3 & 78.6 \\
Claude Sonnet 4.5 & 69.0 & 31.1 & 87 & 62.4 & 75.5 \\
Gemini 3 Flash & 56.8 & 35.2 & 549 & 53.8 & 59.7 \\
Gemini 3 Pro & 56.2 & 35.4 & 61 & 47.3 & 65.1 \\
GPT-4.1 & 53.7 & 35.9 & 68 & 45.2 & 62.2 \\
GPT-4o Mini & 50.3 & 34.0 & 439 & 47.1 & 53.4 \\
Qwen 3.5 Flash & 49.4 & 35.2 & 262 & 45.1 & 53.7 \\
GPT-5.4 & 48.4 & 39.2 & 80 & 39.8 & 57.0 \\
Claude Haiku 4.5 & 39.8 & 39.0 & 528 & 36.5 & 43.1 \\
Claude Haiku 4.5 (Thinking) & 39.3 & 37.9 & 508 & 36.0 & 42.6 \\
GPT-4.1 Mini & 38.5 & 37.8 & 432 & 34.9 & 42.0 \\
Ministral 14B & 37.5 & 18.6 & 293 & 35.4 & 39.7 \\
DeepSeek R1 & 36.6 & 27.1 & 283 & 33.5 & 39.8 \\
Gemini 3.1 Pro & 35.7 & 38.8 & 56 & 25.6 & 45.9 \\
DeepSeek V3 & 31.0 & 30.4 & 328 & 27.8 & 34.3 \\
Gemini 2.5 Flash & 28.5 & 30.7 & 587 & 26.0 & 31.0 \\
Gemini 2.5 Flash (Thinking) & 26.5 & 29.1 & 564 & 24.1 & 28.9 \\
GPT-5.3 & 23.0 & 29.2 & 90 & 17.0 & 29.0 \\
LLaMA 3.3 70B & 19.9 & 9.9 & 346 & 18.8 & 20.9 \\
GPT-4.1 Nano & 15.1 & 22.9 & 466 & 13.0 & 17.1 \\
Gemini 2.0 Flash & 8.3 & 9.9 & 616 & 7.6 & 9.1 \\
GPT-5 Mini & 6.3 & 10.7 & 414 & 5.3 & 7.4 \\
GPT-5 Nano & 1.5 & 4.2 & 378 & 1.0 & 1.9 \\
\end{longtable}
}

N = number of cooperation game trials per model. Rates computed across all 10 cooperation games and all opponent conditions within each model's experimental design.

\newpage

\subsection{Supplementary Table 3: Endgame cooperation by round}\label{supplementary-table-3-endgame-cooperation-by-round}

{\def\LTcaptype{none} % do not increment counter
\begin{longtable}[]{@{}
  >{\raggedright\arraybackslash}p{(\linewidth - 22\tabcolsep) * \real{0.1250}}
  >{\raggedright\arraybackslash}p{(\linewidth - 22\tabcolsep) * \real{0.0714}}
  >{\raggedright\arraybackslash}p{(\linewidth - 22\tabcolsep) * \real{0.0714}}
  >{\raggedright\arraybackslash}p{(\linewidth - 22\tabcolsep) * \real{0.0714}}
  >{\raggedright\arraybackslash}p{(\linewidth - 22\tabcolsep) * \real{0.0714}}
  >{\raggedright\arraybackslash}p{(\linewidth - 22\tabcolsep) * \real{0.0714}}
  >{\raggedright\arraybackslash}p{(\linewidth - 22\tabcolsep) * \real{0.0714}}
  >{\raggedright\arraybackslash}p{(\linewidth - 22\tabcolsep) * \real{0.0714}}
  >{\raggedright\arraybackslash}p{(\linewidth - 22\tabcolsep) * \real{0.0714}}
  >{\raggedright\arraybackslash}p{(\linewidth - 22\tabcolsep) * \real{0.0714}}
  >{\raggedright\arraybackslash}p{(\linewidth - 22\tabcolsep) * \real{0.0714}}
  >{\raggedright\arraybackslash}p{(\linewidth - 22\tabcolsep) * \real{0.1607}}@{}}
\toprule\noalign{}
\begin{minipage}[b]{\linewidth}\raggedright
Model
\end{minipage} & \begin{minipage}[b]{\linewidth}\raggedright
R1
\end{minipage} & \begin{minipage}[b]{\linewidth}\raggedright
R2
\end{minipage} & \begin{minipage}[b]{\linewidth}\raggedright
R3
\end{minipage} & \begin{minipage}[b]{\linewidth}\raggedright
R4
\end{minipage} & \begin{minipage}[b]{\linewidth}\raggedright
R5
\end{minipage} & \begin{minipage}[b]{\linewidth}\raggedright
R6
\end{minipage} & \begin{minipage}[b]{\linewidth}\raggedright
R7
\end{minipage} & \begin{minipage}[b]{\linewidth}\raggedright
R8
\end{minipage} & \begin{minipage}[b]{\linewidth}\raggedright
R9
\end{minipage} & \begin{minipage}[b]{\linewidth}\raggedright
R10
\end{minipage} & \begin{minipage}[b]{\linewidth}\raggedright
Endgame effect
\end{minipage} \\
\midrule\noalign{}
\endhead
\bottomrule\noalign{}
\endlastfoot
Claude Opus 4.6 & 100 & 70.8 & 76.4 & 72.2 & 79.2 & 69.4 & 72.2 & 62.5 & 55.6 & 56.9 & +1.4 \\
Claude Opus 4.5 & 100 & 69.0 & 70.4 & 76.1 & 74.6 & 67.6 & 63.4 & 62.0 & 64.8 & 56.3 & -8.4 \\
Claude Sonnet 4.6 & 100 & 68.9 & 70.0 & 76.7 & 75.6 & 66.7 & 68.9 & 64.4 & 58.9 & 58.9 & 0.0 \\
Claude Sonnet 4.5 & 100 & 66.7 & 75.9 & 79.3 & 73.6 & 72.4 & 63.2 & 65.5 & 60.9 & 32.2 & -28.8 \\
Gemini 3 Pro & 86.9 & 67.2 & 55.7 & 59.0 & 65.6 & 60.7 & 59.0 & 54.1 & 54.1 & 0.0 & -54.1 \\
Gemini 3.1 Pro & 62.5 & 44.6 & 39.3 & 37.5 & 41.1 & 33.9 & 35.7 & 32.1 & 30.4 & 0.0 & -30.4 \\
GPT-4.1 & 89.7 & 66.2 & 61.8 & 57.4 & 57.4 & 57.4 & 48.5 & 50.0 & 47.1 & 1.5 & -45.6 \\
GPT-5 Nano & 2.4 & 2.7 & 1.6 & 1.9 & 1.9 & 1.6 & 0.3 & 0.5 & 1.3 & 0.5 & -0.8 \\
Gemini 2.0 Flash & 12.8 & 18.5 & 12.2 & 11.0 & 7.0 & 5.8 & 4.2 & 5.2 & 5.5 & 1.0 & -4.6 \\
\end{longtable}
}

Selected models shown. Full data for all 25 models available in the OSF repository.

\newpage

\subsection{Supplementary Table 4: Reciprocity statistics}\label{supplementary-table-4-reciprocity-statistics}

{\def\LTcaptype{none} % do not increment counter
\begin{longtable}[]{@{}
  >{\raggedright\arraybackslash}p{(\linewidth - 12\tabcolsep) * \real{0.0909}}
  >{\raggedright\arraybackslash}p{(\linewidth - 12\tabcolsep) * \real{0.1169}}
  >{\raggedright\arraybackslash}p{(\linewidth - 12\tabcolsep) * \real{0.1169}}
  >{\raggedright\arraybackslash}p{(\linewidth - 12\tabcolsep) * \real{0.2468}}
  >{\raggedright\arraybackslash}p{(\linewidth - 12\tabcolsep) * \real{0.1429}}
  >{\raggedright\arraybackslash}p{(\linewidth - 12\tabcolsep) * \real{0.1429}}
  >{\raggedright\arraybackslash}p{(\linewidth - 12\tabcolsep) * \real{0.1429}}@{}}
\toprule\noalign{}
\begin{minipage}[b]{\linewidth}\raggedright
Model
\end{minipage} & \begin{minipage}[b]{\linewidth}\raggedright
P(C\textbar C)
\end{minipage} & \begin{minipage}[b]{\linewidth}\raggedright
P(C\textbar D)
\end{minipage} & \begin{minipage}[b]{\linewidth}\raggedright
Reciprocity index
\end{minipage} & \begin{minipage}[b]{\linewidth}\raggedright
N after C
\end{minipage} & \begin{minipage}[b]{\linewidth}\raggedright
N after D
\end{minipage} & \begin{minipage}[b]{\linewidth}\raggedright
Archetype
\end{minipage} \\
\midrule\noalign{}
\endhead
\bottomrule\noalign{}
\endlastfoot
Claude Opus 4.6 & 92.8 & 4.5 & 88.3 & 469 & 179 & Strict reciprocator \\
Claude Sonnet 4.6 & 91.7 & 4.1 & 87.6 & 588 & 222 & Strict reciprocator \\
Claude Opus 4.5 & 90.4 & 4.1 & 86.3 & 467 & 172 & Strict reciprocator \\
Claude Sonnet 4.5 & 88.7 & 8.8 & 79.9 & 556 & 227 & Strict reciprocator \\
Gemini 3 Flash & 83.6 & 0.3 & 83.3 & 3,066 & 1,875 & Grudge-holder \\
Gemini 3 Pro & 77.5 & 2.8 & 74.7 & 368 & 181 & Grudge-holder \\
GPT-4.1 & 72.2 & 1.5 & 70.7 & 417 & 195 & Strict reciprocator \\
Qwen 3.5 Flash & 70.7 & 7.4 & 63.3 & 1,464 & 894 & Strict reciprocator \\
Claude Haiku 4.5 & 68.5 & 5.1 & 63.4 & 2,382 & 2,370 & Strict reciprocator \\
Claude Haiku 4.5 (Thinking) & 67.9 & 3.7 & 64.2 & 2,283 & 2,289 & Strict reciprocator \\
GPT-4o Mini & 67.2 & 18.3 & 48.9 & 2,225 & 1,726 & Moderate reciprocator \\
GPT-5.4 & 64.8 & 4.8 & 60.0 & 472 & 248 & Strict reciprocator \\
GPT-4.1 Mini & 62.9 & 5.6 & 57.3 & 2,007 & 1,881 & Strict reciprocator \\
Gemini 3.1 Pro & 62.8 & 0.4 & 62.4 & 261 & 243 & Grudge-holder \\
DeepSeek V3 & 54.8 & 8.8 & 46.0 & 1,345 & 1,607 & Moderate reciprocator \\
Gemini 2.5 Flash & 54.0 & 11.1 & 42.9 & 2,077 & 3,206 & Moderate reciprocator \\
DeepSeek R1 & 50.7 & 22.8 & 27.9 & 1,339 & 1,208 & Moderate reciprocator \\
Gemini 2.5 Flash (Thinking) & 50.1 & 6.6 & 43.5 & 2,133 & 2,943 & Moderate reciprocator \\
Ministral 14B & 48.3 & 24.1 & 24.2 & 1,245 & 1,392 & Moderate reciprocator \\
GPT-4.1 Nano & 35.0 & 5.9 & 29.1 & 1,339 & 2,855 & Weak reciprocator \\
GPT-5.3 & 34.0 & 1.8 & 32.2 & 430 & 380 & Weak reciprocator \\
LLaMA 3.3 70B & 14.3 & 22.2 & -7.9 & 1,157 & 1,957 & Anti-reciprocator \\
Gemini 2.0 Flash & 12.5 & 6.1 & 6.4 & 1,503 & 4,041 & Unconditional defector \\
GPT-5 Mini & 10.5 & 2.8 & 7.7 & 1,195 & 2,531 & Unconditional defector \\
GPT-5 Nano & 2.1 & 1.0 & 1.1 & 1,043 & 2,359 & Unconditional defector \\
\end{longtable}
}

All percentages. Reciprocity index = P(C\textbar C) - P(C\textbar D). N after C and N after D are the number of round pairs in each condition.

\newpage

\subsection{Supplementary Table 5: Provider-level summary statistics}\label{supplementary-table-5-provider-level-summary-statistics}

{\def\LTcaptype{none} % do not increment counter
\begin{longtable}[]{@{}
  >{\raggedright\arraybackslash}p{(\linewidth - 12\tabcolsep) * \real{0.1042}}
  >{\raggedright\arraybackslash}p{(\linewidth - 12\tabcolsep) * \real{0.0833}}
  >{\raggedright\arraybackslash}p{(\linewidth - 12\tabcolsep) * \real{0.1771}}
  >{\raggedright\arraybackslash}p{(\linewidth - 12\tabcolsep) * \real{0.1875}}
  >{\raggedright\arraybackslash}p{(\linewidth - 12\tabcolsep) * \real{0.1667}}
  >{\raggedright\arraybackslash}p{(\linewidth - 12\tabcolsep) * \real{0.1146}}
  >{\raggedright\arraybackslash}p{(\linewidth - 12\tabcolsep) * \real{0.1667}}@{}}
\toprule\noalign{}
\begin{minipage}[b]{\linewidth}\raggedright
Provider
\end{minipage} & \begin{minipage}[b]{\linewidth}\raggedright
Models
\end{minipage} & \begin{minipage}[b]{\linewidth}\raggedright
Cooperation mean
\end{minipage} & \begin{minipage}[b]{\linewidth}\raggedright
Cooperation range
\end{minipage} & \begin{minipage}[b]{\linewidth}\raggedright
Coordination CV
\end{minipage} & \begin{minipage}[b]{\linewidth}\raggedright
Trust mean
\end{minipage} & \begin{minipage}[b]{\linewidth}\raggedright
Silhouette mean
\end{minipage} \\
\midrule\noalign{}
\endhead
\bottomrule\noalign{}
\endlastfoot
Anthropic & 6 & 55.2 & 39.3--71.5 & 0.02 & 0.57 & 0.48 \\
OpenAI & 8 & 28.3 & 1.5--53.7 & 0.02 & 0.40 & -0.60 \\
Google & 6 & 32.2 & 8.3--56.8 & 0.07 & 0.44 & -- \\
DeepSeek & 2 & 33.8 & 31.0--36.6 & 0.04 & 0.41 & -- \\
Meta & 1 & 19.9 & -- & -- & 0.59 & -- \\
Mistral & 1 & 37.5 & -- & -- & 0.50 & -- \\
Alibaba & 1 & 49.4 & -- & -- & 0.45 & -- \\
\end{longtable}
}

CV = coefficient of variation. Silhouette scores computed from the hodoscope behavioural embedding space with provider identity as the cluster label.

\newpage

\subsection{Supplementary Table 6: Lexical feature frequencies}\label{supplementary-table-6-lexical-feature-frequencies}

Selected lexical features (per 1,000 words of reasoning text) for representative models. Full data available in the OSF repository.

{\def\LTcaptype{none} % do not increment counter
\begin{longtable}[]{@{}lllll@{}}
\toprule\noalign{}
Term & Claude Opus 4.6 & GPT-5 Nano & Gemini 3 Pro & LLaMA 3.3 70B \\
\midrule\noalign{}
\endhead
\bottomrule\noalign{}
\endlastfoot
cooperate & 8.2 & 1.1 & 6.4 & 2.3 \\
defect & 3.4 & 0.8 & 5.1 & 1.9 \\
fairness & 4.7 & 0.2 & 1.4 & 0.5 \\
trust & 3.9 & 0.3 & 3.8 & 1.1 \\
optimal & 1.2 & 3.1 & 1.8 & 2.4 \\
equilibrium & 0.9 & 2.7 & 1.5 & 0.6 \\
exploit & 0.4 & 0.9 & 1.2 & 0.3 \\
mutual benefit & 3.2 & 0.1 & 0.8 & 0.2 \\
\end{longtable}
}

Note: Anthropic models reference ``fairness'' and ``mutual benefit'' at rates two to three times higher than models from other providers, consistent with the constitutional AI training approach. GPT-5 Nano references ``optimal'' and ``equilibrium'' at elevated rates, consistent with its near-zero cooperation and unconditional defection pattern.

\newpage

\subsection{Supplementary Figures}\label{supplementary-figures}

\pandocbounded{\includegraphics[keepaspectratio]{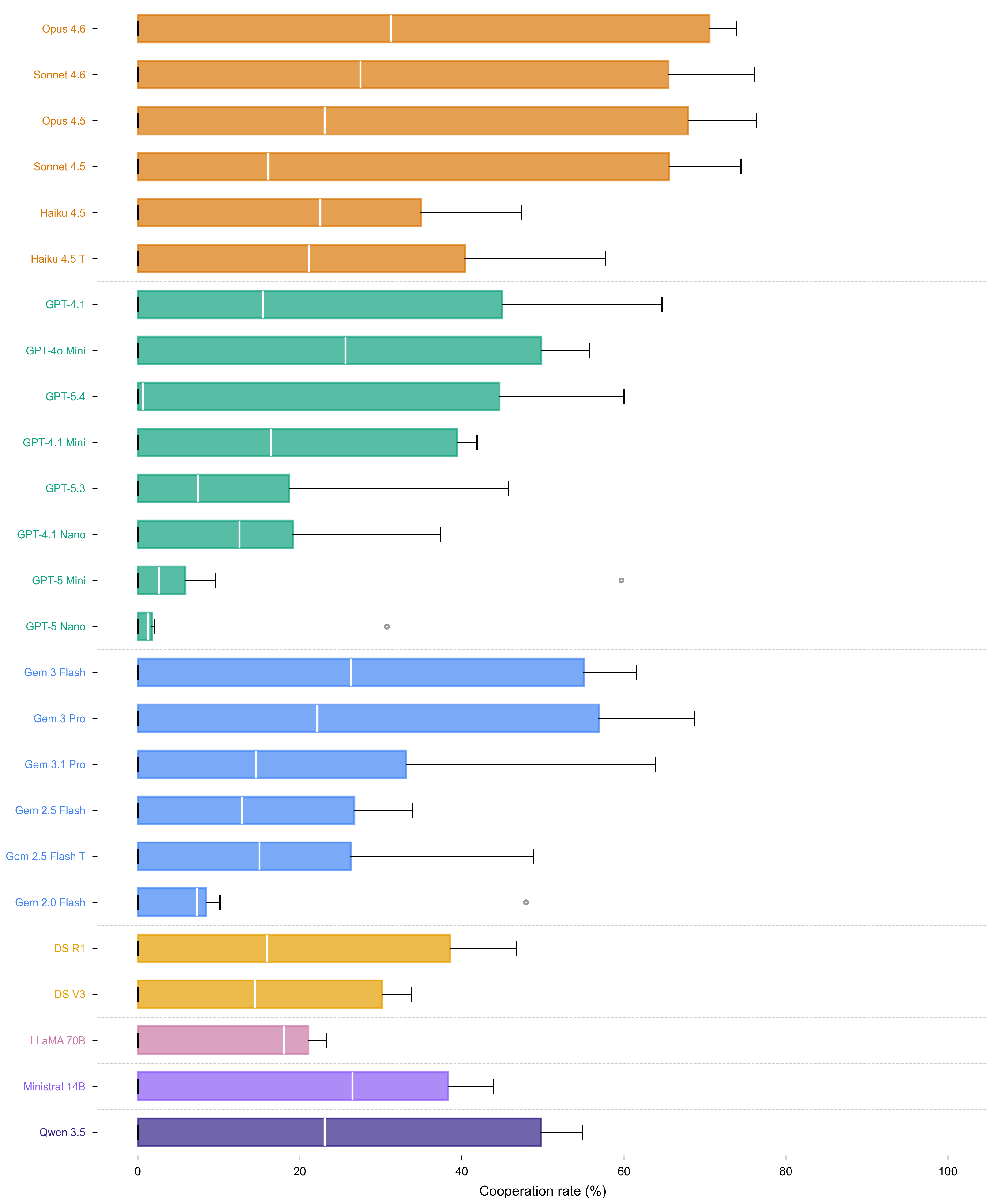}}

\textbf{Supplementary Figure 1 \textbar{} Cooperation rate distributions by game category.} Box plots of cooperation rate across the 10 cooperation games for each of the 25 models, grouped by provider (Anthropic, blue; OpenAI, vermillion; Google, teal; DeepSeek, amber; Meta, pink; Mistral, sky blue; Alibaba, indigo). Games include four prisoner's dilemma variants (canonical, harsh, medium, mild), three public goods games (high, medium, low MPCR), commons dilemma, diner's dilemma, and El Farol bar. Within-model distributions reveal whether cooperation is consistent across game contexts or driven by specific game structures. Anthropic frontier models show high medians with wide interquartile ranges, while unconditional defectors (GPT-5 Nano, GPT-5 Mini, Gemini 2.0 Flash) show compressed distributions near zero.

\pandocbounded{\includegraphics[keepaspectratio]{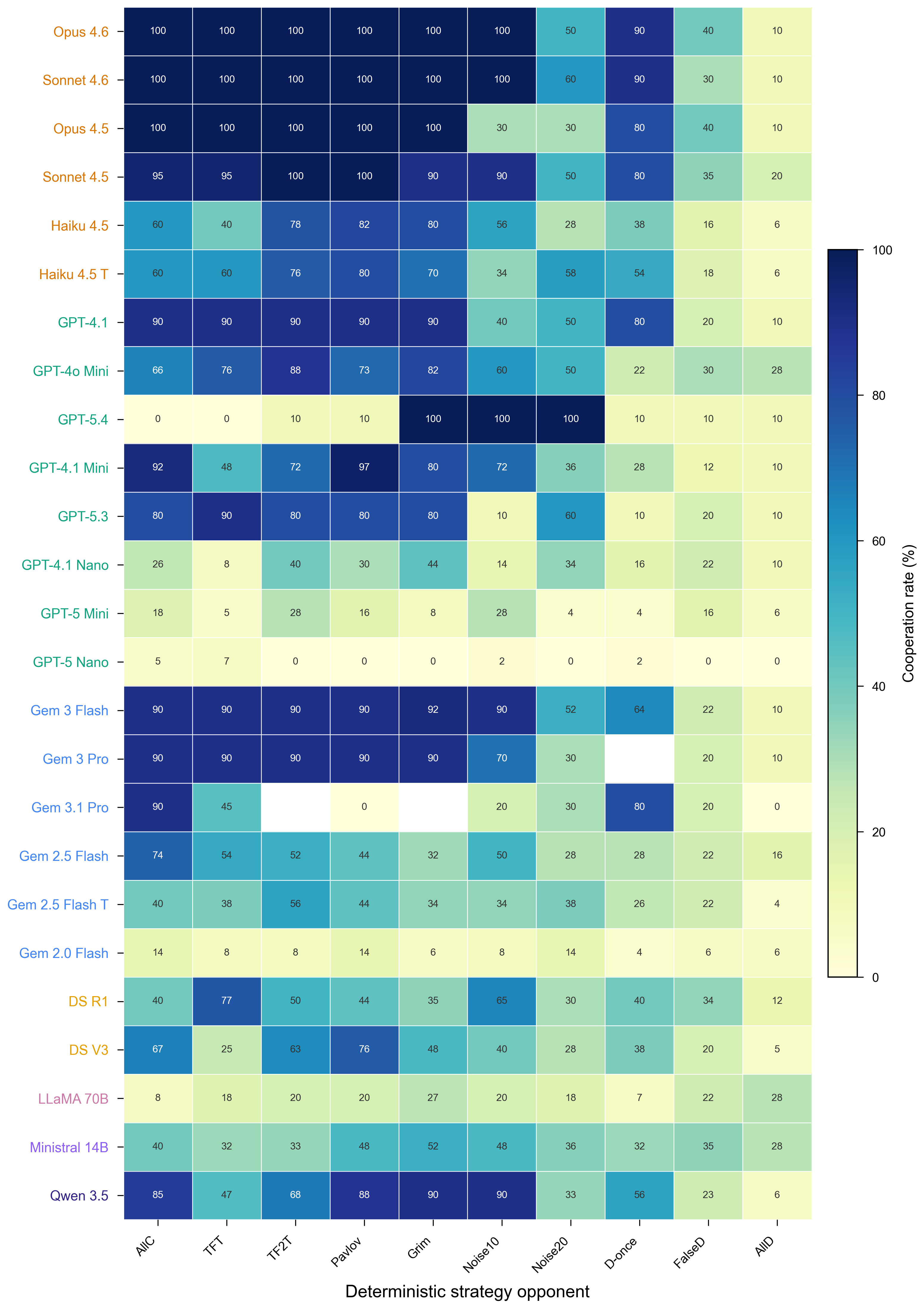}}

\textbf{Supplementary Figure 2 \textbar{} Strategy-play response profiles.} Heatmap of model cooperation rates when playing against each of the 13 deterministic strategy opponents in the canonical prisoner's dilemma. Rows: 25 models, ordered by provider. Columns: 13 strategies, ordered from always cooperate to always defect. Cell values: mean cooperation rate (\%) across all rounds. The tit-for-tat column differentiates reciprocating models (high cooperation) from fixed-strategy models (invariant to opponent). Cooperative models (Anthropic frontier) show strong opponent sensitivity, cooperating at 100\% against always-cooperate but dropping sharply against always-defect. Unconditional defectors show uniformly low cooperation regardless of opponent strategy.

\pandocbounded{\includegraphics[keepaspectratio]{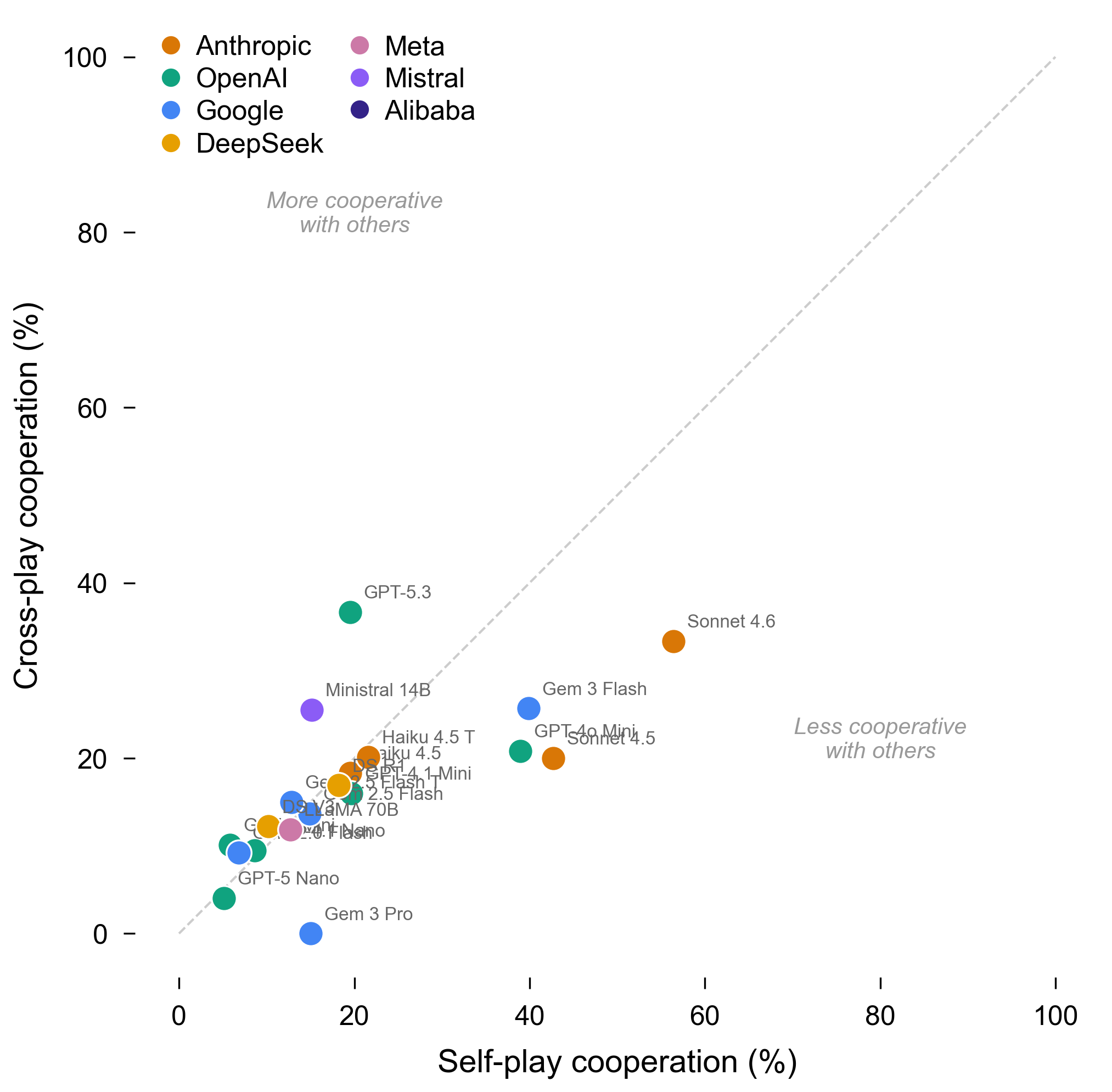}}

\textbf{Supplementary Figure 3 \textbar{} Cross-play interaction effects.} For the 16 Design F models, scatter plot of self-play cooperation rate (x-axis) versus cross-play cooperation rate averaged across all other opponents (y-axis). The dashed diagonal represents equal self-play and cross-play cooperation. Models above the diagonal cooperate more with others than with themselves, while models below cooperate less. Most models cluster near or below the diagonal, indicating that self-play cooperation rates provide an upper bound on cross-play cooperation for most models.

\pandocbounded{\includegraphics[keepaspectratio]{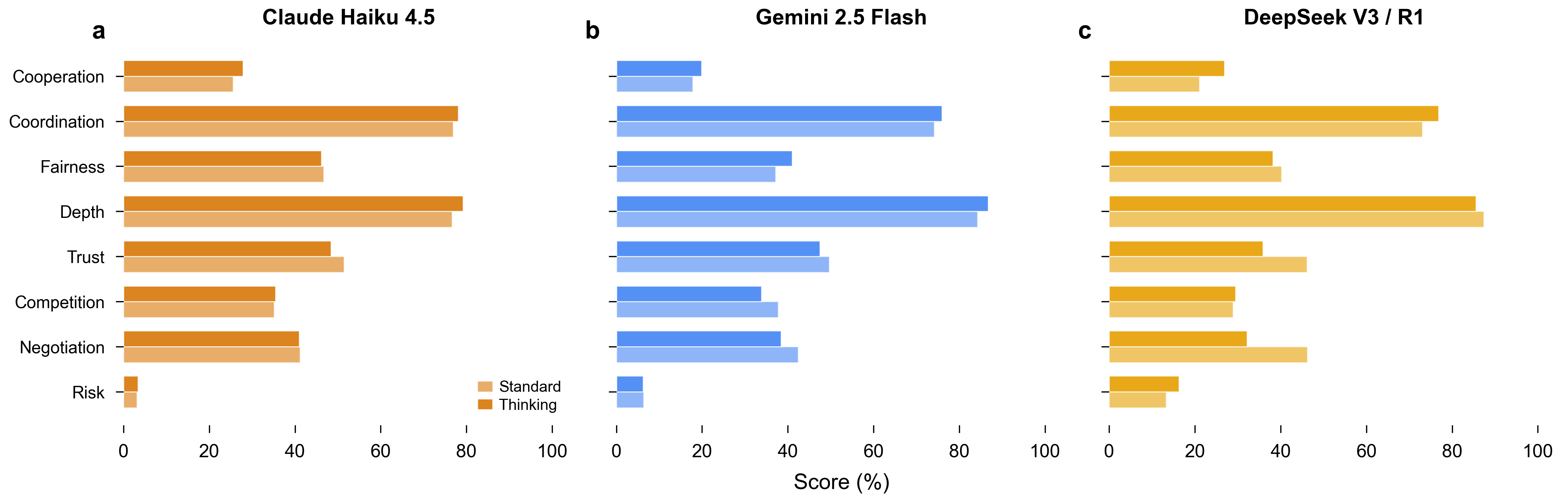}}

\textbf{Supplementary Figure 4 \textbar{} Thinking model comparison.} Paired horizontal bar charts comparing standard (lighter shade) and thinking (darker shade) model variants across all eight behavioural dimensions for three thinking model pairs: (\textbf{a}) Claude Haiku 4.5 versus Claude Haiku 4.5 (Thinking), (\textbf{b}) Gemini 2.5 Flash versus Gemini 2.5 Flash (Thinking), and (\textbf{c}) DeepSeek V3 versus DeepSeek R1 (Thinking). Thinking modes increase strategic depth scores (beauty contest and centipede performance) across all three pairs, while effects on cooperation vary by provider. The DeepSeek pair shows the largest thinking-mode effect on cooperation (R1 more cooperative than V3), while the Gemini pair shows minimal cooperation differences between standard and thinking modes.


\begin{CSLReferences}{0}{0}
\bibitem[\citeproctext]{ref-bansal2025magentic}
\CSLLeftMargin{1. }%
\CSLRightInline{{Bansal, T. \emph{et al.}} Magentic {M}arketplace: A benchmark for two-sided agent interactions in e-commerce. \url{https://arxiv.org/abs/2510.25779} (2025).}

\bibitem[\citeproctext]{ref-allouah2025what}
\CSLLeftMargin{2. }%
\CSLRightInline{{Allouah, A. \emph{et al.}} What is your {AI} agent buying? {A} first look at shopping agent behavior. \url{https://arxiv.org/abs/2508.02630} (2025).}

\bibitem[\citeproctext]{ref-kobis2025delegation}
\CSLLeftMargin{3. }%
\CSLRightInline{{Köbis, N. C. \emph{et al.}} Delegation to AI increases dishonesty. \emph{Nature} \url{https://doi.org/10.1038/s41586-025-09505-x} (2025) doi:\href{https://doi.org/10.1038/s41586-025-09505-x}{10.1038/s41586-025-09505-x}.}

\bibitem[\citeproctext]{ref-betley2026misalignment}
\CSLLeftMargin{4. }%
\CSLRightInline{{Betley, A. \emph{et al.}} Training on narrow tasks can produce broadly misaligned AI models. \emph{Nature} \url{https://doi.org/10.1038/s41586-025-09937-5} (2026) doi:\href{https://doi.org/10.1038/s41586-025-09937-5}{10.1038/s41586-025-09937-5}.}

\bibitem[\citeproctext]{ref-rahwan2019machine}
\CSLLeftMargin{5. }%
\CSLRightInline{{Rahwan, I. \emph{et al.}} \href{https://doi.org/10.1038/s41586-019-1138-y}{Machine behaviour}. \emph{Nature} \textbf{568}, 477--486 (2019).}

\bibitem[\citeproctext]{ref-horton2023large}
\CSLLeftMargin{6. }%
\CSLRightInline{Horton, J. J. Large language models as simulated economic agents: What can we learn from homo silicus? \emph{NBER Working Paper} \url{https://doi.org/10.3386/w31122} (2023) doi:\href{https://doi.org/10.3386/w31122}{10.3386/w31122}.}

\bibitem[\citeproctext]{ref-camerer2003behavioral}
\CSLLeftMargin{7. }%
\CSLRightInline{Camerer, C. F. \emph{Behavioral Game Theory: Experiments in Strategic Interaction}. (Princeton University Press, 2003).}

\bibitem[\citeproctext]{ref-fehr1999theory}
\CSLLeftMargin{8. }%
\CSLRightInline{Fehr, E. \& Schmidt, K. M. \href{https://doi.org/10.1162/003355399556151}{A theory of fairness, competition, and cooperation}. \emph{Quarterly Journal of Economics} \textbf{114}, 817--868 (1999).}

\bibitem[\citeproctext]{ref-guth1982experimental}
\CSLLeftMargin{9. }%
\CSLRightInline{Güth, W., Schmittberger, R. \& Schwarze, B. \href{https://doi.org/10.1016/0167-2681(82)90011-7}{An experimental analysis of ultimatum bargaining}. \emph{Journal of Economic Behavior \& Organization} \textbf{3}, 367--388 (1982).}

\bibitem[\citeproctext]{ref-rand2013human}
\CSLLeftMargin{10. }%
\CSLRightInline{Rand, D. G. \& Nowak, M. A. \href{https://doi.org/10.1016/j.tics.2013.06.003}{Human cooperation}. \emph{Trends in Cognitive Sciences} \textbf{17}, 413--425 (2013).}

\bibitem[\citeproctext]{ref-nowak2006evolutionary}
\CSLLeftMargin{11. }%
\CSLRightInline{Nowak, M. A. \emph{Evolutionary Dynamics: Exploring the Equations of Life}. (Harvard University Press, 2006).}

\bibitem[\citeproctext]{ref-brookins2024playing}
\CSLLeftMargin{12. }%
\CSLRightInline{Brookins, P. \& DeBacker, J. M. Playing games with {GPT}: What can we learn about a large language model from canonical strategic games? \emph{Economics Bulletin} \url{https://doi.org/10.2139/ssrn.4493398} (2024) doi:\href{https://doi.org/10.2139/ssrn.4493398}{10.2139/ssrn.4493398}.}

\bibitem[\citeproctext]{ref-brookins2024strategic}
\CSLLeftMargin{13. }%
\CSLRightInline{Brookins, P. \& DeBacker, J. M. \href{https://doi.org/10.1038/s41598-024-69032-z}{Strategic behavior of large language models: Game structure vs. Contextual framing}. \emph{Scientific Reports} \textbf{14}, 18832 (2024).}

\bibitem[\citeproctext]{ref-akata2025playing}
\CSLLeftMargin{14. }%
\CSLRightInline{Akata, E. \emph{et al.} Playing repeated games with large language models. \emph{Nature Human Behaviour} \url{https://doi.org/10.1038/s41562-025-02172-y} (2025) doi:\href{https://doi.org/10.1038/s41562-025-02172-y}{10.1038/s41562-025-02172-y}.}

\bibitem[\citeproctext]{ref-fan2024canllm}
\CSLLeftMargin{15. }%
\CSLRightInline{Fan, C., Chen, J., Jin, Y. \& He, H. \href{https://doi.org/10.1609/aaai.v38i16.29751}{Can large language models serve as rational players in game theory? {A} systematic analysis}. \emph{Proceedings of the AAAI Conference on Artificial Intelligence} \textbf{38}, 17960--17967 (2024).}

\bibitem[\citeproctext]{ref-fontana2025nicer}
\CSLLeftMargin{16. }%
\CSLRightInline{Fontana, N., Pierri, F. \& Aiello, L. M. \href{https://arxiv.org/abs/2406.13605}{Nicer than humans: How do large language models behave in the {Prisoner's Dilemma}?} in \emph{ICWSM} (2025).}

\bibitem[\citeproctext]{ref-huang2025gamabench}
\CSLLeftMargin{17. }%
\CSLRightInline{{Huang, D. \emph{et al.}} \href{https://arxiv.org/abs/2403.11807}{{GAMA-Bench}: Benchmarking {LLMs}' game-theoretic reasoning abilities}. in \emph{ICLR} (2025).}

\bibitem[\citeproctext]{ref-duan2025llm}
\CSLLeftMargin{18. }%
\CSLRightInline{{Duan, H. \emph{et al.}} \href{https://arxiv.org/abs/2502.20432}{{LLM} strategic reasoning via behavioral game theory}. in \emph{NeurIPS} (2025).}

\bibitem[\citeproctext]{ref-mao2025alympics}
\CSLLeftMargin{19. }%
\CSLRightInline{{Mao, S. \emph{et al.}} \href{https://arxiv.org/abs/2311.03220}{{ALYMPICS}: Language agents meet game theory}. in \emph{COLING} (2025).}

\bibitem[\citeproctext]{ref-guo2024gpt}
\CSLLeftMargin{20. }%
\CSLRightInline{Guo, F. {GPT} in game theory experiments. \emph{arXiv preprint arXiv:2305.05516} \url{https://arxiv.org/abs/2305.05516} (2024).}

\bibitem[\citeproctext]{ref-suzuki2024evolutionary}
\CSLLeftMargin{21. }%
\CSLRightInline{Suzuki, R. \& Arita, T. \href{https://doi.org/10.1038/s41598-024-55903-y}{An evolutionary model of personality traits related to cooperative behavior using a large language model}. \emph{Scientific Reports} \textbf{14}, 5989 (2024).}

\bibitem[\citeproctext]{ref-cheung2025cognitive}
\CSLLeftMargin{22. }%
\CSLRightInline{{Cheung, V. \emph{et al.}} Large language models show amplified cognitive biases in moral decision-making. \emph{Proceedings of the National Academy of Sciences} \url{https://doi.org/10.1073/pnas.2412015122} (2025) doi:\href{https://doi.org/10.1073/pnas.2412015122}{10.1073/pnas.2412015122}.}

\bibitem[\citeproctext]{ref-more2026stake}
\CSLLeftMargin{23. }%
\CSLRightInline{Various. More at stake: {LLM} cooperation in high-stakes games. \emph{arXiv preprint arXiv:2601.19082} \url{https://arxiv.org/abs/2601.19082} (2026).}

\bibitem[\citeproctext]{ref-fairgame2025}
\CSLLeftMargin{24. }%
\CSLRightInline{Various. {FAIRGAME}: A framework for assessing {LLM} fairness in game-theoretic settings. \emph{arXiv preprint arXiv:2512.07462} \url{https://arxiv.org/abs/2512.07462} (2025).}

\bibitem[\citeproctext]{ref-randomness2025llm}
\CSLLeftMargin{25. }%
\CSLRightInline{Various. Playing games with {LLMs}: Randomness and strategy. \emph{arXiv preprint arXiv:2503.02582} \url{https://arxiv.org/abs/2503.02582} (2025).}

\bibitem[\citeproctext]{ref-shin2024emergence}
\CSLLeftMargin{26. }%
\CSLRightInline{{Shin, D. \emph{et al.}} Emergence of strategic reasoning in large language models. \emph{arXiv preprint arXiv:2412.13013} \url{https://arxiv.org/abs/2412.13013} (2024).}

\bibitem[\citeproctext]{ref-calvano2020artificial}
\CSLLeftMargin{27. }%
\CSLRightInline{Calvano, E., Calzolari, G., Denicolò, V. \& Pastorello, S. \href{https://doi.org/10.1257/aer.20190623}{Artificial intelligence, algorithmic pricing, and collusion}. \emph{American Economic Review} \textbf{110}, 3267--3297 (2020).}

\bibitem[\citeproctext]{ref-calvano2021algorithmic}
\CSLLeftMargin{28. }%
\CSLRightInline{Calvano, E., Calzolari, G., Denicolò, V. \& Pastorello, S. \href{https://doi.org/10.1016/j.ijindorg.2021.102712}{Algorithmic collusion with imperfect monitoring}. \emph{International Journal of Industrial Organization} \textbf{79}, (2021).}

\bibitem[\citeproctext]{ref-ijcai2025survey}
\CSLLeftMargin{29. }%
\CSLRightInline{Various. Game theory meets large language models: A survey. \url{https://arxiv.org/abs/2502.09053} (2025).}

\bibitem[\citeproctext]{ref-gao2025scylla}
\CSLLeftMargin{30. }%
\CSLRightInline{{Gao, X. \emph{et al.}} Scylla ex machina: Failures of {LLMs} as human behavioral surrogates. \emph{Proceedings of the National Academy of Sciences} \url{https://doi.org/10.1073/pnas.2501660122} (2025) doi:\href{https://doi.org/10.1073/pnas.2501660122}{10.1073/pnas.2501660122}.}

\bibitem[\citeproctext]{ref-berg1995trust}
\CSLLeftMargin{31. }%
\CSLRightInline{Berg, J., Dickhaut, J. \& McCabe, K. \href{https://doi.org/10.1006/game.1995.1027}{Trust, reciprocity, and social history}. \emph{Games and Economic Behavior} \textbf{10}, 122--142 (1995).}

\bibitem[\citeproctext]{ref-nagel1995unraveling}
\CSLLeftMargin{32. }%
\CSLRightInline{Nagel, R. \href{https://www.jstor.org/stable/2950991}{Unraveling in guessing games: An experimental study}. \emph{American Economic Review} \textbf{85}, 1313--1326 (1995).}

\bibitem[\citeproctext]{ref-mei2024turing}
\CSLLeftMargin{33. }%
\CSLRightInline{Mei, Q., Xie, Y., Yuan, W. \& Jackson, M. O. \href{https://doi.org/10.1073/pnas.2313925121}{A {Turing} test of whether {AI} chatbot behavior is indistinguishable from human behavior}. \emph{Proceedings of the National Academy of Sciences} \textbf{121}, e2313925121 (2024).}

\bibitem[\citeproctext]{ref-serapiogarcia2025psychometric}
\CSLLeftMargin{34. }%
\CSLRightInline{Serapio-Garcı́a, G. \emph{et al.} \href{https://doi.org/10.1038/s42256-025-01115-6}{A psychometric framework for evaluating and shaping personality traits in large language models}. \emph{Nature Machine Intelligence} \textbf{7}, 1954--1968 (2025).}

\bibitem[\citeproctext]{ref-axelrod1984evolution}
\CSLLeftMargin{35. }%
\CSLRightInline{Axelrod, R. \emph{The Evolution of Cooperation}. (Basic Books, 1984).}

\bibitem[\citeproctext]{ref-nash1950bargaining}
\CSLLeftMargin{36. }%
\CSLRightInline{Nash, J. F. \href{https://doi.org/10.2307/1907266}{The bargaining problem}. \emph{Econometrica} \textbf{18}, 155--162 (1950).}

\bibitem[\citeproctext]{ref-mengel2018risk}
\CSLLeftMargin{37. }%
\CSLRightInline{Mengel, F. \href{https://doi.org/10.1111/ecoj.12548}{Risk and temptation: A meta-study on prisoner's dilemma games}. \emph{The Economic Journal} \textbf{128}, 3182--3209 (2018).}

\bibitem[\citeproctext]{ref-zelmer2003linear}
\CSLLeftMargin{38. }%
\CSLRightInline{Zelmer, J. \href{https://doi.org/10.1023/A:1026277420119}{Linear public goods experiments: A meta-analysis}. \emph{Experimental Economics} \textbf{6}, 299--310 (2003).}

\bibitem[\citeproctext]{ref-oosterbeek2004cultural}
\CSLLeftMargin{39. }%
\CSLRightInline{Oosterbeek, H., Sloof, R. \& Van de Kuilen, G. \href{https://doi.org/10.1023/B:EXEC.0000026978.14316.74}{Cultural differences in ultimatum game experiments: Evidence from a meta-analysis}. \emph{Experimental Economics} \textbf{7}, 171--188 (2004).}

\bibitem[\citeproctext]{ref-johnson2011trust}
\CSLLeftMargin{40. }%
\CSLRightInline{Johnson, N. D. \& Mislin, A. A. \href{https://doi.org/10.1016/j.joep.2011.05.007}{Trust games: A meta-analysis}. \emph{Journal of Economic Psychology} \textbf{32}, 865--889 (2011).}

\bibitem[\citeproctext]{ref-selten1978chain}
\CSLLeftMargin{41. }%
\CSLRightInline{Selten, R. \href{https://doi.org/10.1007/BF00131770}{The chain store paradox}. \emph{Theory and Decision} \textbf{9}, 127--159 (1978).}

\bibitem[\citeproctext]{ref-rosenthal1981games}
\CSLLeftMargin{42. }%
\CSLRightInline{Rosenthal, R. W. \href{https://doi.org/10.1016/0022-0531(81)90018-1}{Games of perfect information, predatory pricing and the chain-store paradox}. \emph{Journal of Economic Theory} \textbf{25}, 92--100 (1981).}

\bibitem[\citeproctext]{ref-embrey2018cooperation}
\CSLLeftMargin{43. }%
\CSLRightInline{Embrey, M., Frechette, G. R. \& Yuksel, S. \href{https://doi.org/10.1093/qje/qjx033}{Cooperation in the finitely repeated prisoner's dilemma}. \emph{The Quarterly Journal of Economics} \textbf{133}, 509--551 (2018).}

\bibitem[\citeproctext]{ref-dalbo2018determinants}
\CSLLeftMargin{44. }%
\CSLRightInline{Dal Bó, P. \& Fréchette, G. R. \href{https://doi.org/10.1257/jel.20160980}{On the determinants of cooperation in infinitely repeated games: A survey}. \emph{Journal of Economic Literature} \textbf{56}, 60--114 (2018).}

\bibitem[\citeproctext]{ref-dalbo2005cooperation}
\CSLLeftMargin{45. }%
\CSLRightInline{Dal Bó, P. \href{https://doi.org/10.1257/000282805775014434}{Cooperation under the shadow of the future: Experimental evidence from infinitely repeated games}. \emph{American Economic Review} \textbf{95}, 1591--1604 (2005).}

\bibitem[\citeproctext]{ref-kreps1982rational}
\CSLLeftMargin{46. }%
\CSLRightInline{Kreps, D. M., Milgrom, P., Roberts, J. \& Wilson, R. \href{https://doi.org/10.1016/0022-0531(82)90029-1}{Rational cooperation in the finitely repeated prisoners' dilemma}. \emph{Journal of Economic Theory} \textbf{27}, 245--252 (1982).}

\bibitem[\citeproctext]{ref-zhong2026hodoscope}
\CSLLeftMargin{47. }%
\CSLRightInline{Zhong, Z., Saxena, S. \& Raghunathan, A. Hodoscope: Unsupervised monitoring for AI misbehaviors. \emph{arXiv preprint arXiv:2604.11072} (2026).}

\bibitem[\citeproctext]{ref-bai2022constitutional}
\CSLLeftMargin{48. }%
\CSLRightInline{{Bai, Y. \emph{et al.}} Constitutional {AI}: Harmlessness from {AI} feedback. \emph{arXiv preprint arXiv:2212.08073} \url{https://arxiv.org/abs/2212.08073} (2022).}

\bibitem[\citeproctext]{ref-ouyang2022instructgpt}
\CSLLeftMargin{49. }%
\CSLRightInline{{Ouyang, L. \emph{et al.}} \href{https://arxiv.org/abs/2203.02155}{Training language models to follow instructions with human feedback}. \emph{Advances in Neural Information Processing Systems} \textbf{35}, 27730--27744 (2022).}

\bibitem[\citeproctext]{ref-emergent2025social}
\CSLLeftMargin{50. }%
\CSLRightInline{Various. Emergent social conventions in large language model populations. \emph{Science Advances} \url{https://doi.org/10.1126/sciadv.adu9368} (2025) doi:\href{https://doi.org/10.1126/sciadv.adu9368}{10.1126/sciadv.adu9368}.}

\bibitem[\citeproctext]{ref-llm2025replicate}
\CSLLeftMargin{51. }%
\CSLRightInline{Various. {LLMs} replicate human cooperation in social dilemmas. \emph{arXiv preprint arXiv:2511.04500} \url{https://arxiv.org/abs/2511.04500} (2025).}

\bibitem[\citeproctext]{ref-crandall2018cooperating}
\CSLLeftMargin{52. }%
\CSLRightInline{Crandall, J. W. \emph{et al.} \href{https://doi.org/10.1038/s41467-017-02597-8}{Cooperating with machines}. \emph{Nature Communications} \textbf{9}, 233 (2018).}

\bibitem[\citeproctext]{ref-hle2026}
\CSLLeftMargin{53. }%
\CSLRightInline{{Phan, L. \emph{et al.}} \href{https://doi.org/10.1038/s41586-025-09962-4}{A benchmark of expert-level academic questions to assess AI capabilities}. \emph{Nature} \textbf{649}, 1139--1146 (2026).}

\bibitem[\citeproctext]{ref-wang2025experimental}
\CSLLeftMargin{54. }%
\CSLRightInline{{Wang, Z. \emph{et al.}} When experimental economics meets large language models. \emph{arXiv preprint arXiv:2505.21371} \url{https://arxiv.org/abs/2505.21371} (2025).}

\bibitem[\citeproctext]{ref-gemp2025gamebot}
\CSLLeftMargin{55. }%
\CSLRightInline{{Gemp, I. \emph{et al.}} \href{https://arxiv.org/abs/2412.13602}{{GAMEBoT}: Transparent assessment of {LLM} reasoning in games}. in \emph{ACL} (2025).}

\bibitem[\citeproctext]{ref-bianchi2024negotiation}
\CSLLeftMargin{56. }%
\CSLRightInline{{Bianchi, F. \emph{et al.}} \href{https://arxiv.org/abs/2402.05863}{{NegotiationArena}: A benchmark for language model negotiation}. in \emph{ICML} (2024).}

\bibitem[\citeproctext]{ref-abdelnabi2024llm}
\CSLLeftMargin{57. }%
\CSLRightInline{{Abdelnabi, S. \emph{et al.}} \href{https://arxiv.org/abs/2309.17234}{{LLM}-deliberation: Evaluating {LLMs} with interactive multi-agent negotiation games}. in \emph{NeurIPS} (2024).}

\bibitem[\citeproctext]{ref-spence1973job}
\CSLLeftMargin{58. }%
\CSLRightInline{Spence, M. \href{https://doi.org/10.2307/1882010}{Job market signaling}. \emph{Quarterly Journal of Economics} \textbf{87}, 355--374 (1973).}

\bibitem[\citeproctext]{ref-crawford1982strategic}
\CSLLeftMargin{59. }%
\CSLRightInline{Crawford, V. P. \& Sobel, J. \href{https://doi.org/10.2307/1913390}{Strategic information transmission}. \emph{Econometrica} \textbf{50}, 1431--1451 (1982).}

\bibitem[\citeproctext]{ref-arthur1994inductive}
\CSLLeftMargin{60. }%
\CSLRightInline{Arthur, W. B. \href{https://www.jstor.org/stable/2117868}{Inductive reasoning and bounded rationality (the {El Farol} problem)}. \emph{American Economic Review} \textbf{84}, 406--411 (1994).}

\end{CSLReferences}
\end{document}